\documentclass [10pt,twocolumn]{IEEEtran}
\usepackage{algorithm,algorithmic,amsbsy,amsmath,amssymb,epsfig,bbm,mathrsfs,multirow,amsthm}
\usepackage[T1]{fontenc}
\usepackage[utf8]{inputenc}
\usepackage{amsmath}
\usepackage{algorithm}
\usepackage{array,multirow,graphicx}
\usepackage{color}
\usepackage{cite}
\usepackage{float}
\usepackage{makecell}
\usepackage[x11names,dvipsnames,table]{xcolor} %for use in color links
\usepackage{colortbl}
\usepackage{multirow}
\usepackage{lipsum}
\usepackage{subcaption}
\usepackage{graphicx} % 导言区引入
\usepackage{booktabs}
\usepackage{array}
\usepackage{tabularx} % 自动列宽
\usepackage[colorlinks=true,linkcolor=black,citecolor=colorlinks=true,linkcolor=black,citecolor=blue,urlcolor=blue,urlcolor=blue]{hyperref}

\definecolor{mygray}{gray}{0.6}
\definecolor{myblue}{rgb}{0.8,0.85,1} 
\usepackage{array}
\newcolumntype{L}[1]{>{\raggedright\let\newline\\\arraybackslash\hspace{0pt}}m{#1}}
\newcolumntype{C}[1]{>{\centering\let\newline\\\arraybackslash\hspace{0pt}}m{#1}}
\newcolumntype{R}[1]{>{\raggedleft\let\newline\\\arraybackslash\hspace{0pt}}m{#1}}

\usepackage{array, tabularx, boldline}
\usepackage{eurosym}
\usepackage{amstext} % for \text
\DeclareRobustCommand{\officialeuro}{%
  \ifmmode\expandafter\text\fi
  {\fontencoding{U}\fontfamily{eurosym}\selectfont e}}

\usepackage{tabularx, boldline}
\usepackage{cellspace}
\begin{document}
%\[\title{Pricing Models in Internet of Things: A Survey}
\title{\huge Transformer-Enhanced Reinforcement Learning: Fundamentals and Applications in Communication Networks}

\author{Nguyen Cong Luong, Shaohan Feng, \textit{Member, IEEE}, Nguyen Duc Hai, Zeping Sui, \textit{Member, IEEE}, Bo Ma, \textit{Member, IEEE}, Min Xu, Zhihao Dong, \textit{Student Member, IEEE}, Qiushi Zhao, \textit{Student Member, IEEE}, Nguyen Duc Duy Anh, Nguyen Quoc Khanh, Ngoc Hung Nguyen, Zitian Zhang, \textit{Member, IEEE}, and Jie Cao, \textit{Member, IEEE}
\vspace{-2em}

\thanks{Nguyen Cong Luong, Nguyen Quoc Khanh, and Nguyen Duc Duy Anh are with the Faculty of Computer Science, Phenikaa University, Hanoi 12116, Vietnam. (e-mail: \{luong.nguyencong, khanh.nguyenquoc,21011488\}@phenikaa-uni.edu.vn).}

\thanks{Nguyen Duc Hai is with the Faculty of Artificial Intelligence and Data Science, Phenikaa University, Duong Noi, Hanoi 12116, Vietnam. E-mail: hai.nguyenduc@phenikaa-uni.edu.vn}

\thanks{Shaohan Feng, Bo Ma, Qiushi Zhao, and Zitian Zhang are with the School of Information and Electronic Engineering (Sussex Artificial Intelligence Institute), Zhejiang Gongshang University, Hangzhou 310018,
China. (e-mail: \{feng\_shaohan, mabo, 25020090093, zitian.zhang\}@mail.zjgsu.edu.cn).}

\thanks{Zeping Sui is with the School of Computer Science and Electronics Engineering, University of Essex, Colchester CO4 3SQ, U.K. (e-mail: zepingsui@outlook.com).}

\thanks{Min Xu is with the School of Mathematics, Statistics and Mechanics, Beijing University of Technology, Beijing 100124, China. (e-mail: xm@bjut.edu.cn).}

\thanks{Zhihao Dong and Jie Cao are with the School of Information Science and Technology, Harbin Institute of Technology, Shenzhen 518055, China. (e-mail: \{caojhitsz, zhihaodong\}@ieee.org). Corresponding author: Jie Cao.}

\thanks{Ngoc Hung Nguyen is with the Department of Electrical and Information Technology, Faculty of Engineering (LTH), Lund University, Lund 22100, Sweden. (e-mail: ngoc\_hung.nguyen@eit.lth.se).}

}

\maketitle
%====================================================================
\begin{abstract}
Reinforcement Learning (RL) has long been a powerful solution to various problems in communication networks. However, traditional RL models still face with several limitations. Not only do they rely on large numbers of interactions with the environment, but they are also limited in terms of modeling long-term relationships and tackling partial observability. In recent years, the Transformer model has demonstrated the ability to enhance RL models, allowing them to overcome these issues. Particularly, the self-attention mechanism within the Transformer enables efficient modeling of long-range dependencies and global correlations, as well as accelerates training processes and handles heterogeneous data modalities. In this paper, we present a comprehensive survey of Transformer-based RL algorithms and their applications in communication networks. Specifically, the paper provides the mathematical background of RL and Transformer architectures, along with insights into key issues such as resource allocation, computation offloading, routing, and trajectory control, and network security. We conclude the paper by discussing challenges, open issues, and notable future research directions, including Transformer-enhanced DRL algorithms for semantic communication and network optimization. 
\end{abstract}

%main

%==============================================
\section{Introduction}
%==============================================
Reinforcement Learning (RL) is one of the three grand paradigms of machine learning (ML) alongside supervised learning and unsupervised learning, providing a rigorous mathematical framework for sequential decision-making and interactions between an agent and its environment. Thus, RL has found applications in several important domains, including autonomous driving \cite{kiran2021deep}, robotics \cite{garaffa2021reinforcement} and recommendation systems \cite{afsar2022reinforcement}. In the areas of communication and networking, RL has also been utilized as a powerful tool. As the sixth generation of mobile communication systems (6G) is undergoing extensive development, RL has been extensively applied to a wide range of wireless networking tasks, including resource allocation and power control \cite{alwarafy2021deep}, user scheduling \cite{betalo2023multi} and data sensing and collection \cite{bai2023toward}. The model-free nature of RL makes it particularly attractive for wireless systems such as Internet of Things (IoT), Heterogeneous Networks (HetNets) and Unmanned Aerial Vehicle (UAV) networks, where accurate mathematical models are often unavailable due to nonlinearity, uncertainty, and heterogeneity on a large scale \cite{luong2019applications}. 

Despite these advantages, traditional RL and deep RL (DRL) approaches still suffer from several limitations when utilized for wireless communications. First, many RL algorithms require a large number of interactions with the environment to converge, which is rather costly and impractical in real-world wireless environments \cite{agarwal2023Transformers}. Second, standard DRL architectures, which are based on Recurrent Neural Networks (RNNs) or Convolutional Neural Networks (CNNs) have limited capability in modeling long-term temporal dependencies and global context \cite{chen2022transdreamer}, both of which are highly crucial in dynamic wireless channels and varying network topologies. Third, RL agents often face challenges in partial observability (i.e. the full information of the environment is not available) and poor generalization \cite{esslinger2022deep} across different network settings. These issues can lead to unstable training, suboptimal performance, and limited scalability in large or highly dynamic wireless systems.

Recently, Transformers \cite{vaswani2017attention} have emerged as a promising solution to these challenges. Originally introduced as a sequence modeling method in natural language processing (NLP), Transformers utilize the self-attention mechanism, which allows them to capture long-range dependencies and global correlations, as well as flexibly model complex sequences with varying lengths. Thus, Transformers have been proposed to solve various issues of the traditional RL approaches for communication systems. First, Transformer-based DRL can address resource allocation problems in communication networks by leveraging the ability of the Transformer to capture long-term temporal correlations, multi-agent interactions, and structured decision dependencies that occur in practical resource allocation scenarios such as power control \cite{che2025autonomous} and bandwidth allocation \cite{ma2025dsaf}. Second, Transformer-based RL has been increasingly utilized as a solution to computation offloading problems due to its ability to model high-dimensional system states, handle heterogeneous task requirements and dynamic interactions among computational, communication and network resources \cite{gholipour2023tpto, han2025transformer}. Third, Transformer architectures have been incorporated into DRL frameworks to tackle problems of routing and trajectory decision making in large-scale wireless networks, as they are capable of capturing long-range dependencies, dynamic topologies, and high-dimensional action representations resulting from mobility factors, resource limits, and multi-agent cooperation \cite{abdelkader2025perception, xu2025transformer}. Fourth, Transformer-based RL has shown its effectiveness for network security by enhancing adaptability to dynamic and evolving attacks, as well as handling partial observability in multi-agent and large-scale network security systems \cite{asemian2025anti,ergu2025radar}. 

There have been some relevant surveys and tutorials regarding Transformers and DRL. However, none of them provide a comprehensive view of how Transformers can be integrated into RL for communication systems. Particularly, the work in \cite{rjoub2024enhancing} introduced a new Transformer-based DRL framework that leverages the Transformer's self-attention mechanism to process and interpret the heterogeneous and high-dimensional data of IoT devices, significantly enhancing the state representation for RL agents. In contrast, the authors in \cite{stenhammar2024comparison} compared the performance of the Transformer with other NN architectures in the task of time-domain channel prediction under the standardized 3rd generation partnership project (3GPP) tapped delay line (TDL)-A model. A general framework for AI-enhanced channel prediction was presented by the authors in \cite{kim2025machine}, which also provided insights into the integration of Transformers into this domain. Meanwhile, the work in \cite{sun2025generative} focused on how DRL can be enhanced by Generative AI (GenAI), including Transformer-based DRL, summarizing the advantages and disadvantages of GenAI-enhanced DRL. The authors also discussed a case study comparing several GenAI-enhanced DRL models in near-field communication, including a Transformer-based DRL model. Nevertheless, these works all lack the specialization in summarizing the applications of Transformer-based DRL in communication networks. 

This motivates us to conduct an extensive review of how Transformers are integrated into DRL to address the problems in communication networks. The contributions of this survey are summarized as follows: 
\begin{itemize}
    \item We provide an extensive tutorial on RL and Transformers. Specifically, we elaborate on the mathematical background of RL and Transformer architectures. Moreover, we demonstrate how different types of Transformer can be utilized to enhance RL methods to solve problems in communication networks. 
    \item We review and discuss various applications of Transformer in RL for resource allocation in communication networks. Specifically, we discuss how Transformer-based RL approaches tackle the issues of radio and transmission resource allocation, network access and connectivity control, virtual network function placement, content-aware rate adaptation, and joint computation-communication resource allocation. 
    \item We survey and analyze up-to-date Transformer-based RL methods for computation offloading in communication networks. Specifically, we focus on notable problems such as computation resource allocation, joint computation-communication resource allocation, mobility-aware \& dynamic offloading, AI and GenAI service/large-model-oriented offloading, etc. 
    \item We review and discuss how Transformer-based RL is utilized for routing and trajectory control. Specifically, we discuss Transformer-based DRL methods for long-horizon and multi-agent routing, and trajectory planning, as well as sequence modeling, risk-aware routing, and trajectory control. 
    \item We investigate recent advances in Transformer-based RL for network security, including defenses against several types of attacks such as jamming, data poisoning and injection, adversarial attacks, physical layers, and sensing attacks. 
    \item Finally, we discuss key challenges and open issues, as well as promising research directions for Transformer-based RL in communication networks.
\end{itemize}
The remainder of this paper is organized as follows. Section \ref{sec:funda} provides the fundamental knowledge of Transformers and RL. Section \ref{sec:resource} discusses the applications of Transformer-based RL for resource allocation in communication networks. Section \ref{sec:offloading} reviews the applications of Transformer in RL for computation offloading in communication networks. Section \ref{sec:routing} provides reviews of Transformer-DRL methods for routing and trajectory control in communication networks. Section \ref{sec:security} discusses the integration of Transformers into RL for communication network security issues. Section \ref{sec:conclusion} highlights key challenges and future research directions, as well as concludes this paper. 

%==============================================
\section{Fundamentals of Transformer for DRL}
\label{sec:funda}
%==============================================

In this section, we provide insights into the basic concepts of RL and Transformers, which form the preliminaries of this survey. 

\subsection{Reinforcement Learning}
Formally, RL can be defined by a $T$-timestep Markov decision process (MDP), which is represented by a tuple $\langle \mathcal{S}, \mathcal{A}, P, r, \gamma, \rho_0 \rangle$, where $\mathcal{S}$ is the state space, $\mathcal{A}$ is the action space, $P(s_{t + 1}|s_t, a_t)$ is the transition probability from state $s_t$ to state $s_{t + 1}$ with action $a_t$, $r(s_t, a_t)$ is the reward function for taking action $a_t$ at state $s_t$, $\gamma \in (0, 1)$ is the discount factor, and $\rho_0$ is the distribution of initial states. With an MDP, a trajectory is defined as the set of experiences of the agent, denoted by $\Theta = \{(s_t, a_t, r_t)\}^{T}_{t = 1}$. The typical goal is to learn a policy $\pi(a|s)$ to maximize the expected cumulative return across all possible trajectories, which is formulated as
\begin{equation}
    \label{return}
    J(\pi) = \mathbb{E}_{\pi, P, \rho_0}\bigg[ \sum^T_{t = 1}\gamma^t r(s_t, a_t)\bigg]. 
\end{equation}
To find the optimal policy, RL methods typically estimate three functions: the action-value function $Q_\pi(s_t, a_t)$, the state-value function $V_\pi(s_t)$ and the advantage function $A_\pi(s_t, a_t)$, which is the difference between $Q_\pi(s_t, a_t)$ and $V_\pi(s_t)$. The value function $V_\pi(s_t)$ measures the expected reward when the agent starts at $s_t$ with policy $\pi$, while the action-value function $Q_\pi(s_t, a_t)$ maps the state-action pair to the same total reward value. The advantage function $A_\pi(s_t, a_t)$ is typically utilized by RL algorithms to lower the variance that directly applying $Q_\pi(s_t, a_t)$ creates \cite{wang2016dueling}. As the agent cannot observe the entire state, this is known as partially observable MDPs (POMDPs). Therein, the agent can only access a local observation $o_t \in \mathcal{O}$ in which $p(o_t)$ depends on $s_t$ and $a_{t - 1}$ \cite{hu2024transforming}. 
%POMDPs have been applied in various wireless tasks, such as communication scheduling \cite{jiao2017partially} and handoff management \cite{ammar2024handoffs}. 

% Next, we discuss the primary categories of RL, as well as their advantages and disadvantages. 

\textit{1) Model-based versus model-free:} The difference between these two families of RL algorithms lies in the learning of a model of the environment. Specifically, model-based RL focuses on learning a transition function $p(s_{t + 1}, r_t| s_t, a_t)$ using transitions $(s_{t}, a_t, r_t, s_{t + 1})$, which can be acquired by interacting with the environment. The learned model can be directly applied for planning \cite{schrittwieser2020mastering} or synthesizing trajectories for RL datasets \cite{hafner2020mastering}. This approach can generally achieve high performance using relatively few interactions with the environment \cite{mohanty2021measuring}.  However, learning the model is a computationally expensive task, especially in large environments or POMDPs where a state representation needs to be created first \cite{li2023survey}. In contrast, model-free RL directly learns the policy through interactions with the environment. Thus, it is unable to model transition dynamics, which results in slower convergence and worse sample efficiency compared to model-based RL methods. Nevertheless, model-free RL algorithms are more robust in complex or noisy environments due to their high adaptability \cite{lillicrap2015continuous} and are also less computationally demanding, as the transition model is not needed. 

% Specifically, off-policy RL algorithms use two policies, a behavior policy and a target policy. The behavior policy generates trajectory information, which is then used to update the value functions for improving the target policy. As for on-policy methods, one policy is used for both transition generation and optimization. 

%  For example, in SARSA \cite{sutton1998reinforcement}, tuples in the form of $(s_t, a_t, r_t, s_{t + 1}, a_{t + 1})$ are collected using the policy $\pi$ to estimate the expected return $Q_\pi(s_t, a_t)$. This Q-function is updated as
%     \begin{align}\label{Q_update}
%     &Q_\pi(s_t, a_t)\nonumber\\
%     = & Q_\pi(s_t, a_t) + \alpha \bigg[r_t + \gamma Q_\pi(s_{t + 1}, a_{t + 1})- Q_\pi(s_t, a_t)\bigg],
%     \end{align}
% where $a \in (0, 1]$ is the learning rate. In contrast, Q-learning \cite{watkins1992q}, a typical off-policy algorithm, updates the Q-value via
%     \begin{align}\label{Q_update_off}
%     &Q_\pi(s_t, a_t)\nonumber\\
%     = & Q_\pi(s_t, a_t) + \alpha \bigg[
%  \big(r_t + \gamma \cdot \max_a Q_\pi(s_{t + 1}, a)\big) - Q_\pi(s_t, a_t)\bigg],
%     \end{align}
% where $\max_a Q_\pi(s_{t + 1}, a)$ is the maximum reward that can be obtained from state $s_{t + 1}$.

\textit{2) On-policy versus off-policy:} On-policy RL methods use the current policy to update the value functions, while off-policy RL utilizes two different policies to gather trajectory information and estimate the expected return. The usage of only one policy makes on-policy RL algorithms, while easy to implement and stable, sample-inefficient \cite{agarwal2023Transformers}. Moreover, they cannot explore the environments flexibly, resulting in slower convergence and sub-optimal final policies. Meanwhile, off-policy RL algorithms are more effective for large action spaces, possess a faster learning process and high sample efficiency as they can re-use past experience \cite{sutton1998reinforcement}. 

% \begin{table}[t]
% \caption{Comparison between on-policy and off-policy RL}
% \centering
% \renewcommand{\arraystretch}{1.2}
% \begin{tabular}{|l|l|l|}
% \hline
% \makecell[c]{\textbf{Aspect}}
% & \textbf{On-Policy} 
% & \textbf{Off-Policy} \\ \hline

% \makecell[l]{Data-generating\\ policy}
% & \makecell[l]{Same as target\\ policy}
% & \makecell[l]{Different from target \\ policy} \\ \hline

% \makecell[l]{Policy being\\ improved}
% & \makecell[l]{Current policy\\ ($\pi$)}
% & \makecell[l]{Target policy ($\pi$),\\ data collected from a\\ behavior policy ($\hat{\pi}$)} \\ \hline

% \makecell[l]{Learning from \\ experience}
% & No
% & Yes \\ \hline

% \makecell[l]{Sample efficiency}
% & Lower
% & Higher \\ \hline

% Stability
% & Higher
% & \makecell[l]{Lower without additional\\ stabilization mechanisms} \\ \hline
% \end{tabular}
% \label{tab:on_vs_off_policy}
% \end{table}

\textit{3) Offline RL:} This new discipline of RL has gained much interest in the last few years. In offline RL, the agent learns the optimal policy from a static dataset of transitions instead of interacting with the environment. Formally, the dataset can be denoted by $\mathcal{D} = \{(s^i_t, a^i_t, s^i_{t + 1}, r^i_t)\}$. Offline RL requires the learning algorithm to extract sufficient knowledge of the dynamical system underlying an MDP $\mathcal{M}$ entirely from $\mathcal{D}$, and then construct a policy $\pi(a|s)$ that maximizes the expected return \textit{when it actually guides the agent to interact with $\mathcal{M}$} \cite{levine2020offline}. Action and state sampling are usually conducted by $s \sim \rho^{\pi_\beta}(s)$ and $a \sim \pi_\beta(a|s)$ where $\pi_\beta$ is the behavior policy and $\rho^{\pi_\beta}(\cdot)$ the distribution over states and actions in $\mathcal{D}$. Offline RL has found applications in decision making for healthcare \cite{chen2022diaformer}, autonomous driving \cite{villaflor2022addressing} and robot manipulation \cite{mandlekar2020iris}. However, the agent cannot interact with the environment to gather information, which makes improving the exploration strategy impossible. Moreover, policies learned by offline RL algorithms may encounter states that are outside of their training distribution and keep making mistakes for the rest of the simulation \cite{levine2020offline}. 

\subsection{Fundamentals of Transformer and Attention Mechanism}
\subsubsection{Transformer Architecture}
\begin{figure}
    \centering
    \includegraphics[width=1\linewidth]{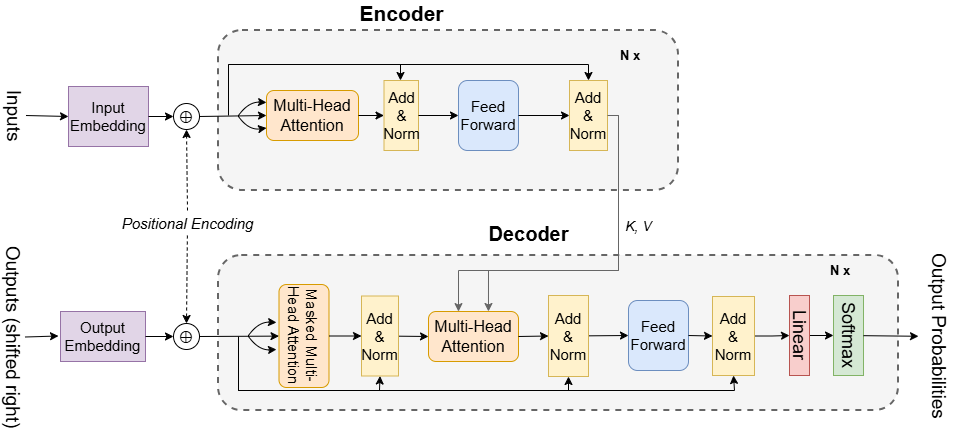}
    \caption{Structure of the original Transformer \cite{vaswani2017attention}.}
    \label{fig:trans_1}
    \vspace{-2em}
\end{figure}

The Transformer \cite{vaswani2017attention} was originally introduced as a deep learning architecture for addressing NLP tasks, but it has been applied in every domain and fundamentally transformed the AI landscape. It consists of an encoder and a decoder fueled by self-attention mechanism, fully-connected networks with residual connections \cite{he2016deep} and layer normalization \cite{ba2016layer}. The encoder first maps an input sequence to latent representations, which are then used by the decoder to generate the desired outputs in an autoregressive manner. Additionally, the previous output is also utilized to generate the next output. The overall structure diagram is shown in Fig.~\ref{fig:trans_1}, and the Transformer architecture is presented in the following parts.

\textbf{Self-Attention Mechanism:} The core component is the Scaled Dot-Product Attention. Given an input $\mathbf{X}$, three linear mappings generate the query $\mathbf{Q}$, key $\mathbf{K}$, and value $\mathbf{V}$ matrices. The attention output and its multi-head (MHA) extension are computed as:
\begin{subequations}
\label{eq:attention_all}
\begin{align}
    \text{Attention}(\mathbf{Q, K, V}) &= \text{Softmax}\left(\frac{\mathbf{QK}^\top}{\sqrt{d^{\rm{k}}}}\right) \mathbf{V}, \label{eq:attn} \\
    \text{MHA}(\mathbf{Q, K, V}) &= \text{Concat}(\mathbf{H}_1, \dots, \mathbf{H}_h) \mathbf{W}^{\rm{o}}, \label{eq:mha}
\end{align}
\end{subequations}
where each head $\mathbf{H}_i = \text{Attn}(\mathbf{QW}^{\rm{q}}_i, \mathbf{KW}^{\rm{k}}_i, \mathbf{VW}^{\rm{v}}_i)$ utilizes distinct learnable projections \cite{vaswani2017attention} to capture information from different representation subspaces.

\textbf{Position-wise Feed-Forward Networks (FFN):} Each layer contains an FFN consisting of two linear transformations with a non-linear activation $\phi$ (e.g., ReLU \cite{agarap2018relu} or GeLU \cite{hendrycks2016gelu}):
\begin{equation}
    \label{eq:ffn}
    \text{FFN}(\mathbf{X}) = \text{max}(0, \mathbf{X}\mathbf{W}_1 + \mathbf{b}_1)\mathbf{W}_2 + \mathbf{b}_2.
\end{equation}

\textbf{Positional Encoding \& Residual Connections:} Since Transformers lack inherent recurrence, sinusoidal positional encodings (PE) are added to input embeddings to inject sequence order:
\begin{equation}
\label{eq:PE}
\begin{cases} 
    \text{PE}_{(pos, 2i)} = \sin(pos / 10000^{2i / d_{\text{model}}}), \\
    \text{PE}_{(pos, 2i+1)} = \cos(pos / 10000^{2i / d_{\text{model}}}),
\end{cases}
\end{equation}
where $pos$ and $i$ denote the position and dimension index, respectively. Finally, to stabilize training, each sub-layer adopts a residual connection \cite{he2016deep} followed by layer normalization \cite{ba2016layer}: $\text{LayerNorm}(\mathbf{X} + \text{SubLayer}(\mathbf{X}))$.

\subsubsection{Transformer and Attention Variants} The survey in \cite{lin2022survey} categorized the variants of attention mechanism into the following groups: (1)  \textit{Sparse attention:} pioneered by the authors in \cite{child2019generating}, this line of work introduced sparsity bias into the attention mechanism, helping to reduce computational complexity; (2)  \textit{Linearized attention:} disentanglement of the attention matrix using kernel feature maps, the attention is then calculated in reverse to achieve linear complexity \cite{katharopoulos2020transformers, choromanski2020rethinking, schlag2021linear}; (3) \textit{Prototype and memory compression:} reducing the size of the attention matrix by lowering the number of queries or key-value memory pairs \cite{liu2018generating, vyas2020fast, zhang2021poolingformer}; (4) \textit{Low-rank self-attention:} utilizing the low-rank property of self-attention, usually through the Nyström method \cite{chen2021compressed, xiong2021nystromformer} or low-rank kernel approximation \cite{choromanski2020masked} (closely related to the linearized attention schemes); (5) \textit{Attention with Prior:} exploring the integration of prior distributions into standard attention modules. The prior attention can be formulated as a trainable attention prior added directly to the unnormalized attention matrix \cite{raffel2020exploring}, generated from positional embeddings \cite{ke2021rethinking} or locality information \cite{yang2018modeling}; (6) \textit{Improved multi-head mechanism:} different alternative multi-head mechanisms are investigated, with prominent examples in head behavior modeling \cite{li2018multi, kovaleva2019revealing}, multi-head with restricted spans \cite{sukhbaatar2019adaptive, guo2020multi} or refined aggregation modules \cite{li2019information, gu2019improving}.

Furthermore, Transformers have been adapted to solve various tasks in computer vision (CV), pioneered by Vision Transformer \cite{dosovitskiy2020an_image}, which divides an image into fixed size patches (e.g. $16 \times 16$), flattening them into token sequences and processing them using global attention. Based on this, several other architectures have been introduced, including Swin Transformer \cite{liu2021swin}, DETR \cite{carion2020end} and SAM \cite{kirillov2023segment}. For graph-structured data, Transformers have also demonstrated great potential, as the self-attention module is also highly capable of capturing and aggregating graphical information. Since the introduction of Graph Attention Network (GAT) \cite{velickovic2018graph}, numerous improvements to the Transformer for graphs have been proposed, with notable examples including GTN \cite{yun2019graph}, HGT \cite{hu2020heterogeneous} and Graphormer \cite{ying2021transformers}. 
\subsection{Transformer-based Reinforcement Learning}
% {\color{red}You should describe in detail the transformer-enabled RL algorithms commonly used for the survey par according to the following organization.}

In recent years, the integration of Transformers into the
RL paradigm has attracted significant attention. This trend is motivated by several fundamental observations. First, the sequential decision-making process of RL can be re-formulated as a sequence modeling process, which can be effectively addressed by the attention mechanism. Self-attention enables RL agents to directly attend to relevant past states, actions, and rewards, while suppressing redundant or less informative features. As a result, Transformer-based RL can accelerate training and avoid vanishing gradient limitations of recurrent architectures such as Long-Short Term Memory networks (LSTMs) \cite{hochreiter1997lstm} in cases of long-term temporal dependencies. Second, many real-world RL tasks require agents to process heterogeneous data modalities. Transformer-based architectures can effectively handle multiple modalities \cite{xu2023multimodal}. Moreover, the parallelizable nature of self-attention allows Transformers to scale efficiently on modern accelerators, supporting large model sizes that benefit from scaling laws \cite{Lee2022MultiGameDT}. In the following sections, we discuss the integration of Transformers into foundational RL algorithms. 

\subsubsection{Transformer-enabled DQN}

Classical Q-learning~\cite{watkins1992q} and its deep learning variant DQN~\cite{mnih2013playing} approximate the action-value function $Q_\theta(s,a)$ by minimizing the temporal-difference (TD) loss:
\begin{equation}
\mathcal{L}_{\text{DQN}}(\theta) 
= \mathbb{E}_{(s,a,r,s')} 
\Big[ 
\big( r + \gamma \max_{a'} Q_{\theta^-}(s',a') 
- Q_\theta(s,a) \big)^2 
\Big],
\end{equation}
where $r$ is the immediate reward after taking action $a$, $\gamma$ is the discount factor, and $\max_{a'} Q_{\theta^-}(s',a')$ is the best estimated future value with next state $s'$. Transformer-enabled DQN replaces the feed-forward encoder with a Transformer that models the trajectory as a sequence $\tau = (s_1,a_1,r_1,\dots,s_t)$, allowing long-range temporal dependency modeling under partial observability~\cite{agarwal2023Transformers,hu2024transforming}. For example, TransDreamer~\cite{chen2022transdreamer} integrates a Transformer-based world model to learn latent dynamics and compute multi-step value targets. The attention mechanism, as shown in Eq.~\ref{eq:attn}, enables global context aggregation across time steps.

In communication systems, Transformer-DQN variants have been applied to resource allocation and network control. Autonomous power control~\cite{che2025autonomous} and bandwidth allocation~\cite{ma2025dsaf} use attention to capture inter-user interference patterns. Computation offloading frameworks~\cite{gholipour2023tpto,han2025transformer} model task queues as sequences to improve decision stability. For routing and trajectory planning, perception-aware attention modules improve spatial generalization~\cite{abdelkader2025perception,xu2025transformer}. Additionally, Transformer-based critics demonstrate better generalization compared to CNN/RNN backbones~\cite{esslinger2022deep}.

\subsubsection{Transformer-enabled A2C}

Advantage Actor-Critic (A2C)~\cite{wang2016dueling,sutton1998reinforcement} optimizes a policy $\pi_\theta(a|s)$ and the value function $V_\phi(s)$ jointly. The policy gradient objective is:
\begin{equation}
\nabla_\theta J(\theta) 
= \mathbb{E} \big[ 
\nabla_\theta \log \pi_\theta(a_t|s_t) 
A_t 
\big],
\end{equation}
where $A_t = r_t + \gamma V_\phi(s_{t+1}) - V_\phi(s_t)$. Transformer-enabled A2C replaces the actor and critic encoders with a Transformer to process trajectory segments: $(s_{t-k},\dots,s_t)$, improving representation learning in POMDP settings ~\cite{agarwal2023Transformers,hu2024transforming}. Self-attention allows the critic to compute value estimates conditioned on long-term context rather than single-step states. Other works extend Transformer-A2C to large-scale network optimization, where attention captures cross-device coupling and non-stationarity~\cite{agarwal2023Transformers}. These methods demonstrate improved stability compared to RNN-based actor-critic architectures.

\subsubsection{Transformer-enabled PPO}

Proximal Policy Optimization (PPO)~\cite{schulman2017proximal} improves policy gradient stability using a clipped surrogate objective:

\begin{equation}
\mathcal{L}_{\text{PPO}}(\theta) =
\mathbb{E}\Big[
\min\big(
r_t(\theta) A_t,
\text{clip}(r_t(\theta),1-\epsilon,1+\epsilon) A_t
\big)
\Big],
\end{equation}
where $r_t(\theta) = {\pi_\theta(a_t|s_t)}/{\pi_{\theta_{\text{old}}}(a_t|s_t)}$ and $\text{clip}()$ is the clipping function that limits a value within a range. Transformer-enabled PPO adopts attention-based encoders for the policy and/or value networks~\cite{agarwal2023Transformers}. The Transformer processes trajectory tokens and produces context-aware action logits, enhancing robustness in high-dimensional and partially observable environments.

In wireless resource allocation, Transformer-based PPO captures user coupling and dynamic Quality-of-Service (QoS) constraints~\cite{che2025autonomous,ma2025dsaf}. For computation offloading, attention-based policies learn task-priority dependencies~\cite{gholipour2023tpto,han2025transformer}. Routing and trajectory planning benefit from spatial-temporal attention mechanisms~\cite{abdelkader2025perception,xu2025transformer}. Recent work comparing generative AI and DRL highlights that Transformer-enhanced PPO offers more stable convergence under dynamic workloads~\cite{sun2025generative}.

\subsubsection{Transformer-enabled MARL}

%. For instance, the multi-agent scheduling framework in~\cite{yuan2021agentformer} employs a shared Transformer encoder across agents. Each agent embedding is treated as a token:
% \begin{equation}
% h_i = \text{Transformer}(e_i, \{e_j\}_{j\neq i}),
% \end{equation}
% where $e_i$ denotes the local observation embedding. 
% The joint loss typically combines individual PPO/A2C objectives:
% \begin{equation}
% \mathcal{L} = \sum_{i=1}^{N} \mathcal{L}_i^{\text{actor}} 
% + \lambda \mathcal{L}^{\text{critic}}.
% \end{equation}
% Attention weights quantify coordination intensity between agents.

In Multi-Agent Reinforcement Learning (MARL), joint policy optimization suffers from scalability and non-stationarity. Transformer-based MARL introduces attention over agents to model inter-agent dependencies explicitly. For instance, the multi-agent coordination framework in QMIX~\cite{rashid2020monotonic} employs a centralized mixing network to integrate individual agent utilities into a joint action-value function. To ensure that global optimization is consistent with local decision-making, QMIX enforces a monotonicity constraint ${\partial Q_{tot}}/{\partial Q_i} \geq 0, \ \forall i \in \{1, \dots, N\}$, where $Q_{tot}$ is the global state-value and $Q_i$ represents the local utility of agent $i$. The joint loss is typically formulated to minimize the TD loss:
\begin{equation}
\mathcal{L} = \mathbb{E} \left[ \left( y_{tot} - Q_{tot}(\mathbf{s}, \mathbf{a}; \theta) \right)^2 \right],
\end{equation}
where $y_{tot}$ is the target value. While QMIX excels at credit assignment through this structural constraint, integrating it with Transformer architectures, as proposed in TransfQMix~\cite{gallici2023transfqmix}, allows the model to leverage the latent graph structure of the environment. In this approach, a Transformer-based Q-mixer treats agent internal states and environmental entities as vertices in a coordination graph. By employing multi-head self-attention (MSA), the mixer generates dynamic weights that adapt to the relative importance of different agents' contributions. This mechanism uses attention weights to quantify coordination intensity and ensures that the framework remains permutation-invariant and transferable to tasks with varying numbers of agents. More generally, Transformer-based MARL improves credit assignment and coordination in partially observable systems~\cite{agarwal2023Transformers,hu2024transforming}. Applications to routing and cooperative trajectory control demonstrate improved scalability due to permutation-invariant attention mechanisms~\cite{abdelkader2025perception,xu2025transformer}.

Improvements of Transformer-based RL can be broadly categorized into two primary streams: (a) Architecture Enhancement, where Transformers replace RNNs in traditional RL loops, and (b) Sequence Modeling, where RL is reformulated as a trajectory generation problem. 
\paragraph{\textbf{Architecture Enhancement}}
Under this paradigm, the fundamental principles of RL (e.g., Bellman equations, policy gradients) remain unchanged. Instead, the focus shifts to upgrading the underlying neural network backbones by integrating Transformer encoders to effectively replace recurrent neural networks (RNNs) and capture complex temporal dependencies \cite{zhou2024Transformer}. This architectural evolution encompasses both value-based methods (e.g., DQN) and Actor-Critic frameworks (e.g., PPO, A2C). It serves as a robust state representation backbon, particularly effective in non-stationary wireless environments \cite{zhao2024trandrl}. The core mechanism involves an MSA layer that computes a context-aware latent embedding $h_t$. It processes a sliding window of historical observations, actions, and rewards to provide a globally informed basis for decision-making:
\begin{equation}
    h_t = \text{Transformer}\big(\{o_{i}, a_{i}, r_{i}\}_{i=t-K}^{t}\big),
\end{equation}
where $K$ denotes the context length. This rich temporal representation $h_t$ is then fed into the respective decision heads depending on the base RL algorithm. 

For instance, in value-based methods, $h_t$ is used directly to estimate the action-value function $Q_{\theta}(h_t, a)$. Conversely, in Actor-Critic setups, this embedding is shared by both the actor $\pi_{\theta}(a_t | h_t)$ and the critic $V_{\phi}(h_t)$ \cite{song2025deep}. The generic joint optimization objective for these actor-critic variants is typically formulated as:
\begin{equation}
    \mathcal{L}(\theta, \phi) = \mathbb{E}_{\tau} \left[ (V_{\phi}(h_t) - \hat{R}_t)^2 - \log \pi_{\theta}(a_t | h_t) A_t \right],
\end{equation}
where $A_t = \hat{R}_t - V_{\phi}(h_t)$ represents the advantage estimate, and $\hat{R}_t$ is the empirical return. This formulation allows the agent to perform more accurate credit assignment over long temporal horizons by adaptively attending to critical past events \cite{ zhang2025robust}.

By decoupling the sequence representation learning from the specific RL update rule, Architecture Enhancement provides a highly flexible solution. It has been extensively adopted in specialized tasks such as traffic prediction and resource optimization in Open Radio Access Networks (O-RANs) \cite{ zhou2024Transformer}, maintaining QoS for 5G Non-terrestrial Networks (NTNs) and satellite-terrestrial networks \cite{zhao2024trandrl, zhang2025robust}, and task scheduling in mobile edge computing O-RAN (MEC-O-RAN) and semantic communication, where capturing the structural dependencies of data is vital \cite{song2025deep, yuan2025learning}. By providing a more stable gradient signal compared to LSTM-based variants, this paradigm has become a standard for enhancing the reliability of DRL agents in the evolving 6G landscape \cite{yuan2025learning, zhao2024trandrl}.

\paragraph{\textbf{Sequence Modeling}}
In stark contrast to the {Architecture Enhancement} paradigm, which preserves traditional RL optimization mechanics via Bellman equations, the {Sequence Modeling} approach fundamentally reformulates the reinforcement learning problem. Pioneered by the \text{Decision Transformer (DT)} \cite{chen2021decision}, this paradigm shifts away from estimating value functions and instead treats offline RL entirely as a conditional sequence generation task.  Particularly, DT models a trajectory not as a sequence of state transitions, but as a sequence of tokens comprising Returns-to-Go (RTG), states, and actions. A trajectory $\tau$ of length $T$ is represented as:
\begin{equation}
    \tau = \big(\hat{R}_1, s_1, a_1, \hat{R}_2, s_2, a_2, \dots, \hat{R}_T, s_T, a_T\big),
\end{equation}
where the return-to-go $\hat{R}_t = \sum_{t'=t}^{T} r_{t'}$ represents the target cumulative reward from timestep $t$ onwards. The core idea is to condition the action generation on the \textit{desired} future return. At timestep $t$, the model takes the sequence of the last $K$ steps as input:
\begin{equation}
    x_t =\big(\hat{R}_{t-K}, s_{t-K}, a_{t-K}, \dots, \hat{R}_t, s_t\big).
\end{equation}
The Transformer then predicts the action $a_t$ autoregressively:
\begin{equation}
    a_t = \text{Transformer}(x_t).
\end{equation}

Unlike traditional RL which maximizes expected return via Bellman updates, DT is trained via supervised learning. For a dataset $\mathcal{D}$ of trajectories, the objective is to minimize the error between the predicted action and the ground-truth action from the dataset:
\begin{equation}
    \mathcal{L}_{DT} = \mathbb{E}_{\tau \sim \mathcal{D}} \left[ \sum_{t=1}^{T} \| a_t - \hat{a}_t \|^2 \right];
\end{equation}
cross-entropy loss is used for discrete actions. By conditioning on high RTG values during inference, the agent generates actions that are correlated with high rewards, effectively performing "hindsight" planning.

\paragraph{\textbf{Variances and Advanced Extensions}}

While the foundational integration of Transformers as state representation encoders (in Architecture Enhancement) and the vanilla Decision Transformer (in Sequence Modeling) are the core baselines for their respective paradigms, the evolving demands of complex wireless networks have spurred further innovations. Recent research has introduced numerous advanced variants to address specific inherent limitations, such as computational overhead, environmental stochasticity, and multi-agent coordination.

% \textbf{Extensions in Architecture and Memory:} 
To address the finite context window of standard Transformers, the \text{Compressive Transformer}~\cite{rae2019compressive} introduces a hierarchical memory mechanism that compresses older activations into a lower-resolution buffer, enabling agents to retain coarse-grained history over extended horizons. For efficiency in latency-constrained systems, hybrid architectures like \textit{CoBERL}~\cite{banino2021coberl} combine Transformers with LSTMs, using contrastive losses to capture long-range dependencies while maintaining efficient online execution.

% \textbf{Extensions in Sequence Modeling (Beyond DT):} 
Moreover, standard DT struggles in stochastic environments where high returns might result from unintentional randomness. To mitigate this, \text{ESPER}~\cite{paster2022you} conditions on expected returns utilizing cluster-based representations, while \text{Q-learning Decision Transformer (QDT)}~\cite{yamagata2023q} leverages conservative Q-learning to relabel RTGs, distinguishing optimal behaviors from suboptimal ones. Furthermore, while DT is model-free, the \text{Trajectory Transformer (TT)}~\cite{janner2021offline} adopts a model-based perspective by learning the joint distribution of states, actions, and rewards, allowing beam-search planning during inference. A brief summary of foundational Transformer-based DRL algorithms can be found in Table \ref{tab:foundational-algorithms}.

\begin{table*}[!h]
\caption{Foundational Transformer-based DRL Algorithms}
\label{tab:foundational-algorithms}
\centering
\renewcommand{\arraystretch}{1.4}

\resizebox{\textwidth}{!}{%
\begin{tabular}{|l|l|l|l|l|}
\hline
\makecell[l]{\textbf{Base} \\ \textbf{Algorithms}}& 
\makecell[l]{\textbf{Transformer} \\ \textbf{Role}} & 
\textbf{Mechanism of Integration} & 
\textbf{Key Advantages} & 
\textbf{Notable Work} \\
\hline

\makecell[l]{\textbf{Transformer-} \\ \textbf{based DQN}} &
\makecell[l]{State \\ representation\\ encoder} &
\makecell[l]{Self-attention processes \\ sequences of 
historical obser-\\ vations $(o_{t-K}, \dots, o_t)$ to estimate \\
Q-values $Q(s,a)$ and capture \\ long-term dependencies.} &
\makecell[l]{Improved partial observability handling;\\
better interference in networking;\\
handles partial observability better than LSTM-DQN;\\
effective for spectrum access \\  and traffic-aware offloading.} &
\makecell[l]{\cite{che2025autonomous, ma2025dsaf};\\
\cite{gholipour2023tpto, abdelkader2025perception};\\
\cite{al2025bandwidth, nerondat2025Transformer, ghosh2025mgco}} \\
\hline

\makecell[l]{\textbf{Transformer-} \\ \textbf{based A2C}} &
\makecell[l]{Shared \\ context\\ encoder} &
\makecell[l]{Transformer backbone \\ shared between Actor ($\pi$) \\
and Critic ($V$) networks; often \\
applied in POMDP settings.} &
\makecell[l]{Enhanced stability and long-term credit\\
assignment vs RNN-based A2C;\\
improves memory retention \\ in continuous control.} &
\makecell[l]{\cite{agarwal2023Transformers, hu2024transforming};\\
\cite{zhang2024handover, aouedi2025hybrid}} \\
\hline

\makecell[l]{\textbf{Transformer-} \\ \textbf{based PPO}} &
\makecell[l]{Policy and \\ value function \\ approximator} &
\makecell[l]{Transformer encoder integrated \\ into policy
and critic networks; \\ trained with clipped\\
surrogate objective for stability.} &
\makecell[l]{Robust under dynamic QoS constraints;\\
effective for resource scheduling and computation\\
offloading under latency \\ and energy constraints.} &
\makecell[l]{\cite{che2025autonomous, ma2025dsaf};\\
\cite{gholipour2023tpto, han2025transformer};\\
\cite{abdelkader2025perception, xu2025transformer};\\
\cite{sun2025generative, zhou2025Transformer, wu2025application}} \\
\hline

\makecell[l]{\textbf{Transformer-} \\ \textbf{based MARL}} &
\makecell[l]{Inter-agent\\interaction \\ modeler} &
\makecell[l]{Agents treated as tokens; \\ self-attention
learns pairwise \\ dependencies and coordination\\
patterns in joint state space.} &
\makecell[l]{Scalable coordination; improved credit\\
assignment in cooperative settings;\\
decouples input size from agent count.} &
\makecell[l]{\cite{yuan2021agentformer};\\
\cite{abdelkader2025perception, xu2025transformer};\\
\cite{gao2025cooperative, ye2025enhanced, li2025hierarchical}} \\
\hline 

\makecell[l]{\textbf{Decision}\\ \textbf{Transformer}} &
\makecell[l]{Sequence-to \\-sequence\\policy \\generator} &
\makecell[l]{Reformulates RL as \\ conditional trajectory
generation \\  using Returns-to-Go tokens;\\
trained via supervised loss.} &
\makecell[l]{High sample efficiency in offline RL;\\
sequence modeling enables \\ long-horizon planning.} &
\makecell[l]{\cite{chen2021decision, agarwal2023Transformers};\\
\cite{wang2025distributed, lu2025attention}} \\
\hline 

\end{tabular}%
}
\vspace{-2em}
\end{table*}

\subsubsection{Transformer-based RL for Wireless Networks: A Case Study}

In this section, we aim to demonstrate the superior performance of Transformer-enabled RL compared to standard offline DRL, thus showing the effectiveness of integrating Transformers into RL. For this, we present a case study investigating the power allocation of two offline RL methods, namely standard offline DRL  \cite{ross2010efficient} and Transformer-enabled RL \cite{chen2021decision}, in learning transmission policies for an energy-harvesting wireless sensor environment modeled as an MDP.

\paragraph{System model and problem formulation}
The system consists of a wireless sensor with a battery that wirelessly and stochastically harvests energy from a power beacon. In this environment, the sensor acts as a decision-maker that must determine its transmission power level at each time step. The sensor's objective is to optimize its transmission policy to maximize the long-term cumulative throughput while maintaining energy sustainability. To achieve this, the agent must learn to balance opportunistic transmissions during high channel quality with the preservation of battery life to avoid costly outages.

The system state at time step $t$ is defined as $s_t = [h_t, E_t]$, in which $h_t$ represents the channel gain and $E_t \in [0, B_{\max}]$ denotes the current battery level. The action space consists  of three discrete power levels: $a_t \in \{0, 0.3, 1.0\}$ W. The channel gain $h_t$ follows Rayleigh fading. The battery dynamics are given by 
\begin{equation}
     E_{t+1} = \min(E_{\max}, \max(0, E_t - E_{tx}(a_t) + E_{harv, t})),
    \label{equ:battery_dynaics}
\end{equation}
where $E_{tx}(a_t)$ is the energy consumed by action $a_t$, and $E_{harv, t}$ is the harvested energy, which is modeled as a random process with arrival probability $p=0.3$ and amount $E_{harv, t} = 0.2$ J. Moreover, we define the reward function as the transmission rate with an outage penalty for battery depletion, as
\begin{equation}
    \label{reward}
    r_t = \begin{cases} \log_2(1 + \frac{h_t a_t}{N_0}), & \text{if } E_t \ge E_{tx}(a_t), \\ 
    -\lambda, & \text{otherwise},
    \end{cases}
\end{equation}
where $N_0 = 0.1$ is the noise, and $\lambda=1.0$ is the penalty term to discourage the agent from taking greedy actions that lead to energy outages. This penalty is designed to discourage greedy, short-term actions that lead to energy depletion, forcing the agent to prioritize long-term efficiency over immediate gains.

\paragraph{Dataset Generation}
As offline RL involves learning from a fixed dataset, one of the primary challenges is learning from suboptimal data. We construct a mixed-quality dataset $\mathcal{D}$ consisting of $1000$ different trajectories, generated from three distinct policies: (1) \textit{Expert Policy $(15\%)$}: A battery-aware heuristic that transmits only when the channel quality is high and sufficient battery is available ($b_t > 0.3$); (2) \textit{Medium Policy ($33\%$}): A moderate policy that transmits at low power when the channel is decent ($h_t > 0.8$); (3) \textit{Aggressive Policy ($50\%$)}: A greedy policy that selects high-power transmission whenever the channel is above average ($h_t \geq 0.5$), often leading to battery depletion and subsequent outages. 
This composition creates a challenging "noisy" dataset where the average behavior is suboptimal, requiring the model to selectively identify and reinforce expert behaviors. 

\paragraph{Performance Analysis}
We compare the performance of Transformer-enabled RL versus the standard DRL framework with a multi-layer perceptron (MLP) backbone. The experiments are conducted using the dataset generation process described in the previous section, consisting of 1000 mixed-quality trajectories. Both the models are trained for 15 epochs using mean-squared error loss with AdamW optimizer \cite{Loshchilov2019DecoupledWD}, using learning rates of $1e^{-4}$ for Transformer-enabled RL and $1e^{-3}$ for standard offline DRL, respectively.  Other parameter settings for the two models are set as follows.
% ($R_{target} \approx \max(\mathcal{D})$)

\begin{itemize}
    \item \textbf{Standard offline DRL (Behavioral Cloning)}: The standard offline DRL backbone is a feed-forward network with a hidden dimension of $128$. It treats the problem as a supervised classification task, mapping states directly to actions without temporal context. Unlike online DRL, which learns through trial and error, this baseline treats the offline policy recovery as a supervised regression task, mapping states directly to actions by minimizing the mean-squared error against the actions in the dataset.
    \item \textbf{Transformer-enabled RL}: In contrast to the single-step mapping of the MLP baseline, the Transformer-enabled RL formulates the problem as a conditional sequence modeling problem. We configure the Transformer-enabled RL with a context length of $K=20$, embedding dimension $d_{model}=64$, $4$ attention heads, and $2$ Transformer layers. The model leverages the return-to-go $\hat{R}_t$, state $s_t$, and action $a_t$ sequences to understand the causal relationship between trajectory history and future rewards. During evaluation, the model is conditioned on the target return-to-go $\hat{R}_0$, set to the maximum return observed in the dataset.
\end{itemize}

% \begin{figure}[htbp]
%     \centering
%     \includegraphics[width=0.9\linewidth]{figure/mlp_dt_benchmark/episodeReturn_mlp_dt.png}
%     \caption{Average episodic return of MLP and Transformer-enabled RL compared to the dataset mean. DT significantly outperforms MLP by effectively filtering out suboptimal behaviors from the mixed dataset.}
%     \label{fig:episode_return}
% \end{figure}

% \begin{figure}[!h]
%     \centering
%   \centering
%   \subcaptionbox{}[0.48\linewidth][c]{
%     \includegraphics[width=\linewidth]{figure/mlp_dt_benchmark/episodeReturn_mlp_dt.pdf}}\quad
%   \subcaptionbox{}[0.48\linewidth][c]{
%     \includegraphics[width=\linewidth]{figure/mlp_dt_benchmark/loss_mlp_dt.pdf}} \quad
%     \caption{(a) Average episodic return of standard offline DRL and Transformer-enabled RL compared to the dataset mean, (b) Training loss convergence of the two models.}
%     \label{fig:analysis}
% \end{figure}
\begin{figure*}[h!]
    \centering
    % Row 1
    \begin{minipage}{5.3cm}
        \centering
        \includegraphics[width=\linewidth]{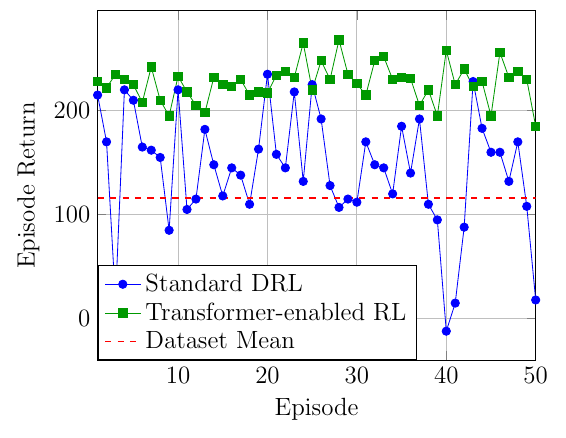}
        \subcaption[]{}
    \end{minipage}
    %\hfill
    \begin{minipage}{5cm}
        \centering
        \includegraphics[width=\linewidth]{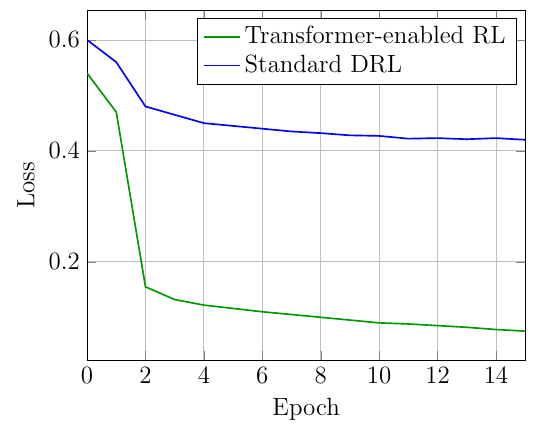}
        \subcaption[]{}
    \end{minipage}
    \begin{minipage}{6.4cm}
        \centering
        \includegraphics[width=\linewidth]{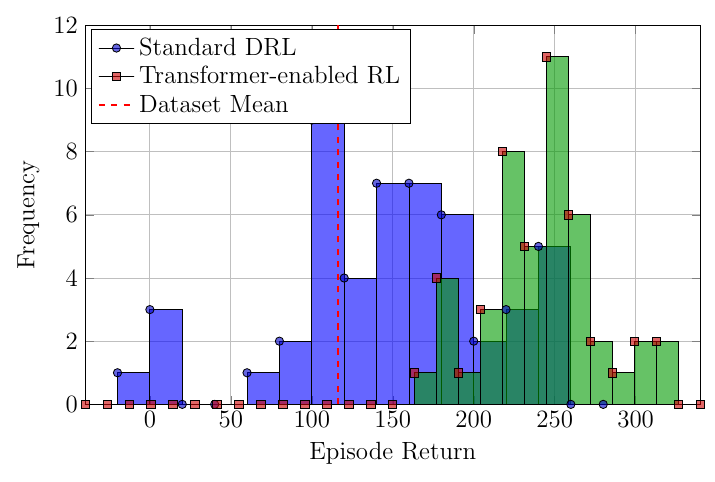}
        \subcaption[]{}
    \end{minipage}
    \caption{(a) Average episodic return of standard offline DRL and Transformer-enabled RL compared to the dataset mean, (b) training loss convergence of the two models, and (c) Distribution of episodic returns of the two algorithms.}
    \vspace{-2em}
    \label{fig:analysis}
\end{figure*}

First, we discuss the performance of the two models via the cumulative reward (episodic return) averaged over $N=50$ evaluation episodes. Fig.~\ref{fig:analysis}(a) compares the episode return of the trained policies against the average dataset quality. As shown in the results, the standard offline DRL baseline tends to learn the "average" behavior of the dataset. Since the dataset is dominated by aggressive and medium policies ($85\%$ of the dataset), the standard offline DRL agent frequently takes greedy actions that lead to battery depletion, resulting in varying performance. In contrast, the Transformer-enabled RL, by conditioning on the highest achievable return, successfully isolates and mimics the expert sub-policy ($15\%$ of the dataset) hidden within the noisy data. Fig.~\ref{fig:analysis}(b) demonstrates the training loss convergence of the two algorithms. As shown, the loss value of Transformer-enabled RL rapidly decreases from $0.47$ at epoch $1$ to around $0.07$ by epoch $15$. Meanwhile, the standard offline DRL baseline starts with a higher loss (around 0.56) and decreases gradually over time, reaching $0.42$ after epoch 10, with only minor fluctuations thereafter. This difference is due to the Transformer-enabled RL being more effective than the standard offline DRL in capturing the underlying structure of the dataset. By leveraging temporal context and return-to-go information, Transformer-enabled RL is able to learn more informative policies, leading to faster convergence and lower training error. Meanwhile, the standard offline DRL model, which relies only on single-step state-action mappings, struggles to model long-term dependencies, resulting in slower learning and suboptimal performance. We further analyze the stability of the learned policies by examining the distribution of collected rewards, as illustrated in Fig.~\ref{fig:analysis}(c). The Transformer-enabled RL policy shifts the probability mass towards higher rewards, demonstrating consistent expert-level performance, whereas standard offline DRL exhibits a multi-modal distribution reflecting the mixed nature of the training data. This demonstrates that Transformer-enabled RL consistently avoids outage states ($b_t=0$) by planning ahead, whereas the standard offline DRL policy exhibits higher variance, occasionally failing to manage energy constraints effectively. 

% \begin{figure}[htbp]
%     \centering
%     \includegraphics[width=0.9\linewidth]{figure/mlp_dt_benchmark/loss_mlp_dt.png}
%     \caption{Training loss convergence. Both models converge, but DT's sequence modeling capability allows it to generalize better during inference.}
%     \label{fig:training_loss}
% \end{figure}

% \begin{figure}[!h]
%     \centering
%     \includegraphics[width=0.9\linewidth]{figure/mlp_dt_benchmark/perDist_mlp_dt.png}
%     \caption{Distribution of episodic returns of the two algorithms.}
%     \label{fig:perf_dist}
% \end{figure}

The above results validate the efficacy of Transformer-based DRL algorithms in offline situations for wireless communication systems utilizing energy harvesting, particularly when expert data is sparse. In the next sections, we discuss various applications of Transformer-based DRL algorithms in wireless domains.

%==============================================
\section{Transformer-based RL for Resource Allocation}
\label{sec:resource}
%==============================================

Resource allocation determines how radio, network, and service resources are dynamically assigned in communication networks in response to time-varying traffic demands, channel conditions, and user behaviors.
These problems often involve sequential decision-making, large and heterogeneous action spaces, and strong coupling across time, users, and resource types, which together pose significant challenges for conventional DRL approaches.
By embedding the Transformer into DRL frameworks, recent studies leverage attention-based sequence modeling and global context modeling to capture long-term temporal correlations, multi-agent interactions, and structured decision dependencies in practical resource allocation.
Accordingly, this section reviews Transformer-enhanced DRL approaches from five representative perspectives: 1) radio and transmission resource allocation, including power, bandwidth, beamforming, and modulation; 2) network access and connectivity control, encompassing access, association, scheduling, and handover; 3) virtual network function placement; 4) content-aware rate adaptation; and 5) joint computation-communication resource allocation. Collectively, these studies suggest that the Transformer improves scalability, long-horizon decision-making, and structural awareness in DRL-based resource allocation.

%==============================================
\subsection{Radio and Transmission Resource Allocation}
%==============================================

Recent works increasingly integrate the Transformer into DRL frameworks for power control, bandwidth allocation, beamforming, and modulation, to handle temporal dynamics, structured dependencies, and high-dimensional decision spaces, as summarized in Table~\ref{tab:SecIII_A}.

\begin{table*}[t] 
\caption{Summary of Transformer-Enhanced DRL for Radio and Transmission Resource Allocation.} 
\label{tab:SecIII_A} 
\centering 
\renewcommand{\arraystretch}{2.2} 

% Bọc bảng trong \resizebox để ép giãn vừa 2 cột (\textwidth) an toàn tuyệt đối
\resizebox{\textwidth}{!}{%
\begin{tabular}{|l|c|l|l|l|l|} 
\hline 
\textbf{Subcategory} & \textbf{Ref.} & \textbf{Scenario} & \textbf{Objective} & \textbf{Proposed Method} & \textbf{Transformer Role} \\ 
\hline 

\multirow{4}{*}{\makecell[l]{Power \\ control}} 
& \cite{che2025autonomous} & \makecell[l]{Multiuser OFDM systems} & \makecell[l]{Long-term BER minimization} & \makecell[l]{DT-aided DDPG} & \makecell[l]{Future channel state generation} \\ 
\cline{2-6}
& \cite{yang2024act} & \makecell[l]{Bidirectional LoRaWAN} & \makecell[l]{Packet delivery rate \\ maximization} & \makecell[l]{Transformer-enhanced \\ Actor-Critic DRL} & \makecell[l]{Global state representation \\ learning} \\ 
\cline{2-6}
& \cite{zhang2025multi} & \makecell[l]{Cellular networks} & \makecell[l]{Sum-rate maximization} & \makecell[l]{Multi-agent decision \\ Transformer} & \makecell[l]{Trajectory-based policy \\ sequence modeling} \\ 
\cline{2-6}
& \cite{xue2024cooperative} & \makecell[l]{Vehicular networks} & \makecell[l]{Reliability enhancement and \\ latency reduction} & \makecell[l]{Cooperative MAPPO} & \makecell[l]{Cooperative feature aggregation} \\ 
\hline 

\multirow{2}{*}{\makecell[l]{Bandwidth \\ allocation}}
& \cite{ma2025dsaf} & \makecell[l]{mmWave IAB networks} & \makecell[l]{Long-term throughput \\ maximization} & \makecell[l]{Transformer-enhanced \\ DQN} & \makecell[l]{Temporal state representation \\ learning} \\ 
\cline{2-6}
& \cite{al2025bandwidth} & \makecell[l]{Multi-operator vehicular \\ networks} & \makecell[l]{Bandwidth reservation cost \\ minimization} & \makecell[l]{Dueling DQN with \\ Transformer} & \makecell[l]{Temporal price dynamics \\ modeling} \\ 
\hline

\multirow{3}{*}{\makecell[l]{Beamforming}}
& \cite{farzanullah2025beam} & \makecell[l]{Indoor mmWave ISAC \\ systems} & \makecell[l]{Spectral efficiency \\ maximization} & \makecell[l]{Multi-agent contextual \\ bandit} & \makecell[l]{Multi-modal feature fusion} \\ 
\cline{2-6}
& \cite{ghassemi2024multi} & \makecell[l]{Downlink OFDM cellular \\ systems} & \makecell[l]{Throughput maximization} & \makecell[l]{Two-stage RL} & \makecell[l]{Multi-modal feature fusion \\ and beam group prediction} \\ 
\cline{2-6}
& \cite{xu2025fully} & \makecell[l]{FD-RAN MIMO-OFDM \\ systems} & \makecell[l]{Throughput maximization} & \makecell[l]{Hierarchical RL} & \makecell[l]{Subcarrier-correlation encoding} \\ 
\hline

\multirow{2}{*}{\makecell[l]{Modulation}}
& \cite{peri2025offline} & \makecell[l]{5G multicellular RAN} & \makecell[l]{Long-term performance \\ improvement} & \makecell[l]{Offline RL with \\ decision Transformer} & \makecell[l]{Sequence-based offline policy \\ learning} \\ 
\cline{2-6}
& \cite{he2025doppler} & \makecell[l]{Highly dynamic FANETs} & \makecell[l]{Bit-error-rate reduction} & \makecell[l]{Transformer-based \\ RL} & \makecell[l]{Long-term temporal dependency \\ modeling} \\ 
\hline

\end{tabular}%
}
\vspace{-2em}
\end{table*}

\subsubsection{Power Control}

Considering Orthogonal Frequency Division Multiplexing (OFDM) systems, \cite{che2025autonomous} studies autonomous transmit power control in time and frequency domains to minimize long-term bit-error rate.
A digital-twin-aided DRL framework pretrains the agent on virtual channels and deploys it for real-world power allocation, where a Transformer-based virtual channel generator predicts future multiuser channel states to enable prediction-then-decision control with improved stability.
Compared with the baseline scheme, the proposed method achieves about a 60\% reduction in bit error rate \cite{che2025autonomous}.
Different from \cite{che2025autonomous}, the work in \cite{yang2024act} considers power control under dense device deployments.
For bidirectional Long Range Wide Area Network (LoRaWAN), \cite{yang2024act} studies uplink transmit power optimization under dense device deployments to improve packet delivery reliability and energy efficiency.
A global Transformer is integrated into an actor-critic DRL framework to encode the joint state of all end devices into a shared representation, supporting coordinated uplink-downlink optimization across devices.
To further enable distributed power optimization under limited information exchange, the work in \cite{zhang2025multi} investigates offline multi-agent transmit power control with local observations, aiming at sum-rate maximization.
An offline MARL framework enables each link to learn a distributed power policy from pre-collected interaction data without online exploration.
A multi-agent decision Transformer performs trajectory-based sequence modeling to capture long-term dependencies, improving power decisions under partial and delayed information.
However, both the works in \cite{yang2024act} and \cite{zhang2025multi} are not tailored to high-mobility scenarios.
For multi-connectivity vehicular networks, \cite{xue2024cooperative} proposes a cooperative Multi-Agent Proximal Policy Optimization (MAPPO) framework for per-link power coordination under high mobility and imperfect Channel State Information (CSI), enabling ultra-reliable low-latency communication with reduced interference and energy cost.
A Transformer enables information sharing among the links serving the same user via attention-based feature aggregation, supporting coordinated power allocation decisions.

\subsubsection{Bandwidth Allocation}

For millimeter-wave Integrated Access and Backhaul (IAB) networks, \cite{ma2025dsaf} studies dynamic subchannel assignment for concurrent access and backhaul links under time-varying traffic demands.
A DQN-based framework is developed to maximize long-term system throughput while accounting for interference and resource constraints.
A Transformer is embedded as a state representation module to capture temporal correlations in traffic demand and link conditions, enabling more effective subchannel allocation in large-scale mmWave environments.
Compared with baseline DRL schemes, the Transformer-based DRL method improves the average throughput by more than 38\% \cite{ma2025dsaf}.
Different from subchannel allocation in IAB networks \cite{ma2025dsaf}, the work in \cite{al2025bandwidth} focuses on bandwidth reservation in multi-operator environments.
For time-critical vehicular applications, \cite{al2025bandwidth} studies bandwidth reservation in advance under time-varying prices and coverage conditions across operators.
A dueling deep Q-learning framework is adopted to minimize long-term reservation cost while meeting latency requirements.
A temporal fusion Transformer models time-dependent price dynamics, supporting cost-aware and reliable operator selection.

\subsubsection{Beamforming}

For indoor millimeter-wave Integrated Sensing and Communication (ISAC) systems, \cite{farzanullah2025beam} studies sensing-assisted beam selection from a predefined codebook to improve multi-user spectral efficiency.
A multi-agent contextual bandit framework adaptively selects beamforming vectors based on environment-aware context, where a multi-modal Transformer fuses ISAC sensing data and user location features into compact representations for accurate and robust beam selection.
Compared with conventional DRL, the Transformer-enhanced framework achieves a 49.6\% improvement in proximity to the optimal spectral-efficiency policy \cite{farzanullah2025beam}.
Different from indoor scenarios considered in \cite{farzanullah2025beam}, the work in \cite{ghassemi2024multi} studies learning-based beam management in OFDM cellular systems, where beam indices are selected from a predefined codebook to improve system throughput.
A two-stage learning framework is proposed, in which beam group selection is decoupled from fine-grained beam index selection to enable efficient beam decision making.
A multi-modal Transformer fuses heterogeneous sensing data to predict an optimal beam group, thereby reducing the RL action space and enabling fast beam selection.
However, \cite{ghassemi2024multi} is limited to a single Base Station (BS) beam management scenario.
To extend beam management to coordinated multi-BS transmission in Multiple-Input Multiple-Output (MIMO)-OFDM systems, \cite{xu2025fully} studies location-aided coordinated precoding via hierarchical RL to maximize user throughput without CSI feedback.
A Transformer encoder learns subcarrier correlation patterns to support fine-grained subband-level precoding decisions in the frequency domain.

\subsubsection{Modulation}

For multicellular Radio Access Networks (RANs), \cite{peri2025offline} studies downlink link adaptation by optimizing the Modulation and Coding Scheme (MCS) selection under time- and frequency-varying radio conditions.
An offline RL framework learns the MCS policy from historical data without online exploration, while a decision Transformer models link adaptation as trajectory-based sequence prediction to capture temporal dependencies.
Compared with the industry baseline, the proposed Transformer-enhanced offline RL framework improves the average spectral efficiency by about 20.7\% \cite{peri2025offline}.
Different from \cite{peri2025offline}, the work in \cite{he2025doppler} considers highly dynamic Flying Ad Hoc Networks (FANETs), where modulation and waveform configurations are adjusted to cope with severe mobility-induced Doppler effects.
An RL-based framework is proposed to select between OFDM and flying-adaptive Orthogonal Time Frequency Space (OTFS) waveforms based on long-term radio-frequency observations, aiming to reduce bit-error rate.
A time-fused Transformer serves as the core policy model, capturing long-range temporal dependencies to guide adaptive waveform and modulation decisions.

%==============================================
\subsection{Network Access and Connectivity Control}
%==============================================

Network access, user association, scheduling, and handover are fundamental control functions in wireless networks that jointly govern connectivity, resource utilization, and service continuity.
Recent studies increasingly integrate the Transformer into DRL frameworks to address high-dimensional, temporally coupled, and structurally constrained decision processes, as summarized in Table~\ref{tab:SecIII_B}.

\begin{table*}[t] 
\caption{Summary of Transformer-Enhanced DRL for Network Access and Connectivity Control.} 
\label{tab:SecIII_B} 
\centering 
\renewcommand{\arraystretch}{2.2} 
\begin{tabular}{|l|l|l|l|l|l|} 
\hline 

\textbf{Subcategory} & \textbf{Ref.} & \textbf{Scenario} & \textbf{Objective} & \textbf{Proposed Method} & \textbf{Transformer Role} \\ 
\hline 

\makecell[l]{Network access} & \cite{chen2025multi} & \makecell[l]{Multiple-access \\ networks} & \makecell[l]{Throughput maximization} & \makecell[l]{Multi-task MARL} & \makecell[l]{Scalable global state \\ aggregation} \\ 
\hline 

\multirow{5}{*}{User association}
& \cite{li2024prompt}  & \makecell[l]{5G cellular networks} & \makecell[l]{Energy consumption minimization} & \makecell[l]{Prompt Decision Transformer} & \makecell[l]{Prompt-guided policy \\ learning} \\ 
\cline{2-6}
& \cite{zhang2024handover}  & \makecell[l]{Fully decoupled \\ RAN} & \makecell[l]{Long-term mismatch minimization} & \makecell[l]{DDQN + Transformer-assisted \\ SAC} & \makecell[l]{Inter-link dependency \\ modeling} \\ 
\cline{2-6}
& \cite{negassa2025dynamic}  & \makecell[l]{UAV-assisted THz \\ networks} & \makecell[l]{Spectral efficiency maximization} & \makecell[l]{Transformer-enhanced DRL} & \makecell[l]{Attention-based feature \\ refinement} \\ 
\cline{2-6}
& \cite{gong2025transcomm}  & \makecell[l]{UAV-satellite \\ networks} & \makecell[l]{Throughput maximization \\ and energy reduction} & \makecell[l]{Transformer-based MARL} & \makecell[l]{Sequential multi-agent \\ coordination} \\ 
\cline{2-6}
& \cite{li2024dynamic}  & \makecell[l]{Terrestrial-satellite \\ networks} & \makecell[l]{Energy efficiency maximization} & \makecell[l]{Transformer-based offline RL} & \makecell[l]{Variable-length sequence \\ policy learning} \\ 
\hline

\multirow{6}{*}{Scheduling}
& \cite{di2024resource}  & \makecell[l]{Multi-user downlink \\ networks} & \makecell[l]{Spectral efficiency maximization} & \makecell[l]{Transformer-enhanced PPO} & \makecell[l]{Structured dependency \\ modeling} \\ 
\cline{2-6}
& \cite{nerondat2025efficient}  & \makecell[l]{5G networks} & \makecell[l]{Packet loss and delay violation \\ minimization} & \makecell[l]{Action-branching Q-learning \\ with Transformer} & \makecell[l]{Global state representation \\ learning} \\ 
\cline{2-6}
& \cite{qiao2024ai}  & \makecell[l]{Open radio access \\ networks} & \makecell[l]{Bandwidth aggregation efficiency \\ maximization} & \makecell[l]{Transformer-assisted \\ Actor-Critic DRL} & \makecell[l]{Per-path throughput \\ upper-bound prediction} \\ 
\cline{2-6}
& \cite{nerondat2025Transformer}  & \makecell[l]{Multi-link packet \\ scheduling} & \makecell[l]{Packet loss minimization} & \makecell[l]{Transformer-enhanced deep \\ Q-learning} & \makecell[l]{Global state aggregation} \\ 
\cline{2-6}
& \cite{li2025early}  & \makecell[l]{Radar work mode \\ recognition} & \makecell[l]{Early and accurate recognition} & \makecell[l]{CNN-Transformer + RL} & \makecell[l]{Global feature extraction} \\ 
\cline{2-6}
& \cite{xu2024optimizing}  & \makecell[l]{Dynamic 5G BS \\ activation} & \makecell[l]{Energy consumption minimization} & \makecell[l]{Transformer-assisted offline \\ RL} & \makecell[l]{Sequence modeling} \\ 
\hline

\multirow{2}{*}{Handover}
& \cite{chaccour2024joint}  & \makecell[l]{RIS-assisted THz \\ networks} & \makecell[l]{QoE improvement and handover \\ cost reduction} & \makecell[l]{Multi-agent hysteretic DRL} & \makecell[l]{Future sensing information \\ prediction} \\
\cline{2-6}
& \cite{aouedi2025hybrid}  & \makecell[l]{5G non-terrestrial \\ networks} & \makecell[l]{QoS improvement and handover \\ reduction} & \makecell[l]{Transformer-assisted A2C} & \makecell[l]{Trajectory prediction} \\ 
\hline
\end{tabular}
\vspace{-2em}
\end{table*}

\subsubsection{Network access}

For dynamic wireless multiple-access systems, \cite{chen2025multi} studies distributed access control where stations make binary transmission decisions under varying network sizes and traffic conditions.
A multi-task MARL framework improves throughput and fairness, with a Transformer-based centralized critic aggregating variable-size station information via attention to guide decentralized decisions.
The resulting Transformer-enhanced DRL approach achieves up to 95\% of the throughput upper bound \cite{chen2025multi}.

\begin{figure}[t]
    \centering
    \includegraphics[width=\linewidth]{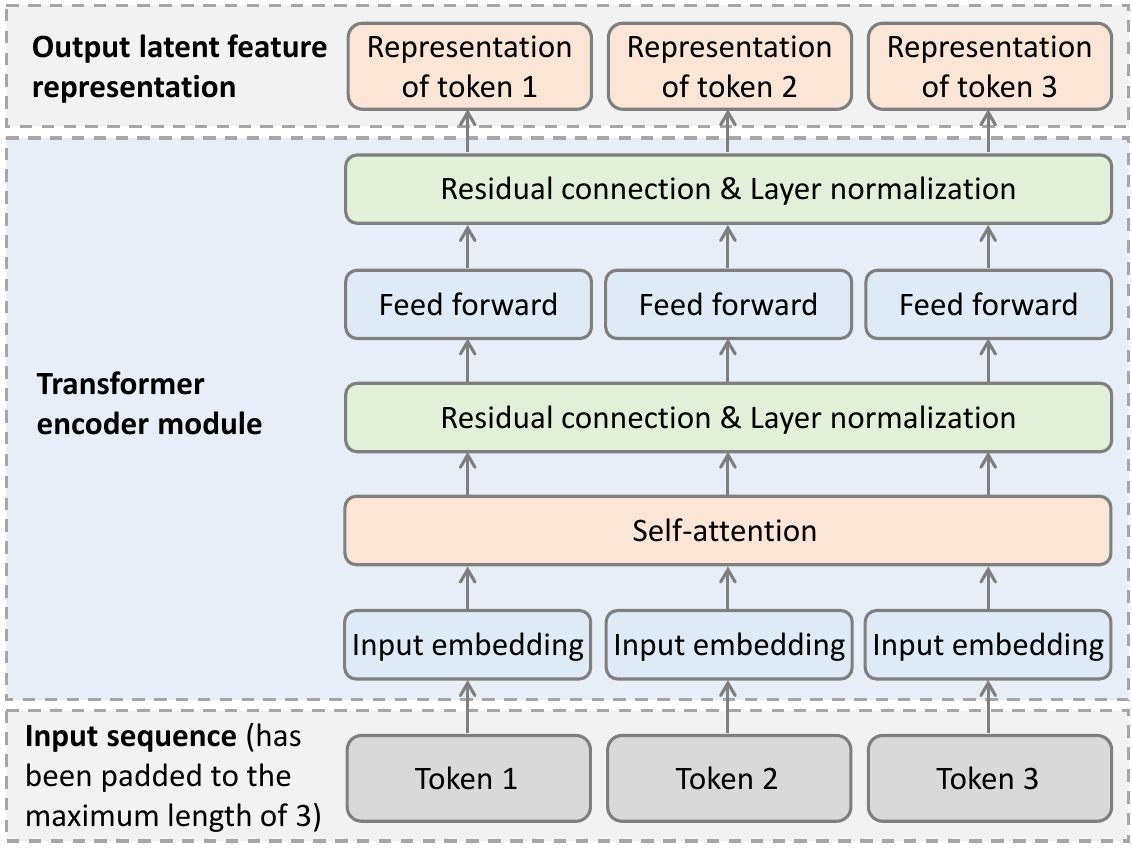}
    \caption{Transformer-based link-state aggregation framework, in which the encoder treats the state vectors of all links associated with a UE as a sequence of tokens to learn holistic latent representations from a variable number of links \cite{zhang2024handover}.}
    \vspace{-2em}
    \label{fig:SecIII_B}
\end{figure}

\subsubsection{User association}

For 5G cellular networks, \cite{li2024prompt} studies energy-efficient BS operation, where user association is jointly considered with sleep mode selection and antenna switching.
An offline learning framework is adopted to minimize overall BS energy consumption while satisfying QoS constraints.
A prompt decision Transformer learns association and operation policies from historical trajectories, enabling robust generalization across different network scales without online retraining.
Compared with conventional DRL baselines, the proposed prompt decision Transformer achieves a 75\% reduction in User Equipment (UE) drop ratio \cite{li2024prompt}.
While \cite{li2024prompt} assumes single-BS association per UE, \cite{zhang2024handover} considers multiple BSs serving each UE.
For fully decoupled RANs, \cite{zhang2024handover} studies handover-free multi-connectivity UE-BS association to support time-varying rate demands.
A hierarchical DRL framework jointly optimizes UE-BS association and downlink power control to enable seamless connectivity without frequent handovers.
A Transformer-assisted actor-critic model captures inter-link dependencies to coordinate power allocation across associated links, as shown in Fig.~\ref{fig:SecIII_B}.
However, \cite{li2024prompt} and \cite{zhang2024handover} focus on terrestrial cellular networks and are not directly applicable to non-terrestrial or hybrid networks involving UAVs and satellites.
For UAV-assisted Terahertz (THz) networks, \cite{negassa2025dynamic} studies dynamic user-cell association under high-mobility conditions.
A DRL-based framework jointly optimizes user association with beam alignment and Doppler compensation to improve link robustness and throughput.
A Transformer-enhanced module refines channel- and beam-related features through attention-based modeling, enabling more reliable and spectrally efficient association decisions.
More generally, for hybrid non-terrestrial networks, \cite{gong2025transcomm} studies UAV-satellite association to improve end-to-end throughput while reducing energy consumption.
A MARL framework is developed to coordinate UAV association and mobility decisions in dynamic environments.
A Transformer-based MARL model encodes joint observations and generates coordinated actions via autoregressive decision modeling, enabling efficient multi-agent association control.
Further, in integrated terrestrial-satellite networks, \cite{li2024dynamic} studies dynamic user association between terrestrial BSs and low-Earth-orbit satellites to improve overall energy efficiency under service constraints.
User association and transmission power control are jointly optimized using an offline RL framework to adapt to dynamic network conditions.
An elastic decision Transformer generates decisions from variable-length trajectories, enabling fast convergence and robust generalization.

\subsubsection{Scheduling}

For multi-user cellular downlink systems, \cite{di2024resource} studies multi-slot resource block scheduling to jointly improve spectral efficiency and user fairness.
A PPO-based DRL framework is developed to optimize multi-slot scheduling policies by modeling user-resource allocation dependencies over time.
A Transformer-based actor captures dependencies between user sequences and resource block assignments via attention mechanisms, enabling scalable and structured downlink scheduling.
Compared with conventional DRL schemes, the Transformer-based PPO improves spectral efficiency by about 10\% \cite{di2024resource}.
Different from \cite{di2024resource}, the work in \cite{nerondat2025efficient} considers 5G heterogeneous traffic demands and studies dynamic resource block assignment to reduce packet losses and delay violations.
A DRL-based framework addresses the large combinatorial action space of per-resource-block scheduling across varying numbers of users.
An encoder-only Transformer provides a global permutation-invariant state representation, enabling scalable action-branching Q-learning.
Further extending scheduling to multipath traffic, \cite{qiao2024ai} investigates dynamic downlink traffic distribution across heterogeneous paths to enhance bandwidth aggregation efficiency.
A DRL-based scheduler adapts packet scheduling ratios to path heterogeneity and time-varying wireless conditions.
A Transformer-based throughput prediction module estimates per-path achievable throughput from historical measurements, thereby improving scheduling efficiency and energy performance.
More generally, for centralized multi-link packet scheduling, \cite{nerondat2025Transformer} considers packet transmission under finite buffers and strict delay constraints, where losses stem from buffer overflow and delay violations.
A deep Q-learning-based framework is developed to minimize the overall packet loss rate through dynamic transmission scheduling across links.
An encoder-only Transformer aggregates per-link buffer and channel states via attention, enabling permutation-invariant and scalable scheduling over a variable number of links.
Different from \cite{di2024resource} --\cite{nerondat2025Transformer}, the studies in \cite{li2025early} and \cite{xu2024optimizing} focus on work-mode scheduling decisions.
In \cite{li2025early}, an early decision scheduling problem is studied for radar work mode recognition, where the system determines whether to continue or stop signal observation to balance recognition accuracy and decision latency.
A Transformer captures long-range temporal dependencies in pulse descriptor word sequences, and RL exploits the encoded representations to guide sequential stop-or-continue decisions.
In \cite{xu2024optimizing}, an offline RL framework is proposed for energy-efficient operation in 5G cellular networks, where base station cells are dynamically activated or deactivated according to traffic conditions.
A decision Transformer learns cell on-off control policies from historical trajectories, capturing long-term temporal dependencies and enabling safe energy-saving BS activation scheduling without online exploration.

\subsubsection{Handover}

Considering Reconfigurable Intelligent Surface (RIS)-assisted THz systems, \cite{chaccour2024joint} investigates sensing-aware handover, where users dynamically associate with RIS subarrays to maintain robust high-frequency links.
A joint sensing-communication-AI framework is developed to reduce handover cost and enhance link reliability while improving Quality-of-Experience (QoE).
A Transformer-based generative model predicts sensing information to augment the RL state, enabling proactive and foresighted handover decisions.
Different from the indoor scenario in \cite{chaccour2024joint}, the work in \cite{aouedi2025hybrid} studies handover management for 5G non-terrestrial networks, where the high mobility of low-Earth-orbit satellites leads to frequent handovers and degrades service continuity.
An A2C framework is employed to improve QoS while reducing unnecessary handovers.
A Transformer-based trajectory prediction module provides short-horizon mobility forecasts that augment the RL state.
Compared with conventional DRL baselines, the proposed method reduces the number of handovers by up to 99\% \cite{aouedi2025hybrid}.

%==============================================
\subsection{Virtual Network Function Placement}
%==============================================

Virtual Network Function (VNF) placement is a fundamental resource allocation problem in Network Function Virtualization (NFV)-enabled systems, as it determines how Service Function Chains (SFCs) are mapped onto physical infrastructures under dynamic constraints. Unlike radio-level resource allocation, VNF placement involves structured decision spaces with combinatorial dependencies across nodes, links, and service graphs, challenging DRL methods that rely on local or myopic state representations. Recently, Transformers have been integrated into DRL frameworks to enhance global state modeling, long-term reasoning, and structure-aware decision-making for VNF placement, as summarized in Table~\ref{tab:SecIII_C}. As illustrative examples, recent Transformer-enhanced DRL works report a 23\% improvement in service request acceptance rate \cite{sahraoui2024energy}, a doubled SFC orchestration success rate \cite{wang2025distributed}, a 66.7\% reduction in end-to-end delay \cite{li2025multimodal}, and a 40\% reduction in system energy consumption \cite{sahraoui2025intelligent}.

\begin{table*}[t] 
\caption{Summary of Transformer-Enhanced DRL for Virtual Network Function Placement.} 
\label{tab:SecIII_C} 
\centering 
\renewcommand{\arraystretch}{2.2} 

\resizebox{\textwidth}{!}{%
\begin{tabular}{|c|l|l|l|l|} 
\hline 
\textbf{Ref.} & \textbf{Scenario} & \textbf{Objective} & \textbf{Proposed Method} & \textbf{Transformer Role} \\ 
\hline 
\cite{sahraoui2024energy} & \makecell[l]{NFV networks} & \makecell[l]{Service acceptance maximization \\ with energy minimization} & \makecell[l]{Transformer-based \\ Actor-Critic DRL} & \makecell[l]{Service graph encoding for one-shot \\ placement} \\ 
\hline 
\cite{wang2025distributed} & \makecell[l]{UAV networks} & \makecell[l]{Orchestration time and resource \\ consumption minimization} & \makecell[l]{Distributed decision \\ Transformer Actor-Critic} & \makecell[l]{Trajectory-based contextual feature \\ extraction} \\ 
\hline 
\cite{li2025multimodal} & \makecell[l]{Satellite-terrestrial \\ networks} & \makecell[l]{Long-term revenue \\ maximization} & \makecell[l]{Multimodal \\ PPO-based DRL} & \makecell[l]{Graph-based network \\ state encoding} \\ 
\hline 
\cite{sahraoui2025intelligent} & \makecell[l]{5G end-to-end \\ network slicing} & \makecell[l]{Request acceptance maximization \\ and energy minimization} & \makecell[l]{Transformer-based \\ Actor-Critic DRL} & \makecell[l]{Global dependency modeling for one-shot \\ slice placement} \\ 
\hline 
\cite{wu2023d3t} & \makecell[l]{5G mobile edge \\ computing networks} & \makecell[l]{End-to-end service delay and \\ rejection minimization} & \makecell[l]{DDQN decision Transformer} & \makecell[l]{Offline trajectory-based \\ action prediction} \\ 
\hline 
\end{tabular}%
}
\vspace{-2em}
\end{table*}

From an energy-aware perspective, \cite{sahraoui2024energy} studies dynamic placement of VNF graphs in NFV networks, jointly mapping all VNFs onto physical servers under resource and energy constraints. 
A Transformer-based actor-critic DRL framework maximizes service acceptance while reducing energy consumption. 
The Transformer captures ordered dependencies in service graphs to enable one-shot placement with improved efficiency and scalability.
Different from the general NFV infrastructure in \cite{sahraoui2024energy}, the work in \cite{wang2025distributed} explores distributed SFC orchestration in UAV swarm networks, embedding sequential VNFs across neighboring UAVs without central control.
A generative RL framework reduces orchestration time and resource usage, improving success rate and efficiency.
The decision Transformer encodes historical local trajectories to extract temporal and contextual features, guiding stable and distributed VNF placement via actor-critic learning.
Extending from UAV swarm networks in \cite{wang2025distributed} to heterogeneous space-air-ground infrastructures, \cite{li2025multimodal} studies dynamic SFC deployment in satellite-terrestrial networks, where VNFs are sequentially placed across heterogeneous nodes under time-varying topology and resource constraints. 
A multimodal RL framework maximizes long-term revenue by considering both deployment rewards and resource costs. 
A graph Transformer encodes network structure and dependencies, which are fused with QoS and global context to guide PPO-based placement decisions.
More broadly, \cite{sahraoui2025intelligent} investigates joint VNF and virtual link placement for 5G-and-beyond network slicing under latency and resource constraints. 
A Transformer-enhanced actor-critic DRL framework improves slice acceptance and reduces energy consumption, where the Transformer models global dependencies within slice requests and enables scalable one-shot placement.
Complementary to online DRL-based approaches, \cite{wu2023d3t} adopts offline learning for VNF placement in mobile edge networks, dynamically mapping incoming SFCs to edge servers.
A Double Deep Q-Network (DDQN)-assisted framework minimizes end-to-end delay and request rejection.
The decision Transformer models historical placement trajectories to capture long-term dependencies and directly infer actions in high-dimensional spaces.

%==============================================
\subsection{Content-Aware Rate Adaptation}
%==============================================

Content-aware rate adaptation adjusts the amount of transmitted information based on user requirements and network dynamics.
To address the resulting decision complexity, Transformers have been increasingly incorporated into DRL to enhance temporal modeling, perceptual awareness, and long-term decision consistency in multimedia transmission systems, as summarized in Table~\ref{tab:SecIII_D}.
For example, compared with conventional DRL, recent Transformer-enhanced DRL schemes have reported a 27\% reduction in analytics latency for edge video analytics \cite{wang2023edge} and a 57\% improvement in QoE for 360° video streaming \cite{wang2024madrl}.

\begin{table*}[t] 
\caption{Summary of Transformer-Enhanced DRL for Content-Aware Rate Adaptation.} 
\label{tab:SecIII_D} 
\centering 
\renewcommand{\arraystretch}{2.2} 

% Bọc bảng trong \resizebox để ép giãn vừa 2 cột (\textwidth) một cách an toàn
\resizebox{\textwidth}{!}{%
\begin{tabular}{|c|l|l|l|l|l|} 
\hline 
\textbf{Ref.} & \textbf{Scenario} & \textbf{Problem Focus} & \textbf{Objective} & \textbf{Proposed Method} & \textbf{Transformer Role} \\ 
\hline 
\cite{wang2023edge} & \makecell[l]{Multi-device edge \\ video analytics} & \makecell[l]{Adaptive frame degradation} & \makecell[l]{Accuracy-latency \\ tradeoff} & \makecell[l]{Transformer-enhanced \\ SAC} & \makecell[l]{Temporal state-action sequence \\ encoding} \\ 
\hline 
\cite{wang2024madrl} & \makecell[l]{360° video streaming} & \makecell[l]{Viewport-aware adaptive \\ bitrate allocation} & \makecell[l]{QoE maximization} & \makecell[l]{Transformer-assisted \\ MAPPO} & \makecell[l]{Multiviewpoint trajectory and \\ probability prediction} \\
\hline 
\cite{wang2025hierarchical} & \makecell[l]{Real-time XR video \\ transmission} & \makecell[l]{Adaptive video quality \\ selection} & \makecell[l]{QoE maximization} & \makecell[l]{Transformer-enhanced \\ Hierarchical MAPPO} & \makecell[l]{Multi-agent coordinator} \\
\hline 
\cite{pan2024quality} & \makecell[l]{Real-time XR video \\ transmission} & \makecell[l]{Video bitrate adaptation} & \makecell[l]{QoE maximization} & \makecell[l]{Transformer-based PPO \\ + multi-step DQN} & \makecell[l]{Temporal feature extraction} \\
\hline 
\cite{zhou2023transabr} & \makecell[l]{UHD video streaming} & \makecell[l]{Adaptive bitrate selection} & \makecell[l]{QoE maximization} & \makecell[l]{Transformer-enhanced \\ SAC} & \makecell[l]{Sequential state representation \\ learning} \\
\hline 
\cite{jin2025t} & \makecell[l]{Underwater wireless \\ sensor networks} & \makecell[l]{Quantization bit allocation} & \makecell[l]{Energy consumption \\ minimization} & \makecell[l]{Transformer-enhanced \\ D3QN} & \makecell[l]{State feature extraction via \\ self-attention} \\
\hline 
\end{tabular}%
}
\vspace{-2em}
\end{table*}

For edge video analytics, \cite{wang2023edge} proposes adaptive frame degradation before transmission to balance accuracy and latency. 
A Soft Actor-Critic (SAC)-based DRL framework jointly optimizes frame quality and bandwidth allocation.
A Transformer-enhanced actor-critic architecture encodes historical state-action sequences to capture long-term dependencies, thereby achieving an improved accuracy-latency tradeoff under time-varying content characteristics.
Different from the video analytics scenario in \cite{wang2023edge}, the work in \cite{wang2024madrl} focuses on QoE-oriented 360° video streaming with viewport-aware bitrate adaptation. 
A MAPPO-based framework coordinates region-level adaptation using predicted user viewpoints. 
A multimodal spatio-temporal attention Transformer forecasts viewpoint trajectories and probabilities to support robust, fine-grained rate control.
Further extending to real-time Extended Reality (XR) video transmission, \cite{wang2025hierarchical} studies joint video quality selection and wireless resource allocation under strict latency constraints.
A hierarchical MAPPO framework performs multi-timescale source-channel optimization.
A multi-agent Transformer models interactions among wireless resource blocks via attention-based coordination, supporting scalable and consistent multi-agent decisions.
Similar to \cite{wang2025hierarchical}, which focuses on XR video transmission, \cite{pan2024quality} explores coordinated bitrate adaptation and wireless scheduling for QoE improvement.
A cross-layer framework decouples optimization across time scales, with a Transformer-enhanced PPO module capturing temporal patterns from historical network states to stabilize rate adaptation.
More generally, for Ultra-High-Definition (UHD) video streaming over mixed-band 5G networks, \cite{zhou2023transabr} investigates adaptive bitrate control for each video chunk under highly variable throughput conditions. 
A SAC-based DRL framework maximizes long-term QoE, while a Transformer encodes historical streaming states to capture temporal dependencies.
Different from the above works, \cite{jin2025t} targets underwater wireless sensor networks and studies quantization bit adaptation for sensor precision control.
A Double Dueling Deep Q-Network (D3QN)-based framework adjusts per-node quantization to minimize energy while preserving tracking performance.
A Transformer extracts structured state features via self-attention, enabling stable and efficient bit allocation in dynamic underwater environments.

%==============================================
\subsection{Joint Computation-Communication Resource Allocation}
%==============================================

Joint computation-communication resource allocation has emerged as a critical problem in dynamic network conditions. Unlike single-domain resource allocation, joint optimization introduces high-dimensional decision spaces and strong inter-resource dependencies, posing significant challenges for conventional DRL. Recently, the Transformers have been integrated into DRL to enhance global representation learning, structured reasoning, and long-term dependency modeling, enabling more effective joint resource management.

In \cite{you2025reacritic}, transmission power, bandwidth, and computation resources are jointly optimized, and a large reasoning Transformer-based critic, namely ReaCritic, is integrated into the DRL framework to enhance state-action value estimation and generalization, yielding up to a 170\% gain in final episodic return over the vanilla DRL baseline.
Different from \cite{you2025reacritic}, the work in \cite{han2024joint} considers dynamic task arrivals in mobile-edge computing networks, jointly optimizing offloading, caching, and resource provisioning.
An online centralized DRL framework minimizes long-term task execution time without prior knowledge of future tasks, where a Transformer-based actor-critic handles the high-dimensional state-action space, enabling scalable joint resource management.
From a long-term scheduling perspective, \cite{zhou2024ec} investigates dynamic resource allocation in edge-cloud environments, where incoming tasks are assigned to distributed edge servers. 
An improved SAC framework is developed to reduce task rejection and improve system efficiency, with Transformer modules embedded in the actor-critic architecture to encode long scheduling sequences and global system states for more adaptive resource allocation.

%==============================================
\subsection{Lessons Learned}
%==============================================

From the surveyed studies, four key factors explain the effectiveness of Transformers in DRL-based resource allocation. 
Transformers can effectively model long-term temporal dependencies, enabling better decisions in trajectory-dependent environments. 
Self-attention enables scalable modeling of inter-agent dependencies in large networks, supporting generalizable policies across dynamic and heterogeneous topologies.
Transformers capture interdependencies among structured actions, making it well suited for combinatorial problems like one-shot placement. Attention-based fusion of sensing, prediction, and twin-state inputs empowers Transformers to support adaptive, context-aware decision making.

%==============================================
\section{Transformer-based RL for Computation Offloading}
\label{sec:offloading}
%==============================================
Computation offloading has been studied under diverse edge intelligence settings, leading to varied formulations and objectives. Conventional RL algorithms effectively address such problems, but they often suffer from unstable training and limited scalability in dynamic, high-dimensional wireless environments. Transformer- and diffusion-enabled RL frameworks have recently been used to improve state representation and decision-making over complex action spaces. This section therefore organizes the literature by application scenarios, including computation resource allocation, joint computation--communication optimization, mobility-aware offloading, and AI-service-oriented offloading. Task dependencies are treated as modeling features within each scenario. This scenario-driven taxonomy provides a coherent view of offloading design across edge intelligence systems.

\subsection{Computation Resource Allocation}\label{Section 4-1}
%Computation resource allocation focuses on scheduling and distributing limited computing resources to optimize system performance, while communication effects are typically abstracted or treated as secondary constraints.
This category primarily addresses computation-centric decisions such as processor assignment and task scheduling in heterogeneous edge-cloud and data center environments, as summarized in Table \ref{Section 4-1}.

\begin{table*}[t]
\centering
\caption{Summary of Transformer-Enhanced DRL for Computation Resource Allocation.}
\label{tab:SecIV_A} 
\renewcommand{\arraystretch}{1.4} 

% Tuyệt chiêu \resizebox giúp kéo giãn toàn bộ bảng vừa khít lề (\textwidth)
\resizebox{\textwidth}{!}{%
\begin{tabular}{|l|c|l|l|l|l|}
\hline
\textbf{Subcat.} & \textbf{Ref.} & \textbf{Scenario} & \textbf{Objective} & \textbf{Method} & \textbf{Transformer Role} \\
\hline

\multirow{4}{*}{\makecell[l]{Computation \\Scheduling \\and Consolidation}} 
& \cite{jeong2023sdn} & VM consolidation (SDN) 
& \makecell[l]{Energy-efficient \\ scheduling} 
& \makecell[l]{Transformer-based \\ DRL} 
& \makecell[l]{Global workload \\ modeling} \\ \cline{2-6}

& \cite{zhou2025Transformer} & \makecell[l]{Data center \\ power control} 
& \makecell[l]{Efficient power \\ allocation} 
& \makecell[l]{Transformer-embedded \\ RL} 
& \makecell[l]{Temporal workload \\ modeling} \\ \cline{2-6}

& \cite{song2025edge} & Holographic edge service 
& \makecell[l]{Latency-energy-\\ QoE tradeoff} 
& Hierarchical RL 
& \makecell[l]{Multimodal spatio-temporal \\ feature extraction} \\
\hline

\multirow{7}{*}{\makecell[l]{Distributed or\\ Cooperative\\ Resource \\Allocation}} 
& \cite{chen2025dynamic} & IoV task allocation 
& \makecell[l]{Delay reduction, \\ stability} 
& \makecell[l]{DRL with \\ feedback control} 
& \makecell[l]{Feature-prioritization \\ module} \\ \cline{2-6}

& \cite{fan2025jpds} & \makecell[l]{Vehicle routing \\ allocation} 
& \makecell[l]{Task-routing \\ optimization} 
& \makecell[l]{RL-based \\ allocation NN} 
& MDP modelling \\ \cline{2-6}

& \cite{zhou2025multi} & Coded ML workload split 
& Worker selection 
& \makecell[l]{Multi-agent \\ Transformer RL} 
& \makecell[l]{Inter-agent \\ dependency} \\ \cline{2-6}

& \cite{wang2025tf} & Edge-cloud IoT scheduling 
& Distributed efficiency 
& TF-DDRL 
& \makecell[l]{Distributed state \\ modeling} \\ \cline{2-6}

& \cite{yao2025eftr} & Fault-aware rescheduling 
& Robust task recovery 
& \makecell[l]{Learning-based \\ rescheduler} 
& \makecell[l]{Two-stage \\ embedding} \\
\hline
\end{tabular}%
}
\end{table*}

\subsubsection{Centralized Computation Scheduling and Consolidation} %This subcategory focuses on centralized computation resource management, where a single controller coordinates task scheduling and resource consolidation. Typical problems include VM consolidation, computing power scheduling, and centralized application placement in data centers or edge-cloud systems.

\begin{figure}[t]
    \centering
    \includegraphics[width=\linewidth]{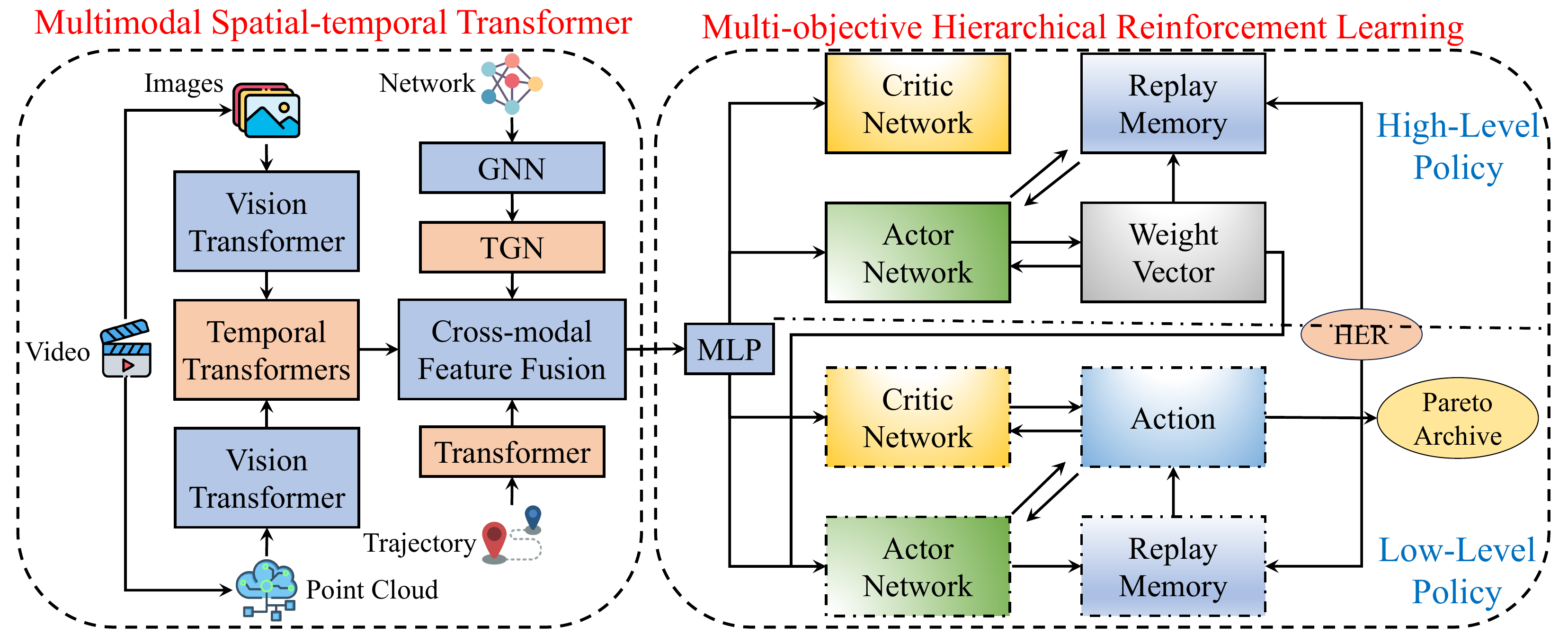}
    \caption{A representative architecture where the multimodal spatio-temporal Transformer provides auxiliary feature representations for hierarchical RL \cite{song2025edge}. TGN: temporal graph networks; GNN: graph neural networks; HER: hindsight experience replay.}
    \vspace{-2em}
    \label{fig:multi}
\end{figure}

From the perspective of computation offloading and workload placement, the authors of \cite{jeong2023sdn} study Virtual Machine (VM) consolidation and resource management in Software-Defined Networking (SDN)-enabled cloud environments. %The consolidation problem is formulated to dynamically migrate and allocate VMs to balance resource utilization and reduce operational overhead under time-varying traffic demands. 
A Transformer-enhanced DRL framework is developed, integrating a Transformer into the policy learning process to capture long-range dependencies and correlations among network and resource states. %Within this framework, the learned policy determines VM migration and consolidation actions based on global system observations provided by the SDN controller. 
%By leveraging Transformer-based representation learning, the proposed approach improves the modeling of complex spatio-temporal dynamics inherent in large-scale cloud infrastructures. %The overall decision-making process follows a learning-based optimization paradigm, where Transformer functions act as expressive function approximators within the RL pipeline. Experimental results demonstrate that the proposed method achieves more efficient VM consolidation and resource utilization than baseline consolidation strategies, thereby supporting adaptive workload management in dynamic cloud environments.
%
In the face of increasing computational demand and complex operational constraints, task-level computing power scheduling for distributed data centers is investigated in \cite{zhou2025Transformer}. %The scheduling problem is modeled as a Markov decision process to jointly account for energy consumption, job latency, and real-time control requirements in edge-enabled cloud environments. 
A Transformer-embedded PPO framework is developed, integrating the Transformer into the policy learning process to capture temporal relationships across tasks and process variable-length system states. %Within this framework, PPO optimizes scheduling decisions while the Transformer enhances the representation of temporally correlated workload and resource dynamics. The proposed approach supports flexible load management and robust control across heterogeneous scheduling conditions. Numerical experiments are conducted using the Alibaba Cluster Trace V2017 dataset, demonstrating compatibility with various lower-level control policies. %The results show that the proposed method effectively reduces energy consumption while maintaining job latency within predefined thresholds, enabling a controllable trade-off between energy cost and latency in distributed data center scheduling.
The authors of \cite{song2025edge} study the holographic video communication in dynamic edge computing environments. %, where stringent bandwidth and computing requirements pose significant challenges to service adaptability and stability. 
As shown in Fig. \ref{fig:multi}, this architecture employs a Transformer for multimodal spatio-temporal feature extraction, while decision-making is carried out by a hierarchical RL controller. %The problem is formulated as a multi-objective optimization task to jointly address network dynamics, computational constraints, and quality-of-service requirements. 
%To this end, a Multi-Objective Hierarchical RL (MOHRL) framework is proposed, in which high-level policies determine stage-based goals and low-level policies execute real-time control decisions. %A multimodal spatio-temporal Transformer model is introduced to capture subtle variations in communication tasks and network conditions, providing informative state representations for the hierarchical RL framework. 
Within this design, RL handles decision-making at different temporal scales. At the same time, the Transformer serves as an auxiliary components that enhance the modeling of spatio-temporal dynamics rather than directly approximating policies or value functions. %The proposed framework supports adaptive holographic video transmission under dynamic edge network conditions. Experimental evaluations demonstrate improved adaptability and performance of holographic video communication services in practical edge computing scenarios.

\subsubsection{Distributed or Cooperative Resource Allocation} %This line of work considers distributed computing environments with multiple decision-making entities, where resource allocation decisions are made cooperatively or in a decentralized manner. %The emphasis is on worker selection, distributed scheduling, and coordination among edge nodes under limited global information.

In Internet of Vehicles (IoV) scenarios under B5G and 6G networks, the authors of \cite{chen2025dynamic} investigate the allocation of computational resources for high-concurrency, delay-sensitive transportation tasks. %The resource allocation problem is formulated by jointly modeling task latency requirements, heterogeneous computing capabilities, and energy consumption in highly dynamic vehicular environments. 
A Dynamic Self-Feedback (DSF) resource allocation framework is proposed, integrating a Transformer-based self-attention mechanism to emphasize critical resource characteristics during allocation. %The framework adaptively adjusts the dimensionality of attention heads based on predicted resource consumption, enabling differentiated handling of light and heavy computational tasks. In addition, attention embeddings are dynamically refined via a self-feedback mechanism, and the resulting feedback guides subsequent allocation rounds. 
%Under this learning-based paradigm, the Transformer serves as an expressive feature-prioritization module during the allocation process. %Simulation results show that the proposed method achieves up to 85\% task execution efficiency and reduces the number of timeout tasks by 38\%. Moreover, more than 50\% of tasks exhibit low energy consumption after allocation, while nearly 100\% of computing units maintain workloads below 40\%, demonstrating effective load balancing under high-concurrency IoV workloads.
%
%In agricultural operations, the entrance-dependent vehicle routing problem requires the consideration of field geometry and working line entrances when planning routes for agricultural vehicles. %Traditional heuristic approaches, however, often overlook these geometric and entrance constraints. 
The authors of \cite{fan2025jpds} introduce an RL-based Joint Probability Distribution Sampling Neural Network (JPDS-NN). The model follows an encoder-decoder architecture enhanced with graph Transformer and attention mechanisms, formulates routing as a Markov decision process, and is trained via proximal policy optimization. %Experimental results show that JPDS-NN reduces travel distance by 48.4-65.4\%, lowers fuel consumption by 14.0-17.6\%, and runs two orders of magnitude faster than baseline methods, while also achieving a 15-25\% performance improvement in dynamic scheduling scenarios.
In addition, the study in \cite{zhou2025multi} focuses on optimizing worker selection and workload allocation in distributed coded machine learning systems operating within large-scale dynamic edge networks. %, addressing challenges such as unstable connectivity, dynamic worker availability, and high-dimensional decision spaces.
A Multi-Agent Transformer-based workload Allocation and worker Selection (MAT-AS) scheme formulates the problem as a partially observable Markov game and integrates multi-agent RL with a Transformer module. %Experimental results demonstrate that MAT-AS achieves significant performance advantages: MAT-AS reduces task completion time by approximately 12\% compared to the best baseline while maintaining superiority across machine learning models of varying dimensions, and lowers overall computation cost by around 15\% relative to top baseline methods.
In \cite{wang2025tf}, the adaptive scheduling of heterogeneous IoT applications in edge and cloud computing environments is studied. %, characterized by dynamic and stochastic system behaviors. The scheduling problem is formulated to jointly address task dependencies, distributed resource coordination, and training efficiency under large-scale IoT workloads. 
To this end, a Transformer-enhanced distributed DRL approach, termed TF-DDRL, is proposed based on an actor-critic architecture. %The framework scales across multiple distributed servers and incorporates an off-policy correction mechanism to stabilize learning. 
Within this design, a Transformer is integrated to capture long-term dependencies among interdependent tasks, thereby enhancing temporal and structural representations in the decision process. %Experimental results from practical evaluations show that TF-DDRL reduces response time, energy consumption, monetary cost, and weighted cost by up to 60\%, 51\%, 56\%, and 58\%, respectively, compared with existing approaches. These results demonstrate the effectiveness of Transformer-enhanced distributed learning for scalable and efficient IoT task scheduling.
%
%In edge computing environments, sudden node failures often lead to task interruptions and performance degradation, while traditional task rescheduling methods struggle to adapt to dynamic faults and effectively model the complex relationships among tasks, resources, and failures. 
The authors in \cite{yao2025eftr} propose a heterogeneous graph neural network-based Directed Acyclic Graph (DAG) task rescheduling method for Edge computing, featuring a Transformer-based architecture and actor-critic RL. %Compared to baseline, such as the heterogeneous earliest finish time algorithm, the proposed scheme demonstrates significant performance improvements, reducing the makespan by at least 11.11\% and increasing the task rescheduling success rate by 4.20\%.

\subsection{Joint Computation-Communication Resource Allocation}\label{Section 4-2}
This category investigates computation offloading problems where computing decisions are tightly coupled with communication resource allocation, and the papers are summarized in Table \ref{tab:SecIV_B}.

\begin{table*}[t]
\footnotesize
\caption{Summary of Transformer-Enhanced DRL for Joint Computation-Communication Resource Allocation.}
\label{tab:SecIV_B} 
\renewcommand{\arraystretch}{1.2} 

% Đã sửa lại lỗi cú pháp ở dòng tabular* này
\begin{tabular*}{\textwidth}{@{\extracolsep{\fill}}|l|l|l|l|l|l|}
\hline
\textbf{Subcat.} & \textbf{Ref.} & \textbf{Scenario} & \textbf{Objective} & \textbf{Method} & \textbf{Transformer Role} \\
\hline

\multirow{10}{*}{\makecell[l]{Joint Offloading and \\Radio Resource\\ Allocation}} 

& \cite{gholipour2023tpto} & \makecell[l]{MEC offloading with\\ wireless coupling} 
& \makecell[l]{Latency-energy\\ minimization} 
& Transformer-based PPO 
& \makecell[l]{Encodes task-channel\\ correlation} \\ \cline{2-6}

& \cite{li2024Transformer} & \makecell[l]{Latency-sensitive\\ distributed MEC} 
& \makecell[l]{Delay minimization\\ under radio coupling} 
& Transformer-based DRL 
& \makecell[l]{Extracts long-range\\ task dynamics} \\ \cline{2-6}

& \cite{ni2025fast} & \makecell[l]{AAV-assisted MEC\\ with constraints} 
& \makecell[l]{Delay minimization\\ with UAV mobility} 
& Decision Transformer 
& \makecell[l]{Return-conditioned\\ policy modeling} \\ \cline{2-6}

& \cite{wu2025application} & \makecell[l]{Satellite edge\\ scheduling} 
& \makecell[l]{Latency and utilization\\ optimization} 
& Attention-enhanced PPO 
& \makecell[l]{Multi-dimensional\\ attention modeling} \\ \cline{2-6}

& \cite{xie2025towards} & \makecell[l]{Task-number adaptive\\ MEC offloading} 
& \makecell[l]{Adaptive offloading\\ under varying load} 
& Transformer-based DRL 
& \makecell[l]{Task-sequence\\ scalability modeling} \\ \cline{2-6}

& \cite{xu2025novel} & \makecell[l]{Multi-neighborhood\\ scheduling} 
& \makecell[l]{Topology-aware\\ scheduling} 
& Self-attention PPO 
& \makecell[l]{Neighborhood-level\\ state modeling} \\ \hline

\multirow{6}{*}{\makecell[l]{Joint Comm.-Comp\\ Coordination}} 

& \cite{han2025transformer} & \makecell[l]{Cloud-edge\\ distributed offloading} 
& \makecell[l]{Joint task and\\ resource management} 
& Transformer-based DRL 
& \makecell[l]{Captures cross-node\\ dependency} \\ \cline{2-6}

& \cite{shi2025collaborative} & \makecell[l]{Multi-agent AGV\\ transmission-computation} 
& \makecell[l]{Joint comm.-comp.\\ coordination} 
& Transformer-based MADRL 
& \makecell[l]{Models inter-agent \\interaction} \\ \cline{2-6} % Đã xóa một dòng \cline{2-6} bị lặp thừa ở đây

& \cite{yuan2025Transformer} & \makecell[l]{Cloud-Edge-End\\ video streaming} 
& \makecell[l]{Joint bandwidth-\\computation allocation} 
& Transformer-based MADRL 
& \makecell[l]{Cross-layer \\dependency modeling} \\ \cline{2-6}

& \cite{zhou2024blockchain} & \makecell[l]{5G private network\\ offloading (blockchain)} 
& \makecell[l]{Secure joint offloading\\ and allocation} 
& Transformer-based MADRL 
& \makecell[l]{Inter-agent coordination\\ modeling} \\
\hline

\end{tabular*}
\vspace{-2em}
\end{table*}

\subsubsection{Joint Offloading and Radio Resource Allocation} %This subcategory focuses on the tight coupling between task offloading decisions and wireless resource allocation. The studies typically optimize offloading ratios along with radio resources, such as bandwidth, transmission power, or decoding order, to mitigate interference and reduce latency.

To handle task dependencies in edge computing environments, the authors of \cite{gholipour2023tpto} propose a computation offloading approach for latency-sensitive IoT applications. %Specifically, the offloading problem is formulated to determine task placement decisions that minimize end-to-end application latency when computation tasks are represented as a directed graph. 
To address the complexity induced by task dependencies and heterogeneous user preferences, a DRL approach, termed TPTO, is proposed by integrating Transformer networks with PPO. %Within this framework, a Transformer is employed to capture structural and sequential dependencies among tasks, while PPO governs the policy optimization process. The proposed method supports wireless offloading from user devices to edge servers and accommodates diverse application graph characteristics. 
Through Transformer-enhanced policy learning, TPTO enables coordinated offloading decisions for dependent tasks under dynamic edge computing conditions. %Performance evaluations demonstrate that, for fat application graphs, TPTO reduces latency by at most 30.24\% compared to other benchmarks. In addition, the proposed approach achieves approximately 2.5$\times$ faster training time than an existing DRL-based offloading method, indicating improved learning efficiency.
The authors in \cite{li2024Transformer} conceive distributed computation offloading for latency-sensitive tasks in mobile edge computing. The offloading problem is formulated to capture time-varying wireless conditions, heterogeneous computing resources, and strict delay requirements across distributed edge nodes. A Transformer-assisted DRL framework is developed, integrating a Transformer into the policy learning process to enhance the representation of high-dimensional, temporally correlated system states. Within this framework, RL determines offloading decisions across distributed nodes, while the Transformer captures long-range dependencies among network, queueing, and computation states. %The proposed design supports decentralized decision-making by enabling each agent to infer offloading actions based on locally observed information enriched through Transformer-based feature extraction. Through this integration, the framework enables coordinated, adaptive offloading under dynamic latency constraints. Experimental evaluations demonstrate improved task latency and resource utilization performance compared to baseline learning-based offloading approaches in distributed mobile edge computing scenarios.
The authors of \cite{ni2025fast} propose a Constrained Decision Transformer (CDT) framework based on offline pre-training and online fine-tuning, jointly optimizing the autonomous aerial
vehicle's trajectory and computing resource allocation to maximize fairness-based throughput under battery capacity and quality of service constraints. CDT formulates the sequential optimization problem as a constrained Markov decision process and inherently embodies a sequence-level policy by transforming RL into a sequence modeling task. %Compared with benchmark deep RL algorithms, the CDT effectively improves fairness-based throughput by 2.1\%-7.4\% under constraints and stably satisfies core constraints across unseen constraint ranges.
Given stringent latency constraints and dynamic resource limitations, the authors of \cite{wu2025application} study task scheduling and resource management in satellite edge computing systems. %The scheduling problem is formulated to capture time-varying satellite resources, heterogeneous task demands, and the coupling between communication and computation processes. 
A PPO-based scheduling framework is developed that integrates multi-dimensional attention mechanisms into the policy learning process. %Within this framework, attention modules enhance the representation of complex system states by selectively emphasizing relevant resource and task attributes across multiple dimensions. 
The RL policy determines scheduling actions based on attention-enhanced state representations, enabling adaptive decision-making in dynamic satellite-edge environments. %By incorporating attention mechanisms into the PPO architecture, the proposed approach enables flexible, scalable scheduling across varying workload conditions. Experimental evaluations demonstrate effective task scheduling and improved resource utilization in satellite-edge computing scenarios.
Considering dynamically varying numbers of tasks and users, the authors in \cite{xie2025towards} investigate computation offloading in mobile edge computing systems. The offloading problem is formulated to accommodate variable-length system states arising from stochastic task arrivals and departures under heterogeneous resource conditions. A Transformer-based DRL framework is proposed to enable task-number-adaptive decision-making within a unified policy model. %In this framework, the Transformer is integrated into the policy learning process to flexibly encode variable-sized task sets and capture interactions among concurrent offloading requests. The learned policy determines offloading actions without requiring a fixed task dimension, thereby enabling scalable, adaptive operation. By leveraging attention mechanisms, the approach maintains consistent decision quality across different workload intensities. 
The overall design follows a learning-based offloading paradigm, where the Transformer enhances state representation while RL optimizes the policy. %Numerical evaluations demonstrate effective adaptation to varying task numbers and stable offloading performance in dynamic mobile edge computing (MEC) environments.
Characterized by heterogeneous service demands and dynamic network conditions, the authors in \cite{xu2025novel} discuss task scheduling and resource management in satellite edge computing systems. %The scheduling problem is formulated to jointly account for satellite resource constraints, time-varying workloads, and the interaction between communication and computation processes. 
A self-attention-enhanced multi-neighborhood PPO framework is proposed to support adaptive scheduling decisions. %Within this framework, self-attention mechanisms are integrated into the policy learning process to emphasize relevant task and resource features across multiple neighborhood representations.
The RL policy determines scheduling actions based on attention-enhanced state representations, enabling flexible adaptation to dynamic satellite-edge environments. By incorporating multi-neighborhood modeling, the proposed approach captures interactions among tasks and resources at different granularities. %The overall design follows a learning-based scheduling paradigm, where self-attention improves state representation while PPO governs policy optimization. Experimental evaluations demonstrate effective task scheduling and improved resource utilization in satellite-edge computing scenarios.

\subsubsection{Joint Computation-Communication Coordination in Distributed Systems} %This line of work considers distributed systems where computation and communication resources are jointly coordinated across multiple nodes or agents. The emphasis is on system-wide cooperation rather than per-link radio optimization.

Consider delay-sensitive applications in Industrial Cyber-Physical Systems (ICPS), where joint task offloading and resource management are proposed in \cite{han2025transformer}, utilizing cloud-edge computing architectures. The problem is formulated to support real-time decision-making and energy-efficient resource utilization in highly dynamic, large-scale ICPS environments. To cope with the hybrid and high-dimensional action space, a Distributed Transformer-based Actor-Critic (DTAC) algorithm is proposed by integrating Transformer models into an actor-critic learning framework. %A centralized model is first trained to capture coordination patterns among user equipment (UE), and the learned knowledge is subsequently transferred to decentralized agents through a decentralized transfer learning mechanism.
%Under this design, each UE independently manages local offloading and resource decisions based solely on local observations, while benefiting from globally learned coordination structures. 
Transformers are employed to enhance the representation and processing of complex system states during policy learning. The proposed framework enables scalable and adaptive computation offloading without incurring excessive signaling overhead. %Simulation results demonstrate that DTAC achieves improved performance compared to existing multi-agent RL and transfer learning schemes across both small- and large-scale scenarios. In addition, DTAC and the decentralized transfer learning strategy reduce training costs by 73\%, highlighting their practicality for ICPS deployment.
Based on MADRL enhanced with a Transformer module, a Transferable-joint Task Offloading and Multi-Channel Access (T2OMCA) algorithm is proposed in \cite{shi2025collaborative}. Simulation results demonstrate that T2OMCA can achieve an average task completion rate exceeding 90\% across scenarios with varying numbers of edge nodes and automated guided vehicles (AGVs). The authors in \cite{yuan2025Transformer} propose a JROC framework that integrates a smart contract-based adaptive incentive mechanism and an adaptive Transformer-based MARL algorithm, leveraging a Transformer-based centralized critic network to capture long-range dependencies and inter-agent interactions. In 5G private networks, the proliferation of terminal devices and diverse service demands may face high-dimensional state spaces and security risks. Blockchain technologies can be introduced, and then the joint resource allocation and computation offloading can be solved by the Transformer-assisted MARL algorithm as presented in  \cite{zhou2024blockchain}. %Experimental results indicate that, compared to conventional MARL (without Transformer) and Lyapunov optimization baselines, their method achieves a QoS satisfaction rate exceeding 94\%, representing an improvement of 4-6 percentage points.

\subsection{Mobility-Aware and Dynamic Offloading}\label{Section 4-3}
Mobility-aware offloading addresses computation decisions under time-varying system dynamics induced by user, vehicle, or UAV mobility. These studies emphasize adaptive task placement and resource allocation in highly dynamic and non-stationary environments.%, as summarized in \ref{tab:SecIV_C}.

%\begin{table*}[t]
%\centering
%\footnotesize
%\caption{Summary of Transformer-Enhanced DRL for Mobility-Aware and Dynamic Offloading.}
%\label{tab:SecIV_C} 
%\renewcommand{\arraystretch}{1.4} 

% Ép giãn toàn bộ bảng vừa khít 2 cột một cách an toàn
%\resizebox{\textwidth}{!}{%
%\begin{tabular}{|l|c|l|l|l|l|}
%\hline
%\textbf{Subcategory} & \textbf{Ref.} & \textbf{Scenario} & %\textbf{Objective} & \textbf{Method} & \textbf{Transformer Role} \\
%\hline

%\multirow{3}{*}{\makecell[l]{User/Vehicle Mobility-\\Aware Offloading}} 
%& \cite{chen2025IFresher} & Mobile AR edge system 
%& Freshness-latency tradeoff 
%& Multi-agent RL 
%& Coordinated decision-making \\ \cline{2-6}

%& \cite{ghosh2025mgco} & Mobility-aware edge-cloud 
%& Dynamic offloading opt. 
%& Generative offloading 
%& Contextual reasoning \\ \cline{2-6}

%& \cite{zou2025deployment} & Vehicular digital twins 
%& Sync. and deployment opt. 
%& Learning framework 
%& Relational modeling \\
%\hline

%\multirow{2}{*}{\makecell[l]{UAV/AAV-Assisted\\ Dynamic Offloading}} 
%& \cite{gao2025cooperative} & AAV-enabled MEC 
%& Partition + offloading opt. 
%& Cooperative RL 
%& Motion features extraction \\ \cline{2-6}

%& \cite{zhu2025heterogeneous} & Multi-UAV task assignment 
%& Spatio-temporal opt. 
%& Graph RL 
%& Graph modeling \\
%\hline

% Xóa \multirow{1} và tự ngắt dòng bằng \makecell để form bảng vuông vức
%\makecell[l]{Time-Varying\\ Offloading} 
%& \cite{hsu2025Transformer} & Dynamic MEC system 
%& Adaptive offloading 
%& DRL-based framework 
%& Representation enhancement \\
%\hline

%\end{tabular}%
%}
%\vspace{-1.5em}
%\end{table*}

\subsubsection{User/Vehicle Mobility-Aware Offloading} %This subcategory focuses on computation offloading decisions under user or vehicle mobility, where task execution and resource availability vary over time. The studies emphasize adaptive offloading and task placement to cope with frequent changes in topology and workload.

Under dynamic wireless and computing conditions, the timeliness-oriented resource control for multi-agent Mobile Augmented Reality (MAR) systems is studied in \cite{chen2025IFresher}. %To explicitly characterize real-time task responsiveness, the concept of the age of analytics information (AoAI) is introduced to jointly capture the effects of video analytics accuracy, transmission delay, and computational efficiency. 
A closed-form expression of AoAI is derived, based on which a centralized control objective is formulated to minimize AoAI through joint bandwidth allocation and video configuration while maintaining accuracy requirements. %Owing to the mixed-integer nonlinear structure of the optimization problem and the presence of partial observations at individual agents, the control problem is reformulated as a decentralized partially observable Markov decision process. 
Within this framework, an MARL algorithm, termed Convex-embedded Transformer QMIX (CTQMIX), is proposed under the centralized training and decentralized execution paradigm. Convex optimization is used to determine bandwidth allocation, while a Transformer-based architecture captures temporal dependencies across observations and actions in dynamic environments. %Through coordinated decision-making enabled by Transformer-enhanced MARL, the proposed framework supports timely and accuracy-aware task execution in MAR systems. Experimental evaluations based on real-world settings demonstrate that CTQMIX outperforms benchmarks in terms of AoAI-related performance.
Considering hierarchical edge-cloud architectures, the authors of \cite{ghosh2025mgco} study mobility-aware computation offloading in multi-access edge computing systems. The offloading problem aims to jointly optimize horizontal inter-edge and vertical edge-to-cloud task placement decisions under realistic conditions of user mobility. %To exploit real trajectory information, a two-dimensional offloading scheme is proposed to enable collaborative task execution among resource-constrained edge nodes. 
A mobility-aware generative computation offloading framework, termed MGCO, is developed based on a Transformer-driven sequence-to-sequence deep Q-network. Within this framework, RL determines action selection, while the Transformer architecture enables parallel contextual reasoning over mobility sequences. %The proposed design supports real-time task placement under strict latency constraints in dynamic edge-cloud environments. Experimental results show that MGCO reduces turnaround time by up to 41.61\% compared with GASTO and achieves improvements of up to 645.40\% and 751.90\% over DMQTO and HMAOA, respectively, under a 48-time-slot prediction horizon. These results demonstrate the effectiveness of Transformer-enhanced generative decision modeling for mobility-aware computation offloading.
%
%In high-density IoV environments, overlapping edge node coverage and heterogeneous RSU reputations can lead to suboptimal DT deployment, resulting in resource imbalance and synchronization failures. 
The study in \cite{zou2025deployment} proposes a digital twin-assisted vehicular edge computing framework that incorporates a reputation-driven digital twin deployment model and employs a Transformer-Critic-based Multi-Agent Proximal Policy Optimization (TC-MAPPO) algorithm to jointly optimize digital twin placement and resource allocation. Experimental results show that the proposed method achieves the highest final average reward, surpassing both the standard MAPPO and the multi-agent deep deterministic policy gradient approaches.

\subsubsection{UAV/AAV-Assisted Dynamic Offloading}
%This line of work considers aerial platforms such as UAVs or AAVs as mobile edge servers, where mobility is inherent to the system design. The main challenge lies in jointly handling aerial mobility, energy constraints, and dynamic task offloading.

In the field of intelligent tasks for Autonomous Aerial Vehicles (AAVs), the authors of \cite{gao2025cooperative} formulate the problem of deep neural network partitioning, edge offloading, and hybrid action decision-making in energy-harvesting AAV networks as a predictive Markov decision process. They propose a Transformer-Enhanced Multi-agent Hybrid Action Proximal Policy Optimization (TE-MHAPPO) framework. Simulation results demonstrate that compared to the benchmark MHAPPO algorithm, TE-MHAPPO reduces the comprehensive cost accounting for task delay and energy consumption by at least 12.1\%. Furthermore, as the prediction time increases, the degradation in TE-MHAPPO’s performance is capped at 55.2\% of the baseline MHAPPO's performance, highlighting its superior stability. 
The authors of \cite{zhu2025heterogeneous} propose a heterogeneous alignment-based spatio-temporal graph RL framework for dynamic multi‑UAV task allocation. The HASTG‑RL framework constructs a dynamic spatio‑temporal graph to continuously update environmental states, employs a Transformer‑based heterogeneous alignment mechanism to handle UAV heterogeneity, and designs independent critic networks for multi‑objective optimization. %Experimental results show that the proposed scheme achieves outstanding performance in both task completion rate and flight distance compared to baselines. It attains a task completion rate of 94\%, which is 32 percentage points higher than mixed‑integer linear programming (62\%) and 2 percentage points higher than MADDPG (92\%).

\subsubsection{Time-Varying Offloading} %This subcategory addresses offloading problems under dynamic environments that are not solely caused by physical mobility, such as fluctuating workloads, time-varying task arrivals, or changing system states.

\begin{figure}[t]
    \centering
    \includegraphics[width=\linewidth]{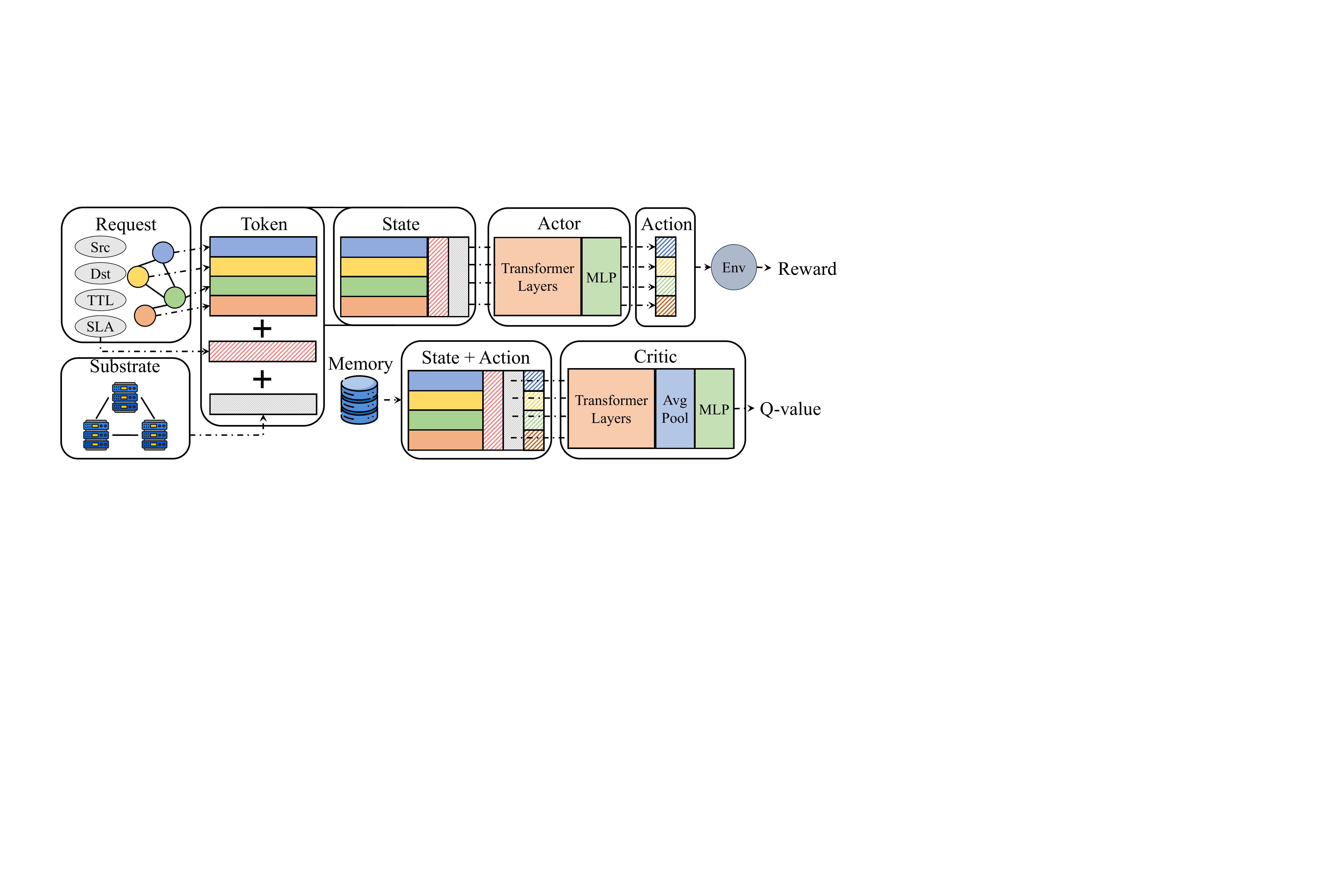}
    \caption{A representative architecture where a Transformer encoder is embedded into the RL decision loop for sequence-aware policy and value learning \cite{hsu2025Transformer}. Src: source; Dst:  destination; TTL: time to live; SLA: service level agreements; MLP: multi-layer perceptron; Env: environment.}
    \vspace{-2em}
    \label{fig:sdac}
\end{figure}

In \cite{hsu2025Transformer}, SFC partitioning in multi-domain 6G network infrastructures is investigated, considering stringent latency and resource constraints. The SFC partitioning problem is formulated to capture sequential dependencies among virtualized network functions while maintaining scalability across heterogeneous domains. As illustrated in Fig.~\ref{fig:sdac}, Transformer layers are embedded into both the actor and critic networks to encode structured state representations and enable sequence-aware policy and value learning within the RL loop. A Transformer-empowered actor-critic RL framework is proposed to enable sequence-aware decision making for SFC partitioning. %Within this framework, self-attention mechanisms are embedded into the policy and value learning process to model inter-dependencies among VNFs and support coordinated and parallelized decisions. RL governs the optimization of partitioning strategies, while the Transformer enhances the representation of sequential dependencies inherent in SFCs. %To improve training stability and convergence, $\epsilon$-LoPe exploration and asymptotic return normalization are incorporated into the learning process. Simulation results demonstrate that the proposed approach achieves improved long-term acceptance rates, higher resource utilization efficiency, and enhanced scalability, while maintaining rapid inference performance. The study further highlights the applicability of Transformer-enhanced RL to intelligent network orchestration in emerging 6G environments. 

\subsection{AI Service/AIGC/Large-Model-Oriented Offloading}\label{Section 4-4}
This section focuses on offloading problems driven by AI services, Artificial Intelligence-Generated Content (AIGC) applications, and large-model inference workloads. The main objective is service-level orchestration, including model selection, inference placement, and resource-aware service migration.%, as summarized in Table \ref{tab:SecIV_D}.

\subsubsection{Large-Model Inference Offloading} %This subcategory focuses on offloading decisions for large-model inference tasks, where computation complexity is dominated by deep neural networks or foundation models. The main objective is to determine appropriate execution locations under stringent latency and resource constraints.

The authors of \cite{zhou2025large} investigate task offloading in IoV systems under large-scale data processing demands and stringent latency constraints. The offloading problem is formulated to minimize energy consumption and task latency by leveraging edge computing resources in multi-access edge computing environments. A Transformer-based large model is integrated with DRL, where DRL manages policy optimization, while the large model enhances action inference and adaptability under complex offloading conditions. Additionally, a task classification framework is introduced to categorize tasks by computational complexity, data size, and time sensitivity, thereby enabling differentiated offloading decisions. %A reward function based on task adaptability is designed to guide the learning process toward energy- and latency-efficient offloading strategies. Experimental results show that the proposed approach effectively determines task offloading decisions that reduce energy consumption and latency in dynamic IoV scenarios.
%
% In \cite{zhuang2025joint}, mobile edge computing-empowered Large Vision Model (LVM) services in wireless networks are proposed, where heterogeneous LVMs are deployed across cloud and edge servers to support delay-sensitive applications. The problem jointly optimizes model inference selection and task offloading for LVM users, aiming to maximize service utility while minimizing delay and energy consumption. %To characterize LVM service utility, a multidimensional video quality metric is proposed based on real measurements, incorporating prompt-video alignment and conventional video quality indicators. 
A decentralized two-stage solution is developed to address real-time decision-making and resource allocation. In the first stage, an RL-based multi-agent proximal policy optimization approach determines model inference and offloading decisions under dynamic conditions. In the second stage, an optimization-based sequential least squares programming method is employed to allocate resources efficiently. %RL governs real-time control, while optimization refines resource distribution to enhance overall system performance. Simulation results show that the proposed solution reduces delay and energy consumption by up to 17.2\% and 21.7\%, respectively, while increasing service utility by up to 3\%, demonstrating effective support for MEC-enabled LVM services.

\subsubsection{AIGC Service-Oriented Orchestration} %This line of work considers AIGC services as system-level entities, where multiple AI service providers or models are orchestrated to serve heterogeneous user demands. The emphasis lies in service-level allocation rather than individual task scheduling.

In \cite{huang2024federated}, the deployment of AIGC services in wireless networks with an emphasis on efficiency, personalization, and privacy preservation is investigated. %The study focuses on addressing the limitations of centralized training and inference pipelines for AIGC models in wireless environments, particularly under privacy constraints.
To this end, federated learning is adopted as a collaborative framework for distributing model training across multiple data owners without explicit data sharing. Within this framework, AIGC model training, fine-tuning, and inference are supported in a decentralized manner while preserving user privacy. %A case study based on federated learning-aided fine-tuning of a stable diffusion model is conducted to illustrate the applicability of the proposed approach. Numerical results show that the proposed scheme effectively reduces communication costs and training latency while maintaining privacy. By integrating federated learning into AIGC service provisioning, the work demonstrates a practical pathway to enabling personalized, high-quality content generation in wireless networks. In addition, several research directions and open issues related to the convergence of federated learning and AIGC are outlined.
In \cite{lai2025optimizing}, a service-provisioning approach for AIGC in vehicular metaverse environments is proposed, in which computation-intensive content-generation tasks are offloaded from resource-constrained vehicles to AIGC service providers. The problem is formulated to address resource scarcity and service quality degradation arising from large-scale AIGC service requests under limited computational capacity. To capture the hierarchical interaction between vehicular metaverse users and service providers, a Stackelberg game framework is established under incomplete information. %Within this framework, a generative diffusion model is adopted to characterize AIGC service behavior, and an empirical relationship between image quality and the number of diffusion steps is revealed. A Transformer-based DRL algorithm is employed to learn optimal decision strategies and reach the Stackelberg equilibrium. 
Explicitly, the strategic interactions are determined by RL, while the Transformer enhances the modeling of complex decision dependencies in the game-theoretic setting. %Numerical results demonstrate that the proposed approach converges efficiently to equilibrium and achieves higher utility than baseline schemes, supporting effective AIGC service management in vehicular metaverse scenarios.

%\begin{table*}[t]
%\centering
%\footnotesize
%\caption{Summary of Transformer-Enhanced DRL for AI Service/AIGC/Large-Model-Oriented Offloading.}
%\label{tab:SecIV_D} 
%\renewcommand{\arraystretch}{1.4} 

% Ép giãn toàn bộ bảng vừa khít 2 cột một cách an toàn
%\resizebox{\textwidth}{!}{%
%\begin{tabular}{|l|c|l|l|l|l|}
%\hline
%\textbf{Subcategory} & \textbf{Ref.} & \textbf{Scenario} & \textbf{Objective} & \textbf{Method} & \textbf{Transformer Role} \\
%\hline
%\multirow{2}{*}{\makecell[l]{Large-model\\ inference offloading}} 

%& \cite{zhou2025large} & \makecell[l]{LLM-assisted\\ MEC offloading} 
%& Large-model task execution 
%& Large-model-aided RL 
%& Implicit feature modeling \\ \cline{2-6}

%& \cite{zhuang2025joint} & \makecell[l]{Vision model\\ MEC inference} 
%& Joint inference-offloading opt. 
%& DRL-based framework 
%& Image and video generation \\
%\hline

%\multirow{2}{*}{\makecell[l]{AIGC service\\ orchestration}} 

%& \cite{huang2024federated} & \makecell[l]{Federated AIGC\\ services} 
%& Collaborative content generation 
%& FL-based framework 
%& Processes sequential data \\ \cline{2-6}

%& \cite{lai2025optimizing} & \makecell[l]{Vehicular\\ metaverse AIGC} 
%& Service pricing \& offloading 
%& \makecell[l]{Learning-based\\ Stackelberg game} 
%& Processes sequential data \\
%\hline

% Đã xóa \multirow{1} ở đây để tránh lỗi chữ đè viền
%\makecell[l]{Digital twin\\ management} 

%& \cite{tong2024multi} & \makecell[l]{Twin migration\\ in metaverse} 
%& Resource auction \& migration opt. 
%& GPT-based DRL 
%& Language-model-assisted policy \\
%\hline

%\end{tabular}%
%}
%\vspace{-1.5em}
%\end{table*}

\subsubsection{AI-Driven Digital Twin Management}
%This subsection addresses offloading problems involving digital twins, service migration, or AI-enhanced twin management. The focus is on adaptive migration or resource allocation under AI-driven virtual environments.

Resource allocation and Vehicle Twin (VT) migration in vehicular metaverse environments are investigated in \cite{tong2024multi}, where high mobility and resource-intensive digital twin updating impose significant computation, communication, and storage demands. The VT migration problem is formulated to address resource scarcity and limited Roadside Unit (RSU) coverage through market-based coordination. To this end, an attribute-aware auction-based mechanism is proposed that jointly considers monetary and non-monetary attributes, such as location and reputation. A two-stage matching framework is developed, where resource attribute matching enables participation in a double-auction for multi-attribute resource trading. %Within this framework, a generative pre-trained Transformer-based DRL algorithm is employed to train the auctioneer and dynamically adjust auction clocks during the auction process. 
In the proposed scheme, RL determines auction control decisions, while the Transformer model facilitates the learning of adaptive mechanisms rather than directly approximating offloading or migration policies. %Performance is evaluated in terms of social welfare and the cost of information exchange in auctions under different system settings. Simulation results demonstrate that the proposed GPT-based auction schemes outperform state-of-the-art baselines, supporting efficient and scalable VT migration in vehicular metaverse systems.

\subsection{Caching, Content Delivery, and Service Continuum}
%Caching-oriented offloading studies aim to reduce latency and improve QoE by proactively placing content or services across the cloud-edge-device continuum. Offloading decisions are closely integrated with caching strategies and content delivery mechanisms.

In \cite{he2025qoe}, proactive caching and content update strategies in cloud-assisted edge computing systems are presented. %, where the objective of enhancing Quality of Experience (QoE) is considered. %The caching problem is formulated as a multi-objective optimization task that jointly maximizes cache hit ratio while minimizing traffic load and time latency in the cloud-to-edge continuum. 
To address prediction and decision-making in proactive caching, a unified framework, HT-PAD, is proposed by combining hyperdimensional computing, Transformer-based prediction, and multi-agent RL. %Within this framework, hyperdimensional encoding is employed to extract informative features from user preferences, historical records, and popularity information. A hyperdimensional Transformer is developed as the prediction module to model temporal patterns and user behavior, thereby providing accurate forecasts of content popularity. For decision making, a prioritized experience replay-based multi-agent deep deterministic policy gradient method is adopted to determine caching actions based on predicted content demand. 
Specifically, RL governs the caching policy, while the Transformer serves as an auxiliary component that enhance demand prediction rather than approximating the policy directly. %Experimental results show that the proposed approach achieves strong performance in terms of edge hit ratio, latency, and traffic load, thereby improving overall QoE in cloud-edge caching systems.
In the face of rapidly increasing wireless data traffic and limited backhaul resources, the authors of \cite{bajpai2022adapting} introduce collaborative content caching in Device-to-Device (D2D) communication environments. %The caching problem is studied in the context of leveraging local user intelligence and historical data to improve cache efficiency without relying on extensive prior knowledge of the environment. 
Two learning-based caching frameworks are proposed, including a recurrent deep neural network approach and a Transformer-based framework that exploits attention mechanisms to model user demand patterns. %Within these frameworks, a Transformer is invoked to capture long-term dependencies in historical data and enhance demand representation, while caching decisions are derived from the learned models. %The proposed designs support adaptive caching behavior by continuously learning from evolving user data. Experimental evaluations demonstrate that the Transformer-based framework achieves an overall 25\% increase in D2D cache hit rate compared with a neural collaborative filtering-based baseline. These results demonstrate the effectiveness of attention-based modeling in enhancing content caching performance in D2D communication systems.

\subsection{Workflow-Aware and Structured Task Offloading}
%This class considers computation offloading with explicit task structures, such as workflows, DAGs, or batch execution models. The emphasis is on coordinating multi-stage task execution while respecting task dependencies and resource constraints.

By explicitly accounting for task dependencies represented as Directed Acyclic Graphs (DAGs), the study in \cite{deng2025task} proposes computation offloading for IoV systems. The task offloading problem is formulated to jointly optimize execution latency and resource utilization under heterogeneous vehicular computing environments. To address the structural complexity introduced by DAG-based task scheduling, a DRL framework is developed, coupled with representation learning techniques to encode task dependencies and execution states. Within this framework, task graphs are transformed into compact representations that facilitate sequential offloading and scheduling decisions. %The learned representations are integrated into the RL policy to guide task placement across local and remote computing resources. %By incorporating DAG-aware state modeling, the proposed approach enables coordinated handling of interdependent subtasks in dynamic IoV scenarios. Simulation results demonstrate that the proposed method effectively reduces task completion latency and improves system-level efficiency compared with baseline offloading strategies under varying vehicular workloads.
In heterogeneous CPU-GPU mobile edge computing environments with batch processing, the authors of \cite{huang2025federated} study joint computation offloading and resource scheduling for dependent tasks. The problem is modeled as a Markov decision process to jointly minimize task latency and energy consumption under heterogeneous resources and task dependencies. A distributed Transformer-based Federated Soft Actor-Critic (TFSAC) framework is proposed to enable privacy-preserving and scalable learning across agents. Within this framework, Transformer encoders learn contextual relationships among agents, while FL is integrated to coordinate distributed training.
% RL governs offloading and scheduling decisions, whereas Transformer promotes inter-agent context modeling within the federated setting. Extensive experiments on real-world trace data demonstrate that TFSAC outperforms benchmark methods in terms of quality of service across multiple configurations. The proposed design also supports seamless onboarding of new agents into the federation, facilitating flexible, scalable deployment across heterogeneous MEC systems.
%
In a parked-vehicle-extended MEC architecture, online task offloading and container scheduling are investigated in \cite{wu2025sequence}, where parked vehicles serve as edge servers for multiple devices. The scheduling problem is formulated to address time-varying task arrivals, container image management, and collaborative resource utilization. %To move beyond predefined arrival assumptions, task arrivals are modeled based on long-term temporal trends observed in practical systems. 
A sequence-aware task-scheduling algorithm is proposed using a policy-gradient-based DRL framework. Within this framework, Transformer and LSTM architectures are integrated to capture temporal patterns in task arrivals and relational dependencies among containerized nodes. %The learned policy determines online offloading and scheduling decisions across devices and parked vehicles in a centralized control setting. %The primary objective is to minimize end-to-end delay and energy consumption for both devices and edge nodes. Extensive numerical evaluations against baseline methods demonstrate the effectiveness of the proposed sequence-aware scheduling strategy in containerized MEC environments.

\subsection{General Cooperative and Distributed Offloading}
General cooperative offloading frameworks address distributed decision making across multiple edge nodes or agents without a dominant system constraint. These works focus on cooperative or decentralized resource management under uncertainty and system heterogeneity.

\subsubsection{Fully Distributed Cooperative Offloading} %This subcategory considers decentralized environments where multiple edge nodes or agents collaboratively make offloading decisions without a centralized controller. The emphasis lies in cooperative learning and distributed resource coordination under partial system information.

Considering the rapidly increasing wireless data traffic and limited backhaul resources, collaborative content caching in D2D communication environments is proposed in \cite{liu2025cooperative}. %The caching problem is studied in the context of leveraging local user intelligence and historical data to improve cache efficiency without relying on extensive prior knowledge of the environment. 
Two learning-based caching frameworks are proposed, including a recurrent deep neural network approach and a Transformer-based framework that exploits attention mechanisms to model user demand patterns. Within these frameworks, a Transformer is employed to capture long-term dependencies in historical data and enhance demand representation, while caching decisions are derived from the learned models. %The proposed designs support adaptive caching behavior by continuously learning from evolving user data. Experimental evaluations demonstrate that the Transformer-based framework achieves an overall 25\% increase in D2D cache hit rate compared to a neural collaborative filtering-based baseline. These results demonstrate the effectiveness of attention-based modeling in enhancing content caching performance in D2D communication systems.
In the context of time-varying task statistics and random access behaviors of mobile devices, the authors of \cite{zhang2025adaptive} propose dynamic task offloading and resource allocation in multi-access edge computing systems. %The offloading problem is formulated as a Markov decision process with time-varying state and action spaces, capturing heterogeneous service requirements and dynamic task completion or discarding. 
To address the limitations of conventional DRL approaches in such settings, a general state-action space adaptive DRL framework, termed SASA, is proposed. The framework exploits the Transformer architecture and its multi-head attention mechanism to flexibly encode variable-dimensional system states and actions. Within this design, existing actor-critic DRL algorithms are seamlessly integrated to operate under dynamic state-action configurations. Building on the SASA framework, a specific SASA-based task offloading and resource allocation algorithm, SASA-TORA, is developed. %The resulting approach supports adaptive decision-making across time-varying network conditions and evolving statistical characteristics of the task. Simulation results demonstrate the effectiveness of SASA-TORA in dynamic MEC environments, enabling robust, adaptive offloading decisions across varying system configurations.

\subsubsection{Federated or Learning-Based Distributed Coordination} %This line of work incorporates federated or distributed learning mechanisms to coordinate offloading and resource allocation across heterogeneous nodes. The system focuses on collaborative model updates or shared policy improvement while maintaining local autonomy.

The optimization of resource allocation and partial task offloading in dynamic IoV networks, targeting core challenges including high latency, limited scalability, and inefficient resource ultilization, is investigated. The authors of \cite{singh2025digital} introduce a digital twin-assisted framework that integrates Adaptive Federated Learning (AdFL), multi-agent deep RL, and generative AI. Specifically, a Conditional Variational AutoEncoder (CVAE) is employed. The AdFL component employs CVAE to generate context-aware representations and utilizes Transformer layers to capture long-range dependencies. %Experimental evaluations confirm the notable performance. For instance, in terms of total cost reduction, which incorporates delay, energy, and service costs. The proposed method achieves the lowest overall cost across diverse task sizes and roadside unit computational capacities (2.5-3.5 GHz). It surpasses baseline approaches by 15-30\% in cost efficiency.

\subsubsection{Hybrid System-Level Offloading under Uncertain Environments} %This subcategory addresses cooperative offloading under non-stationary workloads or heterogeneous statistical conditions without emphasizing explicit mobility or radio coupling. The decision process adapts to fluctuating system characteristics.

Considering time-varying user demand and limited cache capacity, the authors in \cite{kim2024t} study adaptive content caching in dynamic content delivery networks. The caching problem is formulated to jointly determine whether newly requested content should be cached and which existing content should be replaced when storage constraints are present. A Transformer-based RL model, termed T-CacheNet, is proposed to optimize caching and replacement decisions based on real-time content request patterns. %Within this framework, the self-attention mechanism is employed to capture long-term dependencies in content request sequences, enabling informed caching decisions over extended time horizons. %To mitigate the computational overhead associated with Transformer models, we introduce miss-based delayed hit reward learning and partial information learning to improve learning efficiency. 
In the proposed solution, RL determines the caching policy, while the Transformer enhances the modeling of temporal correlations in content demand. %The proposed design supports adaptive caching behavior in non-stationary environments. Experimental results indicate that T-CacheNet improves cache hit performance and resource utilization in dynamic content caching scenarios. 

\subsection{Lessons Learned}
The reviewed studies show a clear shift from isolated computation scheduling to system-level offloading design. As scenarios expand to wireless MEC, cloud-edge-end collaboration, and mobility-aware systems, the key challenge lies in modeling high-dimensional, time-varying, and correlated system states. Transformer-enhanced learning is mainly used to capture task dependencies, temporal dynamics, variable-sized inputs, and inter-agent interactions, thereby improving the scalability and robustness of offloading policies. Meanwhile, AI-oriented services such as large-model inference and AIGC are pushing offloading design toward service-aware orchestration. However, most existing studies remain application-driven, and the theoretical understanding of generalization and convergence remains limited.

%===========================================
\section{Transformer-based RL for Routing and Trajectory Control} %==============================================
\label{sec:routing}
\subsection{Long-Horizon and Multi-Agent Routing and Trajectory}
Routing and trajectory decision-making in large-scale wireless, robotic, and aerial systems involves long-horizon planning, strong spatiotemporal coupling, and complex interactions among multiple agents and network entities. Traditional DRL approaches, typically based on multilayer perceptrons or recurrent architectures, often struggle to capture long-range dependencies, dynamic topological variations, and high-dimensional joint decision spaces induced by mobility, resource constraints, and multi-agent coordination. To address these challenges, recent studies incorporate Transformer architectures into DRL frameworks, using self-attention to model long-term temporal correlations, spatial relationships, and inter-agent dependencies. These studies cover a broad range of routing and trajectory problems, from single-agent motion planning to multi-UAV coordination and network-level routing in dynamic and uncertain environments, and show that Transformers provide an effective representation for scalable and robust long-horizon control. The related literature is summarized in Table \ref{tab:Transformer_DRL_Routing_Trajectory}.

\begin{table*}[t]
\caption{Summary of Transformer-Enhanced DRL for Trajectory Optimization and Network-Level Routing.}
\label{tab:Transformer_DRL_Routing_Trajectory}
\centering
\scriptsize
\setlength{\tabcolsep}{3.2pt}
\renewcommand{\arraystretch}{1.55}

% Sử dụng tabular* với \textwidth và \extracolsep{\fill} để bảng tràn lề 2 cột
\begin{tabular*}{\textwidth}{@{\extracolsep{\fill}}|l|l|l|l|l|l|c|c|c|c|}
\hline
\multicolumn{1}{|l|}{\textbf{Subcategory}} &
\multicolumn{1}{l|}{\textbf{Ref.}} &
\multicolumn{1}{l|}{\textbf{Scenario}} &
\multicolumn{1}{l|}{\textbf{Decision Scope}} &
\multicolumn{1}{l|}{\textbf{Proposed Method}} &
\multicolumn{1}{l|}{\textbf{Transformer Role}} &
\multicolumn{1}{c|}{\makecell[c]{\textbf{Multi-}\\\textbf{Agent}}} &
\multicolumn{1}{c|}{\makecell[c]{\textbf{Resource}\\\textbf{Coup.}}} &
\multicolumn{1}{c|}{\makecell[c]{\textbf{Hierarchical}\\\textbf{Sequence}}} &
\multicolumn{1}{c|}{\makecell[c]{\textbf{Robust}\\\textbf{Gen.}}} \\
\hline

\multirow{11}{*}{\makecell[l]{Long-horizon\\ trajectory}}
& \cite{abdelkader2025perception} & \makecell[l]{Robot\\ navigation} & \makecell[l]{Single-agent\\ trajectory} &
\makecell[l]{SAC with\\ Transformer} &
\makecell[l]{Temporal dependencies\\ from sensor history} &
 & & & $\checkmark$ \\
\cline{2-10}
& \cite{xu2025transformer} & \makecell[l]{FAS-enabled\\ UAV networks} & \makecell[l]{Trajectory and\\ antenna selection} &
\makecell[l]{attention-based\\ recurrent MARL} &
\makecell[l]{Mobility--comm.\\ coupling modeling} &
 & $\checkmark$ & & \\
\cline{2-10}
& \cite{chen2024Transformer} & \makecell[l]{Multi-UAV\\ coverage} & \makecell[l]{Multi-UAV\\ trajectory} &
\makecell[l]{Transformer-\\ based MARL} &
\makecell[l]{Permutation-invariant\\ state encoding} &
$\checkmark$ & & & \\
\cline{2-10}
& \cite{yang2025efficient} & \makecell[l]{Mobile\\ crowdsensing} & \makecell[l]{Trajectory and\\ bandwidth} &
\makecell[l]{Lyapunov DRL\\ with Transformer} &
\makecell[l]{Long-horizon\\ constraint coupling} &
$\checkmark$ & $\checkmark$ & & \\
\cline{2-10}
& \cite{dong2025attention} & \makecell[l]{WPT-assisted\\ IoT} & \makecell[l]{UAV\\ trajectory} &
\makecell[l]{DRL with\\ attention} &
\makecell[l]{Global attention\\ over device graph} &
 & $\checkmark$ & & \\
\cline{2-10}
& \cite{feng2025event} & \makecell[l]{Multi-AAV\\ comm.} & \makecell[l]{Trajectory and\\ channel assignment} &
\makecell[l]{Event-driven\\ Transformer RL} &
\makecell[l]{Event-triggered\\ spatiotemporal modeling} &
$\checkmark$ & $\checkmark$ & & \\
\cline{2-10}
& \cite{li2025uav} & \makecell[l]{Radio map\\ updating} & \makecell[l]{UAV\\ trajectory} &
\makecell[l]{Dueling DQN\\ with Agent Transformer} &
\makecell[l]{State--measurement\\ dependency modeling} &
 & $\checkmark$ & & \\
\cline{2-10}
& \cite{liu2025efficient} & \makecell[l]{Warehouse\\ AGVs} & \makecell[l]{Scheduling and\\ trajectory control} &
\makecell[l]{Hierarchical RL\\ with Transformer} &
\makecell[l]{Two-level planning\\ and execution} &
$\checkmark$ & & $\checkmark$ & \\
\cline{2-10}
& \cite{lu2025attention} & \makecell[l]{UAV-assisted\\ comm.} & \makecell[l]{Trajectory and\\ user scheduling} &
\makecell[l]{Decision\\ Transformer} &
\makecell[l]{Return-conditioned\\ sequence decisions} &
 & $\checkmark$ & $\checkmark$ & $\checkmark$ \\
\hline

\multirow{4}{*}{\makecell[l]{Multi-agent\\ interaction}}
& \cite{ye2025enhanced} & \makecell[l]{Cooperative UAV\\ monitoring} & \makecell[l]{Deployment and\\ trajectory} &
\makecell[l]{Transformer-\\ enhanced MARL} &
\makecell[l]{Spatial dependencies\\ for coordination} &
$\checkmark$ & & & \\
\cline{2-10}
& \cite{li2025hierarchical} & \makecell[l]{Heterogeneous\\ UAV networks} & \makecell[l]{Conflict-aware\\ trajectory} &
\makecell[l]{Hier. Graph\\ Transformer DQN} &
\makecell[l]{Interaction and collision\\ risk encoding} &
$\checkmark$ & & $\checkmark$ & $\checkmark$ \\
\cline{2-10}
& \cite{emami2022intraforce} & \makecell[l]{Crowded\\ environments} & \makecell[l]{Multi-agent\\ prediction} &
\makecell[l]{Social\\ Transformer} &
\makecell[l]{Interaction-aware motion\\ for downstream planning} &
$\checkmark$ & & & \\
\hline

\multirow{5}{*}{\makecell[l]{Network-level\\ routing}}
& \cite{adam2025multi} & \makecell[l]{AAV-assisted\\ transport} & \makecell[l]{Trajectory, routing,\\ and resources} &
\makecell[l]{Primal--dual PPO\\ with Transformer} &
\makecell[l]{Cross-layer\\ dependency modeling} &
$\checkmark$ & $\checkmark$ & & \\
\cline{2-10}
& \cite{chen2025distributed} & \makecell[l]{LEO satellite\\ networks} & \makecell[l]{Distributed\\ routing} &
\makecell[l]{Transformer-MIX\\ MARL} &
\makecell[l]{Inter-node temporal\\ modeling under dynamics} &
$\checkmark$ & & & \\
\cline{2-10}
& \cite{dai2025parallel} & \makecell[l]{UAV inspection\\ (dense)} & \makecell[l]{Large-scale\\ routing} &
\makecell[l]{Parallel RL\\ with Enc--Dec Transformer} &
\makecell[l]{Priority-aware routing\\ structure learning} &
 & & $\checkmark$ & \\
\cline{2-10}
& \cite{li2024cgtr} & \makecell[l]{Dynamic\\ networks} & \makecell[l]{Robust\\ routing} &
\makecell[l]{PPO with\\ C-Graph Transformer} &
\makecell[l]{Topology-invariant\\ representation learning} &
 & & & $\checkmark$ \\
\hline
\end{tabular*}
\end{table*}

\subsubsection{Long-Horizon Trajectory Planning} 
Early Transformer-enhanced DRL studies focus on improving long-horizon trajectory representations for mobile agents in dynamic environments. By leveraging self-attention, the works in \cite{abdelkader2025perception,xu2025transformer,chen2024Transformer} capture temporal correlations and spatial dependencies that are difficult for conventional MLP- or RNN-based policies, enabling smoother and more robust trajectory decisions for robots, UAVs, and AGVs. Particularly, the authors in \cite{abdelkader2025perception} integrate self-attention into a soft actor-critic framework for autonomous robot navigation, where sequential LiDAR observations capture long-range temporal dependencies and improve trajectory smoothness, navigation success rates, and generalization in dynamic and partially observable environments. Extending this idea to aerial networks, \cite{xu2025transformer} studies an attention-based recurrent MARL framework for three-dimensional UAV trajectory and positioning optimization in fluid antenna system (FAS)--enabled wireless networks. Under a collaborative learning framework, the proposed method captures spatial correlations and the coupling between UAV motion and reconfigurable antenna states, enabling more accurate 3D trajectory decisions. Beyond fixed-scale settings, \cite{chen2024Transformer} investigates Transformer-based MARL for scalable multi-UAV trajectory control in area coverage tasks by organizing variable-size swarm states with a Transformer, achieving permutation-invariant decisions and adaptability across swarm scales.

Subsequent studies extend this paradigm to resource-coupled scenarios where trajectory planning is jointly optimized with bandwidth allocation, energy harvesting, or channel assignment. The works in \cite{yang2025efficient,dong2025attention,feng2025event,li2025uav} show that Transformer representations capture long-term coupling between mobility and communication resources under stochastic constraints. The authors in \cite{yang2025efficient} investigate Lyapunov-guided Transformer-enhanced DRL for joint bandwidth allocation and multi-UAV trajectory optimization in energy-harvesting mobile crowdsensing systems. The stochastic optimization problem is reformulated using a Lyapunov drift-plus-penalty framework, enabling tractable online control and efficient long-horizon trajectory planning under energy uncertainty. The authors in \cite{dong2025attention} study UAV trajectory optimization in wireless power transfer--assisted IoT systems and propose an attention-based DRL framework for large-scale resource-constrained planning. A graph-attention encoder captures spatial correlations and heterogeneous service demands from graph-structured IoT inputs, enabling scalable and energy-efficient trajectory generation under joint energy and storage constraints.

Beyond continuous resource allocation, \cite{feng2025event} studies joint trajectory design and channel assignment in multi-UAV communication systems using an event-driven Transformer-based RL framework. An event-triggered mechanism selectively updates trajectory and spectrum allocation actions, capturing long-term spatiotemporal dependencies while reducing control overhead compared with time-driven DRL. The authors in \cite{li2025uav} further study Transformer-enhanced DQN for UAV trajectory optimization in information-driven tasks such as radio map updating under energy and destination constraints. Formulated as a finite-horizon MDP with sparse rewards, the problem is solved by integrating an Agent Transformer into a dueling DQN to capture dependencies among UAV states and measurement locations, enabling long-horizon planning while alleviating sparse-reward issues through reward shaping.

\begin{figure}[t]
    \centering
    \includegraphics[width=0.95\columnwidth]{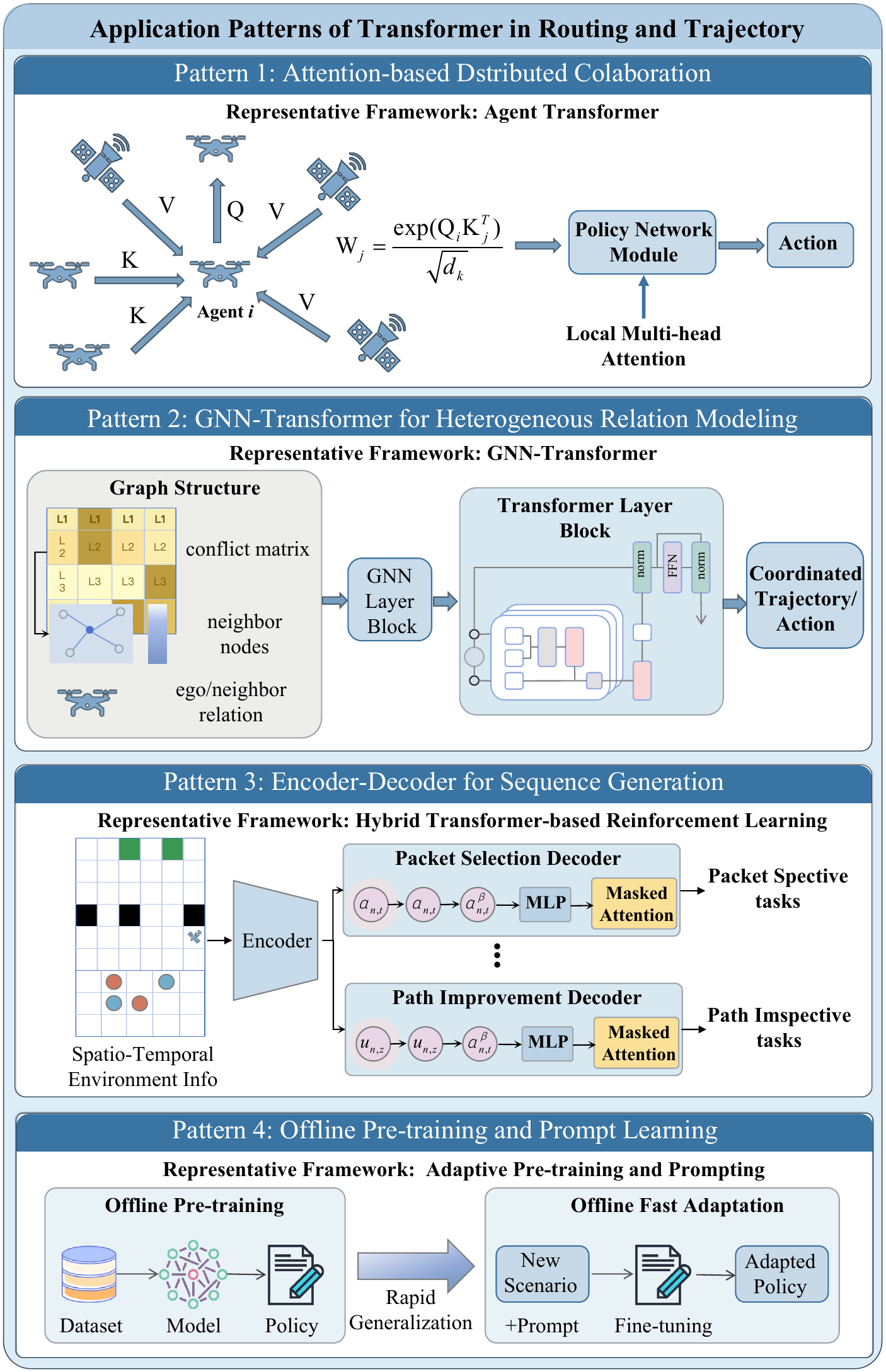}
    \caption{A summary of four application patterns of Transformer in Routing and Trajectory}
    \label{fig:trajectory_routing}
\end{figure}

Hierarchical Transformer-based DRL improves scalability by decoupling long-horizon planning from fine-grained trajectory execution. Representative applications include warehouse AGV scheduling and UAV-assisted communications, where hierarchical decision structures and sequence-based policy learning enable efficient coordination under long planning horizons \cite{liu2025efficient},\cite{lu2025attention}. The authors in \cite{liu2025efficient} propose a hierarchical Transformer-based RL framework for AGV scheduling and trajectory coordination in large-scale warehouses. A two-level structure is adopted, where a high-level Transformer policy captures long-horizon task allocation and congestion patterns while low-level controllers handle fine-grained execution. By modeling inter-AGV interactions and temporal dependencies through self-attention, the framework improves scalability and scheduling efficiency over flat DRL methods. The authors in \cite{lu2025attention} further advance trajectory optimization by adopting a Decision Transformer (DT) for UAV-assisted communications with dynamic user populations. By reformulating joint UAV trajectory planning and user scheduling as sequence modeling and introducing attention mechanisms, prompt-based conditioning, and energy-aware tokens, the framework handles long-horizon Age-of-Information (AoI) optimization and long-term energy constraints while generalizing well to unseen deployment scenarios.

Recent studies indicate a shift toward Transformer architectures as unified representations for long-horizon trajectory optimization, extending from mobility control to resource-coupled and hierarchical decision-making. This paper summarizes four application patterns in Fig.~\ref{fig:trajectory_routing}. However, several gaps remain. Most existing works focus on single-agent or loosely coupled multi-agent settings, leaving fully decentralized and large-scale cooperative trajectory and routing problems underexplored. In addition, many Transformer-enhanced DRL frameworks rely on task-specific architectures and handcrafted state representations, limiting transferability across mobility and network scenarios. Moreover, theoretical understanding of convergence, stability, and sample efficiency for Transformer-based policies under stochastic constraints remains limited, particularly in safety-critical applications. These challenges highlight the need for more generalizable, theoretically grounded, and communication-aware Transformer-based RL frameworks for routing and trajectory optimization.

\subsubsection{Multi-Agent and Conflict-Aware Trajectory Control}
As trajectory optimization becomes increasingly multi-agent and densely coupled, Transformer-enhanced DRL is extended to explicitly model inter-agent interactions and coordination constraints. In cooperative UAV deployment and monitoring, Transformers capture long-horizon spatial dependencies among agents, enabling stable and coverage-aware trajectory generation through sequence-based decision modeling \cite{ye2025enhanced}. To address conflict resolution in heterogeneous UAV networks, hierarchical graph Transformer--based DQN frameworks explicitly encode inter-UAV interactions and collision risks into multi-level decision processes, improving scalability and robustness in dense airspace \cite{li2025hierarchical}. In parallel, social Transformer architectures are explored for multi-agent trajectory prediction, providing interaction-aware motion modeling that supports downstream trajectory planning and decision-making \cite{emami2022intraforce}.

The authors in \cite{ye2025enhanced} study Transformer-enhanced RL for cooperative UAV deployment and trajectory optimization by formulating multi-node monitoring as a sequential decision-making process. An adaptive Transformer-based architecture captures long-horizon spatial dependencies among UAVs and generates coordinated deployment actions; by casting policy learning as sequence prediction, the approach yields more stable deployment and improved coverage than conventional DRL.

To address conflict-aware trajectory control in dense airspace, the authors in \cite{li2025hierarchical} propose a hierarchical graph Transformer--based RL framework for heterogeneous UAV networks. A graph Transformer models inter-UAV interactions, collision risks, and spatiotemporal coupling, while a multi-level decision structure decomposes conflict resolution into hierarchical control layers, improving scalability and robustness over flat multi-UAV DRL methods. At the perception level, the authors in \cite{emami2022intraforce} propose an intra-cluster reinforced social Transformer for multi-agent trajectory prediction in crowded environments. Intra-cluster attention models social interactions explicitly, capturing fine-grained spatiotemporal dependencies and agent-level influence patterns to provide interaction-aware motion predictions for downstream multi-agent planning and conflict-aware decision-making.

\subsubsection{Network-Level Routing and Joint Resource Optimization}
Beyond motion-level control, recent studies extend Transformer-enhanced DRL to network-level routing and joint resource optimization under dynamic topologies and combinatorial decision spaces. In AAV-assisted transportation networks, Transformer-based policies are integrated with primal-dual PPO to jointly optimize trajectory planning, content routing, and radio resource allocation under energy and mobility constraints \cite{adam2025multi}. In large-scale dynamic networks such as LEO satellite systems and ultra-dense UAV inspection, graph-based and encoder--decoder Transformers capture long-range topological dependencies and priority-aware routing structures, enabling scalable routing \cite{chen2025distributed,dai2025parallel}. More recently, contrastive graph Transformer--based PPO improves robustness to topology variations and link failures, showing stronger generalization across heterogeneous networks \cite{li2024cgtr}.

The authors in \cite{adam2025multi} study Transformer-enhanced DRL for joint routing and trajectory optimization in AAV-assisted transportation networks. By embedding a multi-scale generative Transformer into a primal-dual PPO framework, the model jointly optimizes aerial trajectory planning, content routing, and radio resource allocation under energy and mobility constraints. The Transformer captures long-range spatiotemporal dependencies across mobility and network layers, enabling adaptive routing in dynamic transportation environments.

To address large-scale and rapidly changing topologies, the authors in \cite{chen2025distributed} investigate distributed routing in LEO satellite networks using a Transformer-MIX-based multi-agent DQN framework. Integrating this architecture into cooperative multi-agent learning enables the modeling of long-range temporal dynamics and inter-satellite dependencies, achieving scalable routing under frequent topology changes. The work in \cite{dai2025parallel} further studies scalable routing in ultra-dense UAV inspection scenarios by formulating inspection planning as a prioritized traveling salesman problem. Its Transformer-based routing method combines graph segmentation, encoder--decoder Transformers, parallel RL, and adaptive large neighborhood search to capture long-range spatial dependencies and priority-aware routing while reducing complexity and maintaining solution quality.

Beyond scalability, \cite{li2024cgtr} proposes a contrastive graph Transformer--based PPO framework to improve routing robustness under dynamic topologies and link failures. By incorporating an edge-enhanced graph Transformer into a PPO routing policy and introducing contrastive group routing for cross-topology alignment, the method improves generalization and robustness over conventional GNN-based DRL routing approaches.

Existing studies suggest a shift from motion-level trajectory control toward network-level routing and joint resource optimization with Transformer-enhanced deep reinforcement learning methods, particularly multi-agent DQN and PPO, where attention captures long-range topological dependencies and cross-layer mobility--communication couplings. However, most frameworks still rely on centralized training with global state information, leaving decentralized routing and trajectory optimization under partial observability underexplored. In addition, routing, trajectory, and resource decisions are often optimized at a single timescale, while multi-timescale Transformer architectures remain insufficiently studied. Moreover, topology-invariant and distributionally robust routing policies across heterogeneous networks are still lacking. Finally, theoretical guarantees on convergence, stability, and safety under stochastic topology evolution and resource constraints remain limited, hindering deployment in safety-critical scenarios.

\subsection{Sequence-Modeling and Risk-Aware Routing and Trajectory}

\begin{table*}[t]
\caption{Summary of Sequence-Modeling and Risk-/Constraint-Aware Transformer RL Methods.}
\label{tab:seq_risk_transformer}
\centering
\renewcommand{\arraystretch}{2.1}

% Sử dụng tabular* với \textwidth và \extracolsep{\fill} để bảng tràn lề 2 cột
\begin{tabular*}{\textwidth}{@{\extracolsep{\fill}}|l|l|l|l|l|}
\hline
\textbf{Ref.} & \textbf{Scenario} & \textbf{Objective} & \textbf{Proposed Method} & \textbf{Transformer Role} \\
\hline

\cite{sun2024reinforcement} &
\makecell[l]{UAV-RIS \\ control} &
\makecell[l]{Trajectory + phase \\ shift optimization} &
Decision Transformer RL &
Sequence-based action generation \\
\hline

\cite{zhang2025decision} &
\makecell[l]{UAV comm. \\ scheduling} &
\makecell[l]{Path, workload, \\ user association} &
DT adaptive architecture &
Long-horizon decision sequence modeling \\
\hline

\cite{munir2024route} &
\makecell[l]{EV routing \\ (time windows)} &
Distance minimization &
Transformer routing policy &
Cross-node attention decoding \\
\hline

\cite{wu2024multiobjective} &
Vehicle routing &
Multi-objective tradeoff &
Weight-aware DRL + Transformer &
Multi-objective sequence encoding \\
\hline

\cite{mondal2025risk} &
\makecell[l]{UAV-UGV \\ routing} &
\makecell[l]{Risk-aware \\ mission time} &
CMDP + Transformer RL &
Constraint-aware route encoding \\
\hline

\cite{sopegno2025Transformer} &
\makecell[l]{UAV navigation \\ GPS-denied} &
Robust navigation &
PPO + Transformer &
High-dim state dependency modeling \\
\hline

\cite{huang2023goal} &
Goal navigation &
Goal-reaching efficiency &
Goal-guided Transformer RL &
Goal-conditioned attention \\
\hline

\cite{wu2025reconfigurable} &
\makecell[l]{UAV MCS + RIS} &
\makecell[l]{Throughput + \\ energy efficiency} &
Transformer + PER-DDQN &
Joint trajectory-resource encoding \\
\hline

\cite{owusu2025Transformer} &
SDN routing &
Load balancing &
TFT + DQN &
Traffic prediction attention features \\
\hline

\cite{wang2025transal} &
IoT multipath &
Congestion control &
Transformer Actor-Learner RL &
Path-aware temporal encoding \\
\hline

\cite{liu2025uav} &
UAV swarm &
Adversarial decision &
Transformer MARL &
Global swarm dependency modeling \\
\hline

\cite{tong2025rapid} &
UAV swarm &
Resource scheduling &
Transformer RL &
Multi-agent interaction attention \\
\hline

\cite{yu2025hybrid} &
\makecell[l]{Multi-UAV \\ corridors} &
Coordination success &
Hybrid Transformer RL &
Interaction-aware control encoding \\
\hline

\cite{zhao2025integrated} &
\makecell[l]{Swarm multi- \\ target task} &
Cooperative detection &
Transformer graph DRL &
Dual-level interaction attention \\
\hline

\cite{yu2025multi} &
Multi-UAV &
AoI + coverage &
Graph Transformer + diffusion &
Interaction-guided trajectory prior \\
\hline

\end{tabular*}
\end{table*}

\subsubsection{Sequence-Based Routing and Decision}

Recent works further extend Transformer-enhanced reinforcement learning toward sequence-modeling paradigms, where routing and trajectory decisions are reformulated as conditional sequence prediction rather than step-wise policy learning. Decision Transformer and encoder--decoder Transformer architectures directly map state--return--constraint sequences to action trajectories, improving training stability and long-horizon credit assignment. These approaches show strong generalization across varying problem scales and objectives in UAV control, vehicle routing, and wireless resource scheduling.

Early studies mainly explore DT-style learning in joint communication--trajectory control. The authors in \cite{sun2024reinforcement} study UAV-mounted RIS system optimization by modeling joint trajectory and phase-shift control as a sequential decision process and adopting a DT with causal masking to handle non-Markovian state transitions. The sequence-based policy reduces training episodes compared with DDPG while maintaining competitive transmission performance. Similarly, the authors in \cite{zhang2025decision} propose a DT-based adaptive architecture for joint UAV flight path design, workload scheduling, and user association, where sequence modeling enables cloud pre-training and edge fine-tuning, accelerating convergence and improving cross-scenario generalization.

Beyond wireless communication control, sequence-modeling Transformers are also applied to combinatorial routing with structured constraints. The authors in \cite{munir2024route} address the electric vehicle routing problem with time windows by replacing RNN-style decoders with a Transformer using heterogeneous cross-attention between charging stations and customer nodes, improving long-range dependency modeling and generalization across instance sizes. The authors in \cite{wu2024multiobjective} further incorporate a Transformer-based policy into multi-objective vehicle routing, where sequence-aware representations enable simultaneous optimization of travel cost and service satisfaction, reducing training complexity while achieving better Pareto performance than conventional DRL.

Despite these advances, sequence-modeling--based Transformer RL still faces several limitations. Most studies rely on offline or semi-offline trajectory-style training signals and assume relatively stable sequence distributions, limiting adaptability in rapidly changing environments. In addition, constraint handling and safety guarantees are often learned implicitly rather than enforced explicitly, creating feasibility risks in real-time routing and control. Developing online-adaptive DT with explicit constraint integration and stronger robustness to distribution shift remains an open direction.

\subsubsection{Risk- and Constraint-Aware Routing and Trajectory Optimization}

Another emerging direction integrates Transformer-enhanced DRL into risk-aware, constraint-driven, and multi-objective routing and trajectory optimization. These studies formulate routing and control under Constrained Markov Decision Process (CMDP), energy, safety, congestion, or multi-objective trade-offs and use Transformer architectures to capture dependencies among constraints, goals, and agent states, improving robustness and constraint satisfaction across cooperative routing, UAV swarms, and congestion-aware network control.

Early work appears in explicitly risk- and constraint-aware routing. The authors in \cite{mondal2025risk} study cooperative between Unmanned Aerial Vehicles and Unmanned Ground Vehicles (UAV--UGV) routing with stochastic fuel consumption by modeling the task as a constrained MDP and integrating an encoder--decoder Transformer policy, where multi-head attention captures relationships among mission points and resource risks, reducing mission time and constraint violations compared with heuristic baselines.

Transformer-based constraint-aware RL is also applied to goal-driven and perception-constrained navigation. The authors in \cite{sopegno2025Transformer} augment PPO with a Transformer module for UAV navigation in GPS-denied environments, improving decisions under high-dimensional and partially observable states. The work in \cite{huang2023goal} further introduces a goal-guided Transformer that conditions scene encoding on target states, improving data efficiency, robustness, and sim-to-real generalization.

Several studies incorporate communication and energy constraints into Transformer-enhanced control. The authors in \cite{wu2025reconfigurable} jointly optimize UAV trajectories and RIS phase shifts for mobile crowd sensing, improving long-horizon energy--throughput trade-offs. The authors in \cite{wang2025transal} combine Transformer-based time-series feature extraction with RL-based path selection and bandwidth scheduling for multipath congestion control in Power IoT, improving throughput, delay, and packet loss. The work in \cite{owusu2025Transformer} similarly integrates a Temporal Fusion Transformer predictor with DQN routing for congestion-adaptive control in software-defined networks.

Risk- and constraint-aware modeling is further extended to multi-agent and adversarial swarm problems. The authors in \cite{liu2025uav} integrate Transformer self-attention into multi-agent RL for UAV swarm confrontation, mitigating sparse-gradient and local-optimum issues. The authors in \cite{tong2025rapid} apply Transformer RL to adversarial UAV swarm resource scheduling. The authors in \cite{yu2025hybrid,zhao2025integrated,yu2025multi} further extend Transformer coordination to constrained air corridors, cooperative multi-target detection, and AoI-aware multi-UAV trajectory generation, highlighting scalability in complex swarm control.

Despite these advances, several limitations remain. Most constraint-aware Transformer RL methods rely on soft penalties or reward shaping rather than explicit feasibility guarantees, limiting safety in strictly constrained routing and control. Risk and constraint signals are often embedded implicitly within attention features without structured reasoning, reducing interpretability and verifiability. Developing Transformer--RL frameworks with explicit constraint modeling, verifiable safety bounds, and online risk adaptation therefore remains an important research direction.

\subsubsection{Lessons Learned}

From the above studies, several common lessons emerge regarding Transformer architectures in deep reinforcement learning for routing and trajectory control. First, Transformer-based policies improve long-horizon dependency modeling and cross-entity coupling representation, which are critical for tasks with delayed rewards and structured state spaces. Attention mechanisms are effective in capturing spatial--temporal correlations and interaction structures that are difficult to encode with MLP- or RNN-based policies. Second, Transformer-enhanced DRL is advantageous when decisions depend on structured context, including multi-agent interactions, graph topologies, resource constraints, and goal-conditioned states. Graph Transformers, hierarchical Transformers, and Decision Transformers align model structure with problem structure, improving scalability and generalization across varying network sizes and agent populations. Third, DT-style sequence modeling reduces training instability by reframing policy learning as conditional sequence prediction, often improving convergence and cross-scenario transfer, especially with pretraining or offline trajectory data. However, Transformer-enhanced DRL increases model complexity, training cost, and data demand, and constraint handling often relies on reward shaping or soft penalties, which does not guarantee feasibility, safety, or interpretability. Future work should integrate Transformer representations with explicit constraint modeling, structured priors, and lightweight adaptation mechanisms.

%============================================
\section{Transformer-based RL for Network Security}
\label{sec:security}
Collaborative decision-making across distributed entities often requires exchanging large volumes of heterogeneous data over diverse protocols, expanding attack surfaces and complicating management and security \cite{chen2014big, abou2023mec, ranaweera2023novel}. In such environments, security threats can be broadly organized into the taxonomy adopted by this survey: (i) data integrity attacks that compromise the confidentiality, integrity, or provenance of exchanged information (e.g., eavesdropping, injection, and false-data manipulation) \cite{huang2025low, poorvi2024securing}; (ii) learning-oriented adversarial threats that target decision-making and learning/inference pipelines, including adversarial routing/manipulation and adversarial ML behaviors \cite{gottam2025graph, ibrahum2024deep}; (iii) resource and infrastructure attacks that exhaust bandwidth, computation, or energy (e.g., resource depletion) \cite{asemian2025anti-jamming, wang2025resource, ullah2025sdn}; (iv) communication and jamming attacks that degrade link reliability or secrecy under partial observability \cite{asemian2025anti, huang2025low}; and (v) others threats, including faults and microarchitectural leakage that undermine the trust anchor of edge devices \cite{peng2025graph, el2025diwall}.
\iffalse
\begin{figure}[h!]
    \centering
    \includegraphics[width=1.0\columnwidth]{figure/security_survey_taxonomy.pdf}
    \caption{Transformer-Enhanced DRL Defense Overview}
    \label{fig:placeholder}
\end{figure}
\fi
To cope with this dynamic and adversarial landscape, learning-driven defenses \cite{al2024charging, chen2024enhancing, farooq2025robust} can be used. Particulalry, the supervised methods can achieve high accuracy when abundant labels exist but often struggle against novel or evolving attacks, whereas unsupervised and semi-supervised anomaly detection reduces labeling requirements but may incur higher false positives in highly non-stationary settings. RL-based defenses address these limitations by learning adaptive policies through interaction, enabling proactive mitigation and resource-aware decision making at runtime. Recent work further augments DRL with Transformer architectures to better encode high-dimensional observations and capture long-range dependencies, which motivates the Transformer-enhanced DRL taxonomy and comparative summary in Table~\ref{tab:securitysummary}.
% ==========================
\begin{table*}[h!]
\caption{Summary and Classification of Transformer Enhanced RL Approaches for Network Security.}
\label{tab:securitysummary}
\centering
\footnotesize
\renewcommand{\arraystretch}{1.5}
\setlength{\tabcolsep}{4pt}

\resizebox{\textwidth}{!}{%
\begin{tabular}{|c|c|l|l|l|l|l|}
\hline
\textbf{Tax.} & \textbf{Ref.} & \textbf{Scenario} & \textbf{Threat Type} & \textbf{Method} & \textbf{Key Features} & \textbf{DRL+Transformer Role} \\
\hline

\multirow{9}{*}{\rotatebox{90}{\makecell[c]{Learning--Oriented \\ Adversarial}}}
& \cite{gottam2025graph} & \makecell[l]{6G D2D Routing} & \makecell[l]{Adversarial \\ Routing} & \makecell[l]{QMRL, GATN} & \makecell[l]{Graph attention, \\ evolution} & \makecell[l]{Transformer: encode routing context \\ under adversaries; DRL: learn \\ adaptive routing decisions.} \\
\cline{2-7}
& \cite{wang2026Transformer} & \makecell[l]{Spacecraft, \\ Swarm} & Threat Evasion & \makecell[l]{Transformer-DRL} & \makecell[l]{Threat--target \\ highlighting} & \makecell[l]{Transformer: highlight critical \\ threat targets to improve states; \\ DRL: learn evasion policies \\ under constraints.} \\
\cline{2-7}
& \cite{hwang2025adversarial} & \makecell[l]{Adversarial Attacks, \\ DDQN Agents} & \makecell[l]{Adaptive Evasion} & \makecell[l]{Transformer-DDQN} & \makecell[l]{Model robustness, \\ diversified architectures} & \makecell[l]{Transformer: diversified foundation \\ for adversarial training; \\ DRL (DDQN): learn robust \\ evasion policies.} \\
\cline{2-7}
& \cite{xue2025efficient} & \makecell[l]{UAV Path Planning} & \makecell[l]{Adversarial \\ Inference} & \makecell[l]{Attention-DRL \\ (Actor-Critic)} & \makecell[l]{Local attention, \\ multi-agent} & \makecell[l]{Transformer: encode agent context; \\ DRL: learn competitive/cooperative \\ path planning policies.} \\
\cline{2-7}
& \cite{farooq2025robust} & \makecell[l]{Edge-Cloud, IoT} & \makecell[l]{Adversarial \\ Detection} & \makecell[l]{PPO DRL--\\ Transformer} & \makecell[l]{Generic defensive \\ agent} & \makecell[l]{Transformer: enhance adaptive \\defensive decision-making; DRL:\\ actions as input to transformer} \\
\hline

\multirow{9}{*}{\rotatebox{90}{\makecell[c]{Resource \\ \& Infrastructure}}}
& \cite{al2024charging} & \makecell[l]{EV Charging \\ Stations} & \makecell[l]{SoC Manipulation} & \makecell[l]{Transformer-based PPO} & \makecell[l]{Adversarial agent, IDS} & \makecell[l]{Transformer: encode resource/ \\ SoC context; DRL (PPO): jointly \\ learn IDS and robust scheduling.} \\
\cline{2-7}
& \cite{lan2024deep} & \makecell[l]{Multi-objective \\ Offloading} & \makecell[l]{Resource Optimization} & \makecell[l]{Transformer-DRL \\ (PPO)} & \makecell[l]{Large action space} & \makecell[l]{Transformer: handle large \\ action spaces; DRL: \\ optimize offloading under \\ multiple objectives.} \\
\cline{2-7}
& \cite{ni2025scheduling} & \makecell[l]{FL, Wireless \\ Networks} & \makecell[l]{Jamming, Resource} & TS-TEPPO & \makecell[l]{Two-stage scheduling} & \makecell[l]{Transformer encoder: preprocess \\ inputs for actor/critic; \\ DRL: learn scheduling \\ and securing decisions.} \\
\cline{2-7}
& \cite{cao2025deep} & \makecell[l]{Industrial CPS, \\ SDN-based Platform} & \makecell[l]{APTs, \\ Moving Target Defense} & \makecell[l]{Transformer+ \\ Hier-DRL} & \makecell[l]{Attack perception, \\ IP shuffling} & \makecell[l]{Transformer: attack perception \\ and context extraction; \\ Hier--DRL: coordinate multiphase \\ moving target defense.} \\
\cline{2-7}
& \cite{bhutto2022reinforced} & \makecell[l]{SDN Edge Cloud} & \makecell[l]{VSI-DDoS} & \makecell[l]{Transformer-DQN} & \makecell[l]{Temporal attention, \\ burst traffic detection} & \makecell[l]{Transformer: capture short-lived \\ burst traffic; DRL: \\ learn resilient detection policies.} \\
\hline

\multirow{6}{*}{\rotatebox{90}{\makecell[c]{Communication  \\ \& Jamming}}}
& \cite{huang2025low} & \makecell[l]{LEO Sat--Maritime} & \makecell[l]{Eavesdrop, Jamming} & TransSAC & \makecell[l]{Multi-objective, \\ UAV jamming} & \makecell[l]{Transformer: capture global \\ dependencies in sequences; \\ DRL: optimize secrecy \\ rate and energy consumption.} \\
\cline{2-7}
& \cite{asemian2025anti} & \makecell[l]{5G O--RAN, MEC} & Jamming & \makecell[l]{Transformer Actor- \\ Critic (A2C)} & \makecell[l]{Jamming estimator} & \makecell[l]{Transformer: support jamming \\ estimation/temporal context; \\ DRL: learn \\ anti-jamming scheduling.} \\
\cline{2-7}
& \cite{wang2025resource} & \makecell[l]{CIoV, IRS--assisted} & \makecell[l]{Jamming, \\ Physical attack} & \makecell[l]{HMA-TD3QN} & \makecell[l]{Heterogeneous, \\ Self-attention} & \makecell[l]{Transformer: highlight \\ heterogeneous features; \\ DRL: learn robust \\ resource allocation under attacks.} \\
\hline

\multirow{3}{*}{\rotatebox{90}{\makecell[c]{Data \\ Integrity}}}
& \cite{khan2025Transformer} & \makecell[l]{CAVs, Trajectory} & Data Poisoning & \makecell[l]{Transformer + FedQL} & \makecell[l]{Blockchain, \\ Anomaly detection} & \makecell[l]{Transformer: enhance representation \\ learning for detection; \\ DRL: learn collaborative \\ defense policy.} \\
\cline{2-7}

& \cite{chen2024enhancing} & \makecell[l]{IoV, Prod. Planning} & \makecell[l]{Data Security, \\ Scalability} & \makecell[l]{TFRL (FedRL)} & \makecell[l]{Transformer--based \\ Federated learning \\ for RL} & \makecell[l]{Transformer: enhance representation \\ learning for scalability; \\ DRL: learn policies \\ under distributed data.} \\
\hline

\multirow{2}{*}{\rotatebox{90}{\makecell[c]{Others}}}
& \cite{peng2025graph} & \makecell[l]{UAV Swarm, \\ Faults} & \makecell[l]{Imbalanced Faults \\ (in sensor and \\ physical attacks)} & GTDRL & \makecell[l]{Graph conv, \\ Self-attention} & \makecell[l]{Transformer (Graph conv capture): \\ spatial relations and rare-event \\ cues; DRL: learn fault detection \\ under imbalance.} \\
\hline
\end{tabular}%
}
\end{table*}
% ==========================

\subsection{Learning-Oriented Adversarial Attacks}
Deep learning networks are highly vulnerable to adversarial attackers, who employ a variety of techniques to introduce carefully crafted noise that degrades model performance and alters inference outcomes. Numerous attack methods have been proposed to compromise the deep learning and artificial intelligence systems. The primary objective of such attacks is to violate the system integrity through unauthorized interference, either by manipulating the model’s inputs or by altering the signals processed by the system, ultimately leading to incorrect predictions and faulty decision-making \cite{ibrahum2024deep}. 

In the context of D2D communications, the authors in \cite{gottam2025graph} address adversarial routing attacks using a Quantum Multi-agent Reinforcement Learning (QMRL) framework combined with Graph Attention Networks (GATN). The attention mechanism encodes the dynamic routing context and malicious behavior patterns, allowing the DRL agents to learn adaptive routing decisions that bypass compromised nodes and maintain network connectivity. Recently, modern cyber environments have become complicated and distributed, leading to growing number of cyber threats. In another aspect, the work of \cite{hwang2025adversarial} includes a Transformer along with Gated Recurrent Unit (GRU), Long-Short Term Memory (LSTM), and Recurrent Neural Network (RNN) models, providing a diversified architectural foundation for training and evaluating DDQN-based DRL adversarial agents, thereby enabling a more comprehensive assessment of model robustness under adaptive evasion attacks. In the work of \cite{xue2025efficient}, UAVs may be exposed to adversarial agents that attempt to infer their true mission objectives by observing flight trajectories. To address such direct adversarial threats, the authors propose a deceptive path planning framework that integrates RL for the path optimization. Specifically, a local attention mechanism for multiple agents is combined with an actor-critic policy gradient algorithm, enabling UAVs to learn both competitive and cooperative strategies. The trained policies demonstrate strong generalization capability across different environments without requiring additional fine-tuning.

In \cite{wang2026Transformer}, the authors formulate spacecraft threat evasion as a DRL-based decision-making problem. A Transformer-based attention mechanism is integrated to attend to critical threat targets, thereby improving state representation for policy learning. This combination enhances evasion efficiency, policy generalization, and resource utilization in multi-target and swarm-based scenarios.
However, due to the dynamic nature of modern systems, the long-term persistence of attacks, their intermittent and stealthy behavior, and the adaptive learning capabilities of attackers, these methods gradually become less effective when confronting such evolving threats. In this context, Transfomer-based PPO and RL--XGBoost, with their ability to perform continuous online learning and capture the system state before executing defensive actions, have emerged as a highly promising direction for future adversarial detection and mitigation. The actions of the models are first collected as inputs to a Transformer, and then fed into a single agent, referred to as the generic defensive blue agent~\cite{farooq2025robust}. 
%Transformer-enhanced DRL approaches enable adaptive decision-making under uncertainty and offer the potential to proactively respond to sophisticated and evolving attack strategies in distributed edge-cloud environments, especially in IoT and industrial IoT-based systems.

\subsection{Resource and Infrastructure Attacks}
Transformer-enhanced DRL has also been applied to attacks that manipulate or exhaust communication, computation, and energy resources. In EV charging infrastructures, the work in \cite{al2024charging} considers the attack not only as an adversarial learning problem but also as a manipulation of system resource information. Specifically, the authors model state-of-charge (SoC) falsification as a zero-sum game between an adversarial agent and a defender. A PPO-based Transformer policy is employed to jointly train an intrusion detection mechanism and a robust charging scheduler, thereby mitigating malicious exploitation of shared energy resources. In satellite-terrestrial edge systems, the work in \cite{lan2024deep} exploits a high-dimensional PPO agent with self-attention to optimize privacy-preserving task offloading under multiple conflicting objectives including delay, energy consumption, and reliability, making the system resilient to adversarial or misconfigured resource demands.
% In \cite{al2024charging}, the work proposes an adversarial DRL agent scheme designed to evade deception in a hierarchical structure. Another layer of this structure is the LSTM or transformer model, which is responsible for creating complex attack strategies.

Regarding SDN-based security, the study in \cite{bhutto2022reinforced} introduces a reinforced transformer learning framework for detecting Very-Short Intermittent Distributed Denial of Service (VSI-DDoS) attacks in edge clouds. By combining temporal attention mechanisms with DQN, the approach captures short-lived burst traffic patterns that degrade QoS while evading conventional detection methods, thereby enhancing service-layer DDoS resilience in edge environments. In addition, Deep-Shield \cite{cao2025deep} is evaluated on an SDN-based platform that primarily serves as an enabling infrastructure for implementing dynamic reconfiguration (e.g., IP shuffling), rather than addressing SDN-specific control-plane vulnerabilities. The core contribution lies in integrating a Transformer-based attack perception module with hierarchical DRL to coordinate multiphase moving target defense against Advanced Persistent Threats (APTs) in industrial CPS. Therefore, the work is more accurately characterized as an AI-driven adaptive infrastructure defense mechanism rather than an SDN-native or physical-layer security solution.

To enhance the efficiency of asynchronous federated learning, the work in \cite{ni2025scheduling} proposes a Transformer-based PPO that addresses the limited computation and computational resources, as well as energy consumption and security threats. The main proposal is based on a two--stage DRL with transformer--encoder PPO (TS--TEPPO). Therein, both the actor and critic networks receive the preprocessing data handled by a transformer encoder, which may help to provide the most relevant information to generate correct action and state values. Sequentially, the model rapidly converges and reaches the desired utility value in a small number of episodes.

\subsection{Communication and Jamming Attacks}
The work in \cite{asemian2025anti} gains higher performance of anti-jamming task scheduling in MEC-open radio network (MEC-O-RAN) by hierarchical DRL combined with Transformer--based control. In this work, a Transformer-based A2C is used, where the sequential decision making has extremely low predictability errors compared to the standard DRL methods. In the scenario of a UAV with low altitude, the study in \cite{huang2025low} proposes a Transformer-enhanced SAC algorithm, which represents a generative artificial intelligence-enabled approach designed to solve the multi-objective optimization problem of maximizing secrecy rates while minimizing UAV energy consumption. Facing the strong temporal correlations and high-dimensional decision spaces inherent in dynamic maritime environments, the authors integrate a Transformer-based learning strategy that utilizes self-attention mechanisms to capture global dependencies across sequences of states and actions. This architecture employs positional encoding to provide temporal context and multi-head attention to decompose complex action dimensions, effectively preventing the policy from falling into a local optimum. The proposed scheme has been proven to deliver high performance in secrecy rate and energy consumption in UAV operations.
% \cite{poorvi2024securing}. 
Considering  Intelligent Reflecting Surface (IRS)-assisted interweave Connected Internet of Vehicles (CIoV) uplink communication, the work in \cite{wang2025resource} addresses the posed problem by a multi-agent DRL framework enhanced by a Transformer architecture. The system consists of multiple heterogeneous agents belonging to two distinct types, each representing a vehicular user. Every agent incorporates a transformer model that processes summarized experience memory information and feeds it into the IMPALA (Importance Weighted Actor-Learner Architecture) framework. In this approach, the Transformer plays a crucial role in improving the performance of DRL, particularly in enhancing communication security.

\subsection{Data Integrity Attacks}
Many studies have further investigated this attack paradigm, among which deep RL--based approaches have gained significant attention. For instance, the study in \cite{khan2025Transformer} proposes a federated Q-learning (FedQL) framework enhanced by Transformers to detect data poisoning attacks in identifying connected autonomous vehicle (CAV) trajectories. By integrating blockchain for secure model updates and using Transformers to extract temporal features from trajectory data, the method effectively identifies malicious sensing data while preserving privacy. Addressing data security and scalability in the Internet of Vehicles (IoV), the authors in \cite{chen2024enhancing} introduce a Transformer-based Federated Reinforcement Learning (TFRL) approach for production planning. The Transformer module enhances representation learning to handle high-dimensional state spaces, while the federated learning framework enables distributed agents to learn collaborative policies without sharing raw local data, thereby mitigating data leakage risks.

\subsection{Other Security Issues}
For UAV swarms, the study in \cite{peng2025graph} combines graph-based DRL with self-attention to detect imbalanced faults in distributed sensors and actuators, improving rare-event detection and F1-score in safety-critical missions. Beyond wireless channels, microarchitectural threats such as hardware Trojans, side-channel leakage, and speculative-execution abuse compromise the physical trust anchor of edge devices. Recent works explore RL agents that monitor low-level performance and power counters to detect anomalous patterns associated with microarchitectural and physical-layer attacks and dynamically adapt defenses such as cache partitioning, speculation throttling, and execution randomization. Although these approaches face challenges in terms of training cost and deployment overhead on resource-constrained hardware, they highlight how learning-based policies can complement secure-by-design microarchitectures and runtime monitors in safeguarding sensing and computation at the physical layer \cite{el2025diwall, wang2023security, oruma2023security}.

\subsection{Lessons Learned}
From the surveyed studies, four key factors explain the effectiveness of combining DRL and Transformer architectures in network security:
\emph{1) Sequential and Temporal Decision Structures:}
DRL learns adaptive defense policies through interaction with dynamic environments, while Transformer modules model long-term temporal dependencies and extract informative state representations. This synergy enables the detection of persistent and stealthy attacks and supports trajectory-dependent decision making.
\emph{2) Multi-Agent and Large-Scale Interaction:}
DRL enables distributed and multi-agent learning, while the self-attention mechanism of Transformers efficiently models inter-agent relationships, supporting robust and generalizable defense policies in dynamic and heterogeneous environments.
\emph{3) Structured and Combinatorial Decisions:}
DRL solves complex optimization problems, coordinates defense, and allocates resources, while Transformers capture dependencies among structured actions, enabling effective decision making in multi-stage tasks.
\emph{4) Multi-Modal and Context-Aware Control:}
DRL makes decisions based on multi-source data, and attention-based fusion in Transformers integrates information from sensing, prediction, and system states, enhancing adaptability and context-aware security.

%====================================
\section{Conclusions and Future Works}
\label{sec:conclusion}
%====================================
This paper has provided a comprehensive review of the fundamentals of Transformer-based RL schemes, as well as their applications in communication networks. These algorithms are primarily classified into two paradigms: Architecture Enhancement, which utilizes Transformers to replace traditional recurrent units in capturing complex temporal dependencies, and Sequence Modeling, which treats RL as a conditional trajectory generation task. To provide a systematic understanding of these paradigms, we have evaluated and discussed various implementations of Transformer-based RL across important domains of networking, including resource management, computation offloading, routing and trajectory control, and network security. Our analysis highlights that the self-attention mechanism is a pivotal tool for addressing the traditional shortcomings of DRL in dynamic wireless environments, offering several notable advantages as follows. First, unlike traditional recurrent units, Transformers excel at capturing global context and long-term temporal relationships, which are essential for reliable decision-making in time-varying channels. Second, the architecture's inherent flexibility allows it to process diverse data modalities, leading to more comprehensive and richer state representations in complex IoT and 6G scenarios. Third, through permutation-invariant attention, Transformer-based RL models facilitate seamless coordination among numerous network entities, such as UAVs or satellite nodes, without being restricted by fixed input dimensions. Fourth, by adopting offline sequence modeling techniques, Transformer-based RL schemes can extract high-quality policies from noisy datasets, often achieving faster convergence than standard DRL methods. 

By bridging the gap between Transformer architectures and their practical application in networking, this survey establishes a solid foundation for future comparative research and the development of hybrid RL designs in large-scale communication networks. Moreover, based on our review, we identify promising research directions as follows. 

\subsection{Transformer-Based RL for Latency-Sensitive and Reliability-Critical Applications}
Latency-sensitive and reliability-critical applications, such as 
ultra-reliable low-latency communications (URLLC), require fast and 
accurate decision making under dynamic network conditions. Transformer-based learning architectures can provide a promising solution for such scenarios. By leveraging attention mechanisms, Transformer architectures are able to capture complex dependencies among users and network states, enabling more informed decision making \cite{xue2024cooperative}. Moreover, their highly parallelizable structure supports fast inference compared with sequential models, making them attractive for latency-sensitive wireless applications in 
future networks.

\subsection{Transformer-Based RL for Intelligent Surface-Assisted Wireless Systems}
Future wireless networks are expected to widely deploy intelligent 
surfaces, such as reconfigurable intelligent surfaces (RIS) and stacked 
intelligent metasurfaces (SIM), to enhance signal propagation and 
coverage. However, optimizing the configurations of these surfaces is 
challenging due to the large number of controllable elements and the 
strong coupling between surface parameters and wireless channels. 
Transformer-based RL provides an effective solution by leveraging attention mechanisms to model complex interactions 
between network states and surface configurations, thereby enabling more 
efficient optimization in intelligent surface-assisted wireless systems \cite{adam2025generative}.

\subsection{Transformer-Based RL for Vehicular Networks}
Future vehicular systems are expected to exploit multimodal sensory data, 
such as radar, vision, LiDAR, and localization information, to enhance 
communication and sensing capabilities. Transformer-based RL frameworks are particularly useful in such environments, as they can effectively model temporal correlations in highly dynamic vehicular scenarios and leverage historical observations to improve prediction and decision making \cite{wu2024traffic}. This capability allows for more accurate target tracking and resource allocation in high-mobility vehicular networks envisioned for future wireless systems.

\bibliographystyle{IEEEtran}
\bibliography{Transformer_REF}{}
% \addbibresource{DM_REF.bib}

%\end{IEEEbiography}

%\]
\end{document}